\documentclass[a4paper,10pt]{article}
\pdfoutput=1
\usepackage[utf8x]{inputenc}
\usepackage{graphicx}
\usepackage{amsmath}
\usepackage{amssymb}
\usepackage{bbold}
\usepackage{geometry}
\usepackage{afterpage}
\usepackage{cite}
\def\lsim{\;\raise0.3ex\hbox{$<$\kern-0.75em\raise-1.1ex\hbox{$\sim$}}\;}
\def\gsim{\;\raise0.3ex\hbox{$>$\kern-0.75em\raise-1.1ex\hbox{$\sim$}}\;}

\title{A New Tool for the study of the CP-violating NMSSM}
\author{Florian Domingo}
\date{{\em Deutsches Elektronen Synchrotron (DESY),\\ Notkestraße 85, D--22607 Hamburg, Germany}}

\begin{document}

\maketitle
\vspace{-5cm}\rightline{DESY 15-041}\vspace{5cm}
\begin{abstract}
Supersymmetric extensions of the Standard Model open up the possibility for new types of CP-violation. We consider the case of the Next-to-Minimal 
Supersymmetric Standard Model where, beyond the phases from the soft lagrangian, CP-violation could enter the Higgs sector directly at tree-level through
complex parameters in the superpotential. We 
develop a series of Fortran subroutines, cast within the public tool \verb|NMSSMTools| and allowing for a phenomenological analysis of the 
CP-violating NMSSM. This new tool performs the computation of the masses and couplings of the various new physics states in this model: leading 
corrections to the sparticle masses are included; the precision for the Higgs masses and couplings reaches the full one-loop and leading two-loop order. 
The two-body Higgs and top decays are also attended. We use the public tools \verb|HiggsBounds| and \verb|HiggsSignals| to test the Higgs sector. 
Additional subroutines check the viability of the sparticle spectrum in view of LEP-limits and constrain the phases of the model via a confrontation to 
the experimentally measured Electric Dipole Moments. These tools will be made publicly available in the near future. In this paper, we detail the 
workings of our code and illustrate its use via a comparison with existing results. We also consider some consequences of CP-violation for the NMSSM 
Higgs sector.
\end{abstract}

\section{Introduction}

After the discovery of a signal at a mass of about $125$~GeV in the LHC Higgs searches \cite{Aad:2012tfa,Chatrchyan:2012ufa}, the question of 
the identification of the associated state(s) and the underlying physics remains open. While the general properties are consistent so far with 
those expected for the Higgs boson of the Standard Model (SM), a wide range of alternatives could equally well fit the experimental data. In 
particular, softly-broken supersymmetric (SUSY) extensions of the SM \cite{SUSY} count among the appealing options to solve the Hierarchy Problem 
\cite{HierProb} and allow for a smooth transition to higher energy physics (e.g.\ Grand Unification, neutrino physics or weakly-coupled dark matter). 
The scarceness of evidence for new physics effects in precision physics or direct searches should also be weighed by the considerations that 
SUSY-inspired models offer a SM-like decoupling regime, but also that complex mechanisms in the Higgs or SUSY sectors -- see e.g.\ 
\cite{Ellwanger:2014hia} -- may account for this relative invisibility thus far.

The Next-to-Minimal Supersymmetric Standard Model (NMSSM), a singlet-extension of the simplest viable SUSY-inspired extension of the SM \cite{NMSSM}, 
has raised renewed interest ever since the Higgs discovery, notably due to its properties in the Higgs sector, e.g.\ allowing for an uplift of
the mass of the SM-like Higgs related to F-terms or to the mixing of this state with a lighter singlet \cite{natsusy}. The original motivation
for this singlet extension rests with the `$\mu$-problem' of the MSSM \cite{Kim:1983dt}, which can be solved elegantly if this $\mu$-term is generated 
dynamically, via a singlet vacuum expectation value (v.e.v.) \cite{fayet}. Correspondingly, the $\mathbb{Z}_3$-conserving version of the NMSSM 
-- allowing only cubic terms in the superpotential -- is the most studied form of this model, while more general singlet couplings can be justified
by higher-energy considerations -- see e.g.\ \cite{Ellwanger:2008py,Lee:2011dya}. Another usual feature in SUSY extensions of the SM is R-parity,
which both constrains the possibility of baryon-number violation and provides a stable SUSY particle, hence a dark-matter candidate.

A troubling fact rests with the observation that several NMSSM parameters -- especially in the Higgs sector -- can take complex values, hence lead 
to CP-violation beyond that in the quark sector. On the one hand, CP-violation is known as a cosmological necessity for baryogenesis. On the other, 
it receives severe limits at the phenomenological level, from the non-observation of Electric Dipole Moments (EDM's; see e.g.\ \cite{Arbey:2014msa}). In this paper,
we aim at presenting a tool which allows to study the NMSSM with complex parameters within the framework of the public code \verb|NMSSMTools|
\cite{NMSSMTools}. In a first step, we will focus on the $\mathbb{Z}_3$-conserving version although we plan on a generalization to tadpole and
quadratic couplings of the singlet in the future.

The current version of \verb|NMSSMTools| allows to perform several operations in connection with the spectrum of the CP-conserving NMSSM: in particular,
it computes radiative corrections to the Higgs and SUSY spectrum, calculates the widths of Higgs decays or confronts the NMSSM parameter space to 
theoretical -- e.g.\ vacuum stability -- or phenomenological -- e.g.\ Higgs searches, $B$-physics -- limits. Several other tools aiming at the 
calculation of radiative corrections to the NMSSM spectrum have been developed in the past few years, e.g.\ \verb|NMSSMCALC| \cite{NMSSMCALC} or 
\verb|SoftSUSY| \cite{SoftSUSY}. The latter focuses on the CP-conserving NMSSM, while \verb|NMSSMCALC| allows for CP-violation but specializes in 
corrections to the Higgs spectrum. Multi-purpose programs can also be used in connection to the (complex) NMSSM and  allow for a number of similar 
manipulations, provided the implementation of a model-file as input: this applies to \verb|SPHENO| \cite{Spheno} or \verb|FlexibleSUSY| 
\cite{FlexibleSUSY} -- which are usually coupled to \verb|SARAH| \cite{SARAH} to produce their input file. The 
Higgs-sector of the complex NMSSM and its phenomenology have been previously considered in several earlier works \cite{CNMSSM_Higgs}.

Our goal consists in generalizing \verb|NMSSMTools| to the CP-violating case. While this task remains far from complete, the tools which we present here 
already allow for numerous operations: radiative corrections to the SUSY and the Higgs masses are implemented -- so far, only the leading double-log 
corrections (beyond the full one-loop) are taken into account at two-loop order in the Higgs spectrum -- ; Higgs and top two-body decays are computed; 
phenomenological limits from LEP SUSY searches or Higgs physics are tested -- the latter via an interface with the public tools \verb|HiggsBounds| 
\cite{HiggsBounds} and \verb|HiggsSignals| \cite{HiggsSignals} --; finally we designed a subroutine to estimate the EDM's. All these routines should 
become available on the \verb|NMSSMTools|  website \cite{NMSSMTools} in the near future. This paper is intended to serve as a presentation of the calculations
implemented in our tool, as well as a short illustration of its uses. In the following section, we will detail the characteristics of the model
under study, the underlying assumptions and the tree-level spectrum. The third section will present the chain of subroutines that we designed and the 
operations which they carry out. Finally, we will consider phenomenological consequences and compare some of our results to the predictions of existing tools, 
before we conclude.

\section{Model, Phase-counting and Tree-level}\label{secmod}
In this section, we present the details of the model under consideration, our notations as well as the spectrum at tree-level.

\subsection{The CP-violating NMSSM}
The NMSSM is a supersymmetry-inspired extension of the SM with soft SUSY-breaking terms. It differs from the minimal supersymmetric extension 
of the SM, the MSSM, in that it includes, in addition to the two Higgs $SU(2)_L$-doublet superfields $\hat{H}_u$ and $\hat{H}_d$ with opposite
hypercharge $\pm1$, a supplemental gauge-singlet chiral superfield $\hat{S}$. While the couplings of this singlet may take a more complex form
in the general case, we will be considering only the R-parity and $\mathbb{Z}_3$-conserving NMSSM here, which is characterized by the following 
superpotential and SUSY-breaking terms:
\begin{align}
 & W=\lambda e^{\imath\phi_{\lambda}}\hat{S}\hat{H}_u\cdot \hat{H}_d+\frac{\kappa}{3} e^{\imath\phi_{\kappa}}\hat{S}^3-\hat{H}_u\cdot \hat{Q}_L[Y_u]\hat{U}_R^c+\hat{H}_d\cdot \hat{Q}_L[Y_d]\hat{D}_R^c+\hat{H}_d\cdot \hat{L}_L[Y_e]\hat{E}_R^c\label{supo}\\
 & -{\cal L}_{\mbox{\tiny soft}}=-M_1 e^{\imath\phi_{M_1}}\tilde{b}\tilde{b}-M_2 e^{\imath\phi_{M_2}}\tilde{w}_{\alpha}\tilde{w}_{\alpha}-M_3 e^{\imath\phi_{M_3}}\tilde{g}_{a}\tilde{g}_{a}+h.c.\label{soft}\\
 &\null\hspace{0.8cm}+m_{H_u}^2|H_u|^2+m_{H_d}^2|H_d|^2+m_{S}^2|S|^2+Q_L[m_{Q}^2]Q^{\dag}_L+U_R^{c\,\dag}[m_{U}^2]U_R^c+D_R^{c\,\dag}[m_{D}^2]D_R^c+L_L[m_{L}^2]L^{\dag}_L+E_R^{c\,\dag}[m_{E}^2]E_R^c\nonumber\\
 &\null\hspace{0.8cm}+\lambda A_{\lambda}e^{\imath\phi_{A_{\lambda}}}SH_u\cdot H_d+\frac{\kappa}{3} A_{\kappa}e^{\imath\phi_{A_{\kappa}}}S^3-H_u\cdot Q_L[Y_uA_u]U_R^c+H_d\cdot Q_L[Y_dA_d]D_R^c+H_d\cdot L_L[Y_eA_e]E_R^c+h.c.\nonumber
\end{align}
The `matter' (super)fields\footnote{We will omit the $\hat{}$ distinguishing the superfields from their scalar component, from now on.} 
$Q_L$, $U_R^c$, $D_R^c$, $L_L$, $E_R^c$ should be understood as summed over generations and the parameters within brackets should correspondingly 
be seen as (complex) matrices. `$\cdot$' denotes the usual $SU(2)_L$ product. $\tilde{b}$, $\tilde{w}_{\alpha}$ and $\tilde{g}_{a}$
stand for the $U(1)_Y$, $SU(2)_L$ and $SU(3)_c$ gauginos, respectively. In the following $g'$, $g$ and $g_S$ will denote the corresponding gauge 
couplings and $\alpha_S\equiv g_S^2/4\pi$. While the $\mathbb{Z}_3$-conserving NMSSM offers the simplest solution to the $\mu$-problem of the MSSM, 
the inclusion of $\mathbb{Z}_3$-violating terms can be justified from higher-energy considerations \cite{Ellwanger:2008py,Lee:2011dya} and turns up 
as a phenomenological necessity in view of the domain-wall problem. Our restriction to the $\mathbb{Z}_3$-conserving lagrangian follows considerations 
of simplicity and our work shall be extended to the $\mathbb{Z}_3$-violating case in the near future: we discuss in appendix \ref{Z3violation} how this 
can be easily achieved.

The minimization of the scalar potential will generate Higgs vacuum expectation values (v.e.v.'s) so that we may write the Higgs (super)fields in
terms of their (real and positive) v.e.v.'s $s$, $v_u$, $v_d$, and their charged and neutral components:
\begin{equation}
S=e^{\imath\phi_s}\left(s+\frac{h_s^0+\imath a_s^0}{\sqrt{2}}\right)\hspace{2cm}
H_u= e^{\imath\phi_u}\begin{pmatrix}H_u^+\\v_u +\frac{h_u^0+\imath a_u^0}{\sqrt{2}}\end{pmatrix}\hspace{2cm}
H_d= e^{\imath\phi_d}\begin{pmatrix}v_d +\frac{h_d^0+\imath a_d^0}{\sqrt{2}}\\H_d^-\end{pmatrix}
\end{equation}
The three `dynamical' phases $\phi_s$, $\phi_u$ and $\phi_d$ add to the `static' phases appearing in the lagrangian density (Eqs.\ref{supo},\ref{soft}). 
From now on, we will make the following replacements in our notations (which amounts to a redefinition of the Higgs fields):
\begin{align}
 &S\leftarrow \left(s+\frac{h_s^0+\imath a_s^0}{\sqrt{2}}\right)&H_u\leftarrow \begin{pmatrix}H_u^+\\v_u +\frac{h_u^0+\imath a_u^0}{\sqrt{2}}\end{pmatrix}\hspace{1cm}&H_d\leftarrow \begin{pmatrix}v_d +\frac{h_d^0+\imath a_d^0}{\sqrt{2}}\\H_d^-\end{pmatrix}\nonumber\\
 &\phi_{\lambda}\leftarrow \varphi_{\lambda}\equiv\phi_{\lambda}+\phi_s+\phi_u+\phi_d &[Y_u]\leftarrow [Y_u]e^{\imath\phi_u} \hspace{2cm}&[Y_uA_u]\leftarrow [Y_uA_u]e^{\imath\phi_u}\nonumber\\
 &\phi_{\kappa}\leftarrow \varphi_{\kappa}\equiv\phi_{\kappa}+3\phi_s &[Y_d]\leftarrow [Y_d]e^{\imath\phi_d} \hspace{2cm}&[Y_dA_d]\leftarrow [Y_dA_d]e^{\imath\phi_d}\nonumber\\
 &\phi_{A_{\lambda}}\leftarrow \varphi_{1}\equiv\phi_{A_{\lambda}}+\phi_s+\phi_u+\phi_d &[Y_{e}]\leftarrow [Y_{e}]e^{\imath\phi_d} \hspace{2cm}&[Y_{e}A_e]\leftarrow [Y_{e}A_e]e^{\imath\phi_d}\nonumber\\
 &\phi_{A_{\kappa}}\leftarrow \varphi_{2}\equiv\phi_{A_{\kappa}}+3\phi_s & & 
\end{align}

The Yukawa matrices may be written in terms of (real and positive) matrices $Y_u$, $Y_d$, $Y_e$, diagonal in flavour space, using unitary 
transformations:
\begin{equation}
 [Y_u]=X_L^uY_uX_R^u\hspace{1cm};\hspace{1cm}[Y_d]=X_L^dY_dX_R^d\hspace{1cm};\hspace{1cm}[Y_e]=X_L^eY_eX_R^e
\end{equation}
Redefining the quark and lepton (super)fields accordingly, 
\begin{equation}
 Q_L\leftarrow\begin{pmatrix}U_LX^{u\,\dag}_L\\D_LX^{d\,\dag}_L\end{pmatrix}\hspace{0.5cm};\hspace{0.5cm}U_R^c\leftarrow X_R^{u\,\dag}U_R^c\hspace{0.5cm};\hspace{0.5cm}D_R^c\leftarrow X_R^{d\,\dag}D_R^c
\hspace{0.5cm};\hspace{0.5cm}L_L\leftarrow L_LX_L^{e\,\dag}\hspace{0.5cm};\hspace{0.5cm}E_R^c\leftarrow X_R^{e\,\dag}E_R^c
\end{equation}
and introducing the Cabibbo-Kobayashi-Maskawa (CKM) matrix $V_{CKM}\equiv X^{u\,\dag}_LX^{d}_L$, the superpotential of Eq.\ref{supo} now reads:
\begin{equation}
 W=\lambda e^{\imath\varphi_{\lambda}}SH_u\cdot H_d+\frac{\kappa}{3} e^{\imath\varphi_{\kappa}}S^3-H_u\cdot \begin{pmatrix}U_L\\D_LV_{CKM}^{\dag}\end{pmatrix}Y_uU_R^c+H_d\cdot \begin{pmatrix}U_LV_{CKM}\\D_L\end{pmatrix}Y_dD_R^c+H_d\cdot L_LY_eE_R^c\label{supot}
\end{equation}

Finally, we make the following assumptions to ensure minimal flavour violation in the sfermion sector:
\begin{itemize}
 \item $X_L^{u\,\dag}[m^2_{Q}]X_L^u\simeq X_L^{d\,\dag}[m^2_{Q}]X_L^d\equiv m^2_Q$, where $m^2_Q$ is a diagonal (and, without loss of generality, real) 
matrix in flavour space. The 
approximation `$\simeq$' only holds for a matrix proportional to the identity, in the strict sense, but is viable, considering that the CKM matrix is 
hierarchical. Note that we will assume degeneracy for the first two generations of sfermions.
\item $X_R^u[m^2_{U}]X_R^{u\,\dag}\equiv m^2_{U}$, $X_R^d[m^2_{D}]X_R^{d\,\dag}\equiv m^2_{D}$, $X_L^{e\,\dag}[m^2_{L}]X_L^e\equiv m^2_{L}$, $X_R^e[m^2_{E}]X_R^{e\,\dag}\equiv m^2_{E}$ 
are assumed diagonal.
\item $X_L^{u\,\dag}[Y_uA_u]X_R^{u\,\dag}\equiv Y_uA_ue^{\imath\varphi_{A_u}}$, $X_L^{d\,\dag}[Y_dA_d]X_R^{d\,\dag}\equiv Y_dA_de^{\imath\varphi_{A_d}}$
and $X_L^{e\,\dag}[Y_eA_e]X_R^{e\,\dag}\equiv Y_eA_ee^{\imath\varphi_{A_e}}$ are also treated as diagonal in flavour-space.
\end{itemize}
Consequently, the soft SUSY-breaking lagrangian of Eq.\ref{soft} reduces to:
\begin{align}
 & -{\cal L}_{\mbox{\tiny soft}}=-M_1 e^{\imath\phi_{M_1}}\tilde{b}\tilde{b}-M_2 e^{\imath\phi_{M_2}}\tilde{w}_{\alpha}\tilde{w}_{\alpha}-M_3 e^{\imath\phi_{M_3}}\tilde{g}_{a}\tilde{g}_{a}+h.c.\label{softlag}\\
 &\null\hspace{1.5cm}+m_{Q}^2\left(U^{\dag}_LU_L+D^{\dag}_LD_L\right)+m_{U}^2U_R^{c\,\dag}U_R^c+m_{D}^2D_R^{c\,\dag}D_R^c+m_{L}^2\left(N^{\dag}_LN_L+E^{\dag}_LE_L\right)+m_{E}^2E_R^{c\,\dag}E_R^c\nonumber\\
 &\null\hspace{1.5cm}+m_{H_u}^2|H_u|^2+m_{H_d}^2|H_d|^2+m_{S}^2|S|^2+\left[\lambda A_{\lambda}e^{\imath\varphi_1}SH_u\cdot H_d+\frac{\kappa}{3} A_{\kappa}e^{\imath\varphi_2}S^3+h.c.\right]\nonumber\\
 &\null\hspace{1.5cm}-Y_uA_ue^{\imath\varphi_{A_u}}H_u\cdot Q_LU_R^c+Y_dA_de^{\imath\varphi_{A_d}}H_d\cdot Q_LD_R^c+Y_eA_ee^{\imath\varphi_{A_e}}H_d\cdot L_LE_R^c+h.c.\nonumber
\end{align}

Eqs.\ref{supot} and \ref{softlag} fully characterize the model that we will be considering from now on -- note that the three latter terms of Eq.\ref{supot}
as well as the second and fourth lines of Eq.\ref{softlag} are still implicitly summed over fermion generations. All the phases have been explicited and reduce, at 
this level, to four phases in the Higgs sector -- $\varphi_{\lambda}$, $\varphi_{\kappa}$, $\varphi_1$, $\varphi_2$ ; we will see that the minimization
conditions further constrain these, as could be expected from the `dynamical' nature of some phases --, three gaugino phases -- $\phi_{M_1}$, $\phi_{M_2}$, 
$\phi_{M_3}$ --, three sfermion phases per generation -- $\varphi_{A_u}$, $\varphi_{A_d}$, $\varphi_{A_e}$ -- and the CKM phase finally. Given that
we will neglect the Yukawa couplings of the first two generations, only the sfermion phases of the third generation will intervene in practice.

\subsection{The tree-level Higgs sector}
The Higgs potential collects terms from the soft lagrangian (Eq.\ref{softlag}), F-terms from the superpotential (Eq.\ref{supot}) and D-terms
from the gauge interactions. We obtain:
\begin{multline}
{\cal V}_{H}= m_{H_u}^2|H_u|^2+m_{H_d}^2|H_d|^2+m_{S}^2|S|^2+\lambda A_{\lambda}\left[e^{\imath\varphi_1}SH_u\cdot H_d+h.c.\right]+\frac{\kappa}{3}A_{\kappa}\left[e^{\imath\varphi_2}S^3+h.c.\right]\\
+\lambda^2\left[|S|^2\left(|H_u|^2+|H_d|^2\right)+|H_u\cdot H_d|^2\right]+\kappa\lambda\left[e^{\imath(\varphi_{\lambda}-\varphi_{\kappa})}S^{*\,2}H_u\cdot H_d+h.c.\right]+\kappa^2|S|^4\\
+\frac{g'^2+g^2}{8}\left[|H_u|^2-|H_d|^2\right]^2+\frac{g^2}{2}|H_d^{\dag}H_u|^2\label{tlHiggspot}
\end{multline}
The neutral part reduces to:
\begin{multline}
{\cal V}_{H^0}= m_{H_u}^2|H_u^0|^2+m_{H_d}^2|H_d^0|^2+m_{S}^2|S|^2-\lambda A_{\lambda}\left[e^{\imath\varphi_1}SH_u^0 H_d^0+h.c.\right]+\frac{\kappa}{3}A_{\kappa}\left[e^{\imath\varphi_2}S^3+h.c.\right]+\kappa^2|S|^4\\
+\lambda^2\left[|S|^2\left(|H_u^0|^2+|H_d^0|^2\right)+|H_u^0|^2 |H_d^0|^2\right]-\kappa\lambda\left[e^{\imath(\varphi_{\lambda}-\varphi_{\kappa})}S^{*\,2}H_u^0 H_d^0+h.c.\right]+\frac{g'^2+g^2}{8}\left[|H_u^0|^2-|H_d^0|^2\right]^2\label{tlHiggspot2}
\end{multline}

At tree level, the Higgs v.e.v.'s are assumed to minimize this potential. A consequence is the cancellation of first derivatives with respect to
the neutral Higgs fields at the minimum, which provides us with the minimization conditions:
\begin{align}
 &m_{H_u}^2=\lambda s\left[A_{\lambda}\cos\varphi_1+\kappa s\cos(\varphi_{\lambda}-\varphi_{\kappa})\right]\frac{v_d}{v_u}-\lambda^2(s^2+v_d^2)-\frac{g'^2+g^2}{4}(v_u^2-v_d^2)\nonumber\\
 &m_{H_d}^2=\lambda s\left[A_{\lambda}\cos\varphi_1+\kappa s\cos(\varphi_{\lambda}-\varphi_{\kappa})\right]\frac{v_u}{v_d}-\lambda^2(s^2+v_u^2)+\frac{g'^2+g^2}{4}(v_u^2-v_d^2)\nonumber\\
 &m_{S}^2=\lambda\left[A_{\lambda}\cos\varphi_1+2\kappa s\cos(\varphi_{\lambda}-\varphi_{\kappa})\right]\frac{v_uv_d}{s}-\kappa s[A_{\kappa}\cos\varphi_2+2\kappa s]-\lambda^2(v_u^2+v_d^2)\label{minpot}\\
 &A_{\lambda}\sin\varphi_1=-\kappa s\sin(\varphi_{\lambda}-\varphi_{\kappa})\nonumber\\
 &A_{\kappa}\sin\varphi_2=\frac{\lambda}{\kappa}\left[A_{\lambda}\sin\varphi_1-2\kappa s\sin(\varphi_{\lambda}-\varphi_{\kappa})\right]\frac{v_uv_d}{s^2}=-3\lambda\frac{v_uv_d}{s}\sin(\varphi_{\lambda}-\varphi_{\kappa})\nonumber
\end{align}
Here we see that the four phases of the Higgs sector are not independent but that, on the contrary, the minimization conditions relate $\varphi_1$
and $\varphi_2$ to $\varphi_{\lambda}-\varphi_{\kappa}$, the latter being the one and only `observable' phase in the Higgs sector. Note that
$\varphi_{\lambda}$ and $\varphi_{\kappa}$ intervene independently in other parts of the spectrum however. We will make an explicit use of the 
minimization conditions of Eq.\ref{minpot} in the following lines, replacing $m_{H_u}^2$, $m_{H_d}^2$, $m_{S}^2$, $A_{\lambda}\sin\varphi_1$ and 
$A_{\kappa}\sin\varphi_2$ by their expressions in terms of the v.e.v.'s.

The terms of Eq.\ref{tlHiggspot}, bilinear in the charged Higgs fields, define the $2\times2$ (hermitian) mass-matrix of the charged-Higgs states:
\begin{align}
 {\cal V}_{H}&\ni(H_u^-,H_d^-)\left<{\cal M}^2_{H^{\pm}}\right>\begin{pmatrix}H_u^+\\H_d^+\end{pmatrix}\hspace{2cm}\longrightarrow\nonumber\\
&\left<{\cal M}^2_{H^{\pm}}\right>=\left\{\lambda s\left[A_{\lambda}\cos\varphi_1+\kappa s\cos(\varphi_{\lambda}-\varphi_{\kappa})\right]-\left(\lambda^2-\frac{g^2}{2}\right)v_uv_d\right\}\begin{pmatrix}\frac{v_d}{v_u}&1\\1&\frac{v_u}{v_d}\end{pmatrix}\label{charhiggstreemass}\\
&\null\hspace{1.3cm}=\begin{pmatrix}-\sin\beta&\cos\beta\\\cos\beta&\sin\beta\end{pmatrix}\begin{pmatrix}0&0\\0&m_{H^{\pm}}^2\end{pmatrix}\begin{pmatrix}-\sin\beta&\cos\beta\\\cos\beta&\sin\beta\end{pmatrix}\nonumber\\
&m_{H^{\pm}}^2\equiv\left\{\frac{\lambda s}{v_uv_d}\left[A_{\lambda}\cos\varphi_1+\kappa s\cos(\varphi_{\lambda}-\varphi_{\kappa})\right]-\left(\lambda^2-\frac{g^2}{2}\right)\right\}(v_u^2+v_d^2)\ \ \ \ ;\ \ \ \ \tan\beta\equiv\frac{v_u}{v_d}\nonumber
\end{align}
which determines the charged Goldstone boson $G^{\pm}=-\sin\beta\,H_u^{\pm}+\cos\beta\,H_d^{\pm}$ and the physical charged Higgs state 
$H^{\pm}=\cos\beta\,H_u^{\pm}+\sin\beta\,H_d^{\pm}$.

Similarly, the terms bilinear in the neutral Higgs fields provide the $6\times6$ (symmetric) mass-matrix of the neutral Higgs: 
$\left<{\cal M}^2_{H^{0}}\right>_{ij}=\frac{1}{2}\left<\frac{\partial^2{\cal V}_{H^0}}{\partial S^0_i/\sqrt{2}\partial S^0_j/\sqrt{2}}\right>$, with the notation 
$\left<\ \ \right>$ meaning that fields are frozen to their v.e.v.'s. In the base $(h_u^0,h_d^0,h_s^0,a_u^0,a_d^0,a_s^0)$, these entries read:
\begin{align}
 \left<{\cal M}^2_{H^{0}}\right>_{11}& =\lambda s\left[A_{\lambda}\cos\varphi_1+\kappa s\cos(\varphi_{\lambda}-\varphi_{\kappa})\right]\frac{v_d}{v_u}+\frac{g'^2+g^2}{2}v_u^2\nonumber\\
 \left<{\cal M}^2_{H^{0}}\right>_{12}& =-\lambda s\left[A_{\lambda}\cos\varphi_1+\kappa s\cos(\varphi_{\lambda}-\varphi_{\kappa})\right]+2\left(\lambda^2-\frac{g'^2+g^2}{4}\right)v_uv_d\nonumber\\
 \left<{\cal M}^2_{H^{0}}\right>_{22}& =\lambda s\left[A_{\lambda}\cos\varphi_1+\kappa s\cos(\varphi_{\lambda}-\varphi_{\kappa})\right]\frac{v_u}{v_d}+\frac{g'^2+g^2}{2}v_d^2\nonumber\\
 \left<{\cal M}^2_{H^{0}}\right>_{13}& =-\lambda v_d\left[A_{\lambda}\cos\varphi_1+2\kappa s\cos(\varphi_{\lambda}-\varphi_{\kappa})\right]+2\lambda^2sv_u\nonumber\\
 \left<{\cal M}^2_{H^{0}}\right>_{23}& =-\lambda v_u\left[A_{\lambda}\cos\varphi_1+2\kappa s\cos(\varphi_{\lambda}-\varphi_{\kappa})\right]+2\lambda^2sv_d\nonumber\\
 \left<{\cal M}^2_{H^{0}}\right>_{33}& =\kappa s\left[A_{\kappa}\cos\varphi_2+4\kappa s\right]+\lambda A_{\lambda}\cos\varphi_1\frac{v_uv_d}{s}\label{neuthiggstreemass}
\end{align}
\begin{align}
& \left<{\cal M}^2_{H^{0}}\right>_{14}=0\hspace{1.5cm}&\left<{\cal M}^2_{H^{0}}\right>_{24}=0\hspace{3.5cm}&\left<{\cal M}^2_{H^{0}}\right>_{34}=\lambda\kappa sv_d\sin(\varphi_{\lambda}-\varphi_{\kappa})\nonumber\\
& \left<{\cal M}^2_{H^{0}}\right>_{15}=0\hspace{1.5cm}&\left<{\cal M}^2_{H^{0}}\right>_{25}=0\hspace{3.5cm}&\left<{\cal M}^2_{H^{0}}\right>_{35}=\lambda\kappa sv_u\sin(\varphi_{\lambda}-\varphi_{\kappa})\nonumber\\
& \left<{\cal M}^2_{H^{0}}\right>_{16}=-3\lambda\kappa sv_d\sin(\varphi_{\lambda}-\varphi_{\kappa})\hspace{0.3cm}&\left<{\cal M}^2_{H^{0}}\right>_{26}=-3\lambda\kappa sv_u\sin(\varphi_{\lambda}-\varphi_{\kappa})\hspace{0.3cm}&\left<{\cal M}^2_{H^{0}}\right>_{36}=4\lambda\kappa v_uv_d\sin(\varphi_{\lambda}-\varphi_{\kappa})\nonumber
\end{align}
\begin{align}
 \left<{\cal M}^2_{H^{0}}\right>_{44}& =\lambda s\left[A_{\lambda}\cos\varphi_1+\kappa s\cos(\varphi_{\lambda}-\varphi_{\kappa})\right]\frac{v_d}{v_u}\nonumber\\
 \left<{\cal M}^2_{H^{0}}\right>_{45}& =\lambda s\left[A_{\lambda}\cos\varphi_1+\kappa s\cos(\varphi_{\lambda}-\varphi_{\kappa})\right]\nonumber\\
 \left<{\cal M}^2_{H^{0}}\right>_{55}& =\lambda s\left[A_{\lambda}\cos\varphi_1+\kappa s\cos(\varphi_{\lambda}-\varphi_{\kappa})\right]\frac{v_u}{v_d}\nonumber\\
 \left<{\cal M}^2_{H^{0}}\right>_{46}& =\lambda v_d\left[A_{\lambda}\cos\varphi_1-2\kappa s\cos(\varphi_{\lambda}-\varphi_{\kappa})\right]\nonumber\\
 \left<{\cal M}^2_{H^{0}}\right>_{56}& =\lambda v_u\left[A_{\lambda}\cos\varphi_1-2\kappa s\cos(\varphi_{\lambda}-\varphi_{\kappa})\right]\nonumber\\
 \left<{\cal M}^2_{H^{0}}\right>_{66}& =-3\kappa sA_{\kappa}\cos\varphi_2+\lambda\frac{v_uv_d}{s} \left[A_{\lambda}\cos\varphi_1+4\kappa s\cos(\varphi_{\lambda}-\varphi_{\kappa})\right]\nonumber
\end{align}
As in the charged case, the neutral Goldstone boson can be singled out via a $\beta$-angle rotation $G^0\equiv-\sin\beta\,a_u^0+\cos\beta\,a_d^0$.
The remaining $5\times5$ symmetric block spanning the space $(h_u^0,h_d^0,h_s^0,a^0\equiv\cos\beta\,a_u^0+\sin\beta\,a_d^0,a_s^0)$ may be 
diagonalized via an orthogonal matrix $X^{H^0}$:\begin{equation}
 \left<\tilde{\cal M}^2_{H^{0}}\right>=X^{H^0\,T}\mbox{diag}(m_{S_i^0}^2,i=1,\ldots,5)X^{H^0}\label{diagneuthiggs}
\end{equation}
which defines the mass eigenstates:
\begin{equation}
 S_i^0=X^{H^0}_{i1}h_u^0+X^{H^0}_{i2}h_d^0+X^{H^0}_{i3}h_s^0+X^{H^0}_{i4}a^0+X^{H^0}_{i5}a_s^0\equiv X^{R}_{iu}h_u^0+X^{R}_{id}h_d^0+X^{R}_{is}h_s^0+X^{I}_{ia}a^0+X^{I}_{is}a_s^0
\end{equation}
We will use the second notation which allows more clarity in the identification of the components. Additionally, we define $X^{I}_{iu}\equiv\cos\beta\,X^{I}_{ia}$ and 
$X^{I}_{id}\equiv\sin\beta\,X^{I}_{ia}$. 

Note that the positivity of the squared Higgs-masses is a stability condition of the vacuum. Remember also that, at $0^{\mbox{\tiny th}}$ order in the 
electroweak v.e.v.'s, one can isolate the CP-even and CP-odd sectors and 
diagonalize their doublet subspaces via rotations of angle $-\beta$ / $\beta$ (the singlet states are then unmixed), which disentangles the `light' 
(then fully massless) `SM-like' doublet states from the `heavy' states with approximate squared-mass $M_A^2\equiv
\lambda s\left[A_{\lambda}\cos\varphi_1+\kappa s\cos(\varphi_{\lambda}-\varphi_{\kappa})\right]\frac{v_u^2+v_d^2}{v_uv_d}$ (degenerate at this order with the charged state).

\subsection{The supersymmetric spectrum at tree-level}
The whole tree-level spectrum will be treated with further details in appendix \ref{sectl}. Here we simply summarize, for the sake of
notations, the basic ingredients concerning the treatment of masses and mixings of SUSY particles.

\vspace{0.2cm}
{\em i) Gluinos}\newline
The gluinos are the fermionic partners of the gluons and, as such, form a color octet. Their bilinear terms originate in the soft lagrangian: 
$-{\cal L}_{\mbox{\tiny soft}}\ni-M_3 e^{\imath\phi_{M_3}}\tilde{g}_{a}\tilde{g}_{a}$.
The mass states $\tilde{\cal G}_a$, with mass $M_3$ (which we assume positive), then relate to the eigenstates of the SUSY vector superfield 
$\tilde{g}_{a}$ as $\tilde{\cal G}_a\equiv-\imath e^{\frac{\imath}{2}\phi_{M_3}}\tilde{g}_{a}$. The phase shift then affects the couplings of
the gluinos to coloured matter.

\vspace{0.2cm}
{\em ii) Charginos}\newline
The charginos are composed of the charged components of the electroweak gauginos and higgsinos. Their bilinear terms originate from both 
supersymmetry-conserving and violating terms and may be cast into the following form:
\begin{multline}
{\cal V}_{\chi^{\pm}}\ni\frac{1}{2}\chi_C^T\begin{pmatrix}
 0 & \left<{\cal M}_{\chi^{-+}}\right>\\ \left<{\cal M}_{\chi^{+-}}\right> & 0 
\end{pmatrix}\chi_C+h.c.\ \ \ \ ;\ \ \ \ \chi_C^T\equiv(-\imath \tilde{w}^-,\tilde{h}_d^-,-\imath \tilde{w}^+,\tilde{h}_u^+)\label{masscha}\\
 \left<{\cal M}_{\chi^{-+}}\right>=\begin{pmatrix}M_2e^{\imath\phi_{M_2}}&gv_u\\gv_d&\lambda e^{\imath\varphi_{\lambda}}s\end{pmatrix}=\left<{\cal M}_{\chi^{+-}}\right>^T
\end{multline}
We may diagonalize $\left<{\cal M}_{\chi^{-+}}\right>$ with the help of two unitary matrices $U$ and $V$:
$\left<{\cal M}_{\chi^{-+}}\right>=U^T\mbox{diag}(m_{\chi^{\pm}_1},m_{\chi^{\pm}_2})V$. The mass eigenvalues may be assumed real and positive without 
any loss of generality and the mass eigenstates ($i=1,2$) relate to the gauge ones as:
\begin{equation}
 \chi^+_i=V_{i1}(-\imath\tilde{w}^+)+V_{i2}\tilde{h}_u^+\equiv V_{iw}(-\imath\tilde{w}^+)+V_{iu}\tilde{h}_u^+\ \ \ \ ;\ \ \ \ \chi^-_i=U_{i1}(-\imath\tilde{w}^-)+U_{i2}\tilde{h}_d^-\equiv U_{iw}(-\imath\tilde{w}^-)+U_{id}\tilde{h}_d^-
\end{equation}

\vspace{0.2cm}
{\em iii) Neutralinos}\newline
The neutralinos are combinations of the neutral components of the electroweak gauginos and higgsinos. Their bilinear terms, resulting from both 
supersymmetry-conserving and violating terms, form a Majorana mass matrix:
\begin{multline}
{\cal V}_{\chi^{0}}\ni\frac{1}{2}\chi_N^T\left<{\cal M}_{\chi^{0}}\right>\chi_N+h.c.\ \ \ \ ;\ \ \ \ \chi_N^T\equiv(-\imath \tilde{b},-\imath \tilde{w}^3,\tilde{h}_u^0,\tilde{h}_d^0,\tilde{h}_s^0)\label{massneu}\\
\left<{\cal M}_{\chi^{0}}\right>=\begin{pmatrix} M_1 e^{\imath\phi_{M_1}} & 0 & \frac{g'}{\sqrt{2}}v_u & -\frac{g'}{\sqrt{2}}v_d & 0\\
 0 & M_2 e^{\imath\phi_{M_2}} & -\frac{g}{\sqrt{2}}v_u & \frac{g}{\sqrt{2}}v_d & 0 \\
 \frac{g'}{\sqrt{2}}v_u & -\frac{g}{\sqrt{2}}v_u & 0 & -\lambda e^{\imath\varphi_{\lambda}}s & -\lambda e^{\imath\varphi_{\lambda}}v_d\\
 -\frac{g'}{\sqrt{2}}v_d & \frac{g}{\sqrt{2}}v_d & -\lambda e^{\imath\varphi_{\lambda}}s & 0 & -\lambda e^{\imath\varphi_{\lambda}}v_u\\
 0 & 0 & -\lambda e^{\imath\varphi_{\lambda}}v_d & -\lambda e^{\imath\varphi_{\lambda}}v_u & 2\kappa e^{\imath\varphi_{\kappa}}s
\end{pmatrix}=\left<{\cal M}_{\chi^{0}}\right>^T
\end{multline}
$\left<{\cal M}_{\chi^{0}}\right>$ being symmetric, it can be diagonalized by a single unitary matrix $N$ according to: \newline
$\left<{\cal M}_{\chi^{0}}\right>=N^T\mbox{diag}(m_{\chi^{0}_i},i=1,\ldots,5)N$. Without loss of generality the eigenvalues $m_{\chi^{0}_i}$
can be chosen real and positive (remember that $N$ is complex) and the mass eigenstates relate to the gauge ones in the following fashion:
\begin{equation}
 \chi^0_i=N_{i1}(-\imath \tilde{b})+N_{i2}(-\imath \tilde{w}^3)+N_{i3}\tilde{h}_u^0+N_{i4}\tilde{h}_d^0+N_{i5}\tilde{h}_s^0\equiv N_{ib}(-\imath \tilde{b})+N_{iw}(-\imath \tilde{w}^3)+N_{iu}\tilde{h}_u^0+N_{id}\tilde{h}_d^0+N_{is}\tilde{h}_s^0
\end{equation}

\vspace{0.2cm}
{\em iv) Sfermions}\newline
The scalar partners of the SM fermions receive hermitian mass matrices. Due to our assumptions with respect to flavour violation, the three
generations decouple. We keep a generic notation although only the Yukawa couplings of the third generation ($u=t$, $d=b$, $e=\tau$) will
be treated as non-vanishing in practice:
\begin{align}
& {\cal V}_{\tilde{F}}\ni(U_L^{\dag},U_R^c)\left<{\cal M}^2_{U}\right>\begin{pmatrix}U_L\\U_R^{c\,\dag}\end{pmatrix}+(D_L^{\dag},D_R^c)\left<{\cal M}^2_{D}\right>\begin{pmatrix}D_L\\D_R^{c\,\dag}\end{pmatrix}
+N_L^{\dag}\left<{\cal M}^2_{N}\right>N_L+(E_L^{\dag},E_R^c)\left<{\cal M}^2_{E}\right>\begin{pmatrix}E_L\\E_R^{c\,\dag}\end{pmatrix}\label{masssf}\\
& \left<{\cal M}^2_{U}\right>= \begin{pmatrix}m_{Q}^2+Y_u^2v_u^2+\frac{1}{4}\left(\frac{g'^2}{3}-g^2\right)(v_u^2-v_d^2)&Y_u\left[A_ue^{-\imath\varphi_{A_u}}v_u-\lambda e^{\imath\varphi_{\lambda}}sv_d\right]\\
Y_u\left[A_ue^{\imath\varphi_{A_u}}v_u-\lambda e^{-\imath\varphi_{\lambda}}sv_d\right]&m_U^2+Y_u^2v_u^2-\frac{g'^2}{3}(v_u^2-v_d^2)\end{pmatrix}\nonumber\\
& \left<{\cal M}^2_{D}\right>= \begin{pmatrix}m_{Q}^2+Y_d^2v_d^2+\frac{1}{4}\left(\frac{g'^2}{3}+g^2\right)(v_u^2-v_d^2)&Y_d\left[A_de^{-\imath\varphi_{A_d}}v_d-\lambda e^{\imath\varphi_{\lambda}}sv_u\right]\\
Y_d\left[A_de^{\imath\varphi_{A_d}}v_d-\lambda e^{-\imath\varphi_{\lambda}}sv_u\right]&m_D^2+Y_d^2v_d^2+\frac{g'^2}{6}(v_u^2-v_d^2)\end{pmatrix}\nonumber\\
& \left<{\cal M}^2_{N}\right>=m_L^2-\frac{g'^2+g^2}{4}(v_u^2-v_d^2)\nonumber\\
& \left<{\cal M}^2_{E}\right>= \begin{pmatrix}m_{L}^2+Y_e^2v_d^2+\frac{-g'^2+g^2}{4}(v_u^2-v_d^2)&Y_e\left[A_ee^{-\imath\varphi_{A_e}}v_d-\lambda e^{\imath\varphi_{\lambda}}sv_u\right]\\
Y_e\left[A_ee^{\imath\varphi_{A_e}}v_d-\lambda e^{-\imath\varphi_{\lambda}}sv_u\right]&m_E^2+Y_e^2v_d^2+\frac{g'^2}{2}(v_u^2-v_d^2)\end{pmatrix}\nonumber
\end{align}
Each mass matrix $\left<{\cal M}^2_{F}\right>$ -- $F=U,D,N,E$ -- can be diagonalized via a special-unitary matrix $X^F$, according to: \newline
$\left<{\cal M}^2_F\right>=X^{F\,\dag}\mbox{diag}(m^2_{F_1},m^2_{F_2})X^F$. The positivity of the squared masses $m^2_{F_i}$ is a stability
condition of the vacuum. The mass eigenstates are then defined as: $F_i=X^F_{iL}F_L+X^F_{iR}F_R^{c\,\dag}$.

This completes this short presentation of the tree-level spectrum. More details are presented in appendix \ref{sectl}, together with the Higgs couplings.

\section{A short walk-through the code}

In this section, we shall describe the operations which are conducted throughout our subroutines from the perspective of the phenomenology of the
CP-violating NMSSM. 

\subsection{Interface with NMSSMTools}

Before coming to the actual computations of our code, let us remind the reader that we embed it within the \verb|NMSSMTools| package. We actually
use the \verb|NMHDECAY| routines to define its input. In particular, we do not alter the running of parameters -- such as the Yukawa, gauge or soft 
couplings --, e.g.\ to the average scale of the squarks of third generation. We simply use the corresponding quantities as calculated by \verb|NMHDECAY|
as our input and introduce the complex phases at this level. This is justified as the renormalization group equations (RGE's) of the superpotential 
parameters leave the phases unaffected (at least up to two-loop order). A short subroutine \verb|init_CPV.f| defines this interface and stores all the 
relevant quantities within commons of the code. The case of the parameters $A_{\lambda}$ and $A_{\kappa}$ is somewhat more subtle: given that the 
phases $\varphi_1$ and $\varphi_2$ are not free but, in our approach, determined by the minimization conditions of the potential (see Eq.\ref{minpot}),
we will only be using the quantities $A_{\lambda}\cos\varphi_1$ and $A_{\kappa}\cos\varphi_2$ as degrees of freedom in practice. Therefore,
we identify the \verb|NMHDECAY| input for $A_{\lambda}$ and $A_{\kappa}$ as ours for $A_{\lambda}\cos\varphi_1$ and $A_{\kappa}\cos\varphi_2$. The 
one-loop RGE's are correspondingly corrected. The wave-function scaling factors for the Higgs fields are also defined slightly differently from the 
original implementation in \verb|NMSSMTools|, as we shall describe in section \ref{secwavefunc}.

Given our discussion in section \ref{secmod}, the following eight phases are added as new degrees of freedom: $\varphi_{\lambda}$, $\varphi_{\kappa}$,
$\phi_{M_1}$, $\phi_{M_2}$, $\phi_{M_3}$, $\varphi_{A_t}$, $\varphi_{A_b}$, $\varphi_{A_{\tau}}$.

\subsection{Supersymmetric spectrum}
The first actual operations which are carried out in connection to the CP-violating NMSSM consist in the calculation of the masses of the 
supersymmetric matter content. Similarly to the evaluation by \verb|NMSSMTools| in the CP-conserving case, we take into account the leading 
radiative corrections to the masses. In the following, we list the new subroutines and provide relevant information concerning the calculations 
which are performed.

\vspace{0.2cm}
{\em i)} \verb|mcha_CPV.f|\newline
The purpose of this subroutine rests in diagonalizing the chargino mass matrix (Eq.\ref{masscha}) according to 
$\left<{\cal M}_{\chi^{-+}}\right>=U^T\mbox{diag}(m_{\chi^{\pm}_1},m_{\chi^{\pm}_2})V$. Similarly to the corresponding implementation
within \verb|NMSSMTools| for the CP-conserving case, the entries of the mass matrix receive one-loop radiative corrections which are calculated
in the approximation where mass and gauge eigenstates coincide. The corresponding effects are presented in section 4.2 of \cite{Pierce:1996zz} --
in the context the MSSM and still in the CP-conserving case. Small modifications appear in the CP-violating NMSSM, as gaugino and higgsino scalar
couplings are rotated by phase factors of $e^{-\imath\phi_{M_i}/2}$ and $e^{-\imath\varphi_{\lambda}/2}$. Nevertheless, the factors of $B_1$ functions
as well as the corrections involving gauge bosons are immune to this phase shift, so that only the scalar interactions resulting in a $B_0$ function 
-- in the approximations of \cite{Pierce:1996zz}, this reduces to the Higgs / higgsino loops -- are affected. Another difference with respect to ref. 
\cite{Pierce:1996zz} originates from the presence of singlets and singlinos in the higgsino self-energies. A summary of these corrections is 
explicited in appendix \ref{aploopchaneu}.

The following steps are essentially identical to their counterparts in the tree-level case, which is treated into details in appendix \ref{tlchaneumass}: 
we define two special-unitary matrices $U_0$ and $V_0$ diagonalizing the hermitian matrices 
$\left<{\cal M}_{\chi^{-+}}\right>\left<{\cal M}_{\chi^{-+}}\right>^{\dag}$ and 
$\left<{\cal M}_{\chi^{-+}}\right>^{\dag}\left<{\cal M}_{\chi^{-+}}\right>$ respectively. $U_0^*\left<{\cal M}_{\chi^{-+}}\right>V_0^{\dag}$
is then a diagonal matrix with, in general, non-real entries. We thus define the unitary matrices $U$ and $V$ via a phase-shift of $U_0$ and $V_0$,
where the phase of the lightest state is absorbed in $U$ while that of the heavier one is absorbed in $V$: the resulting chargino masses are real and
positive.

\vspace{0.2cm}
{\em ii)} \verb|mneu_CPV.f|\newline
The case of the neutralinos follows the same principles as that of the charginos. The tree-level gaugino and higgsino masses are corrected in accordance
with the one-loop effects presented in appendix \ref{aploopchaneu}. We then diagonalize the complex symmetric neutralino mass matrix according to
$\left<{\cal M}_{\chi^{0}}\right>=N^T\mbox{diag}(m_{\chi^{0}_i},i=1,\ldots,5)N$. For that purpose, we consider the $10\times10$ real symmetric 
matrix $\begin{pmatrix} \mbox{Re}&\mbox{Im}\\-\mbox{Im} &\mbox{Re}\end{pmatrix}\left(\left<{\cal M}_{\chi^{0}}\right>^{\dag}\left<{\cal M}_{\chi^{0}}\right>\right)$, which can be diagonalized numerically
by an orthogonal matrix $\tilde{N}_0$. We then extract a special unitary matrix $N_0$ so that $N_0^*\left<{\cal M}_{\chi^{0}}\right>N_0^{\dag}$
is diagonal. We finally absorb the remaining phases in a phase-shift of $N_0$, which defines the real and positive neutralino masses as well as
the mixing matrix $N$. Details are provided in appendix \ref{tlchaneumass}.

\vspace{0.2cm}
{\em iii)} \verb|msferm_CPV.f|\newline
We now turn to the sfermion masses. The hermitian tree-level mass matrices are diagonalized via special-unitary matrices $X^F$, according to 
$\left<{\cal M}^2_F\right>=X^{F\,\dag}\mbox{diag}(m^2_{F_1},m^2_{F_2})X^F$. We remind the reader that the parameters entering the matrices, 
e.g.\ the top and bottom Yukawa couplings, have been run to the average squark scale. The Yukawa couplings of the first two generation are neglected,
so that the corresponding diagonalizing matrices are trivial. Details can be found in appendix \ref{tlsfermmass}.

We then apply $O(\alpha_S)$ corrections to the squark squared masses (consistently with what was implemented in the original CP-conserving treatment 
in \verb|NMSSMTools|). Gluons, gluinos as well as the quartic sfermion D-term contribute to the sfermion self energy at this order. CP-phases -- here, 
$\phi_{M_3}$ and $\varphi_{A_f}$ -- intervene in the gluino-sfermion couplings leading to a $B_0$ function. A summary is proposed in appendix
\ref{aploopsferm}.

Finally, we check the positivity of the sfermion squared masses, a vacuum-stability requirement.

\vspace{0.2cm}
{\em iv)} \verb|mgluino_CPV.f|\newline
\verb|mgluino_CPV.f| computes the gluino mass, including the $O(\alpha_S)$ radiative corrections, which are obtained in a similar manner to the 
discussion in section 4.1 of \cite{Pierce:1996zz}. Relevant corrections include the gluon / gluino and the quark / squark loops. Complex phases again 
enter the couplings of gluinos to squarks. Details are provided in appendix \ref{aploopgluino}.

\subsection{Higgs masses and radiative corrections}
The following series of subroutines aim at computing the Higgs masses and mixing, including full one-loop and leading two-loop corrections. 
Consistently with the original approach in \verb|NMHDECAY|, we will consider the effective Higgs potential at the average scale of the  squarks of the 
third generation -- denoted as $Q$ --, where the running parameters are thus evaluated.

\subsubsection{Wave-function renormalization}\label{secwavefunc}
Momentum-dependent radiative corrections can be included in two fashions within the effective potential evaluation: one may reject them to the end of
the calculation, as `pole-corrections', or one may take them into account -- at least partially -- into the effective lagrangian as corrections to 
the kinetic terms. The latter choice leads to wave-function renormalization factors. While the two methods are formally equivalent, they lead to 
slightly divergent results at the numerical level, as we will discuss later. Following the original approach in \verb|NMHDECAY| -- presented e.g.\ in
appendix C of \cite{Ellwanger:2004xm} or appendix C of \cite{NMSSM} --, we decide to include the leading $p^2$ terms -- where $p$ stands for the 
external energy-momentum of the Higgs self-energies --, originating in fermion or gauge effects, into the kinetic term of the effective lagrangian. 
Nevertheless, since we aim at a full computation at one-loop, all the missing momentum-dependent parts will be added as pole-corrections (see below).

In the general case, the modified Higgs kinetic terms involve a hermitian (non-degenerate) matrix $Z_H(p^2)$ as follows (here and below $S_i$ 
denotes any Higgs field; we work in momentum space and omit the factor $1/2$ which should appear if the considered field is real):
\begin{equation}
 \hat{\cal L}_{\mbox{\tiny kin}}^{\mbox{\tiny\em eff}}=\sum_{i,j}p^2\left.Z_H(p^2)\right|_{ij}S_i^*S_j
\end{equation}
The normal procedure then consists in rotating and scaling $Z_H$ via an invertible matrix $O_H$ in order to recover the identity -- 
$Z_H(p^2)=O_H^{\dag}\mathbb{1}O_H$ --, then considering the `new' set of fields with standard kinetic term $\tilde{S}_i\equiv O_{H\,ij}S_j$.

Yet, Eqs.(C.1) of \cite{Ellwanger:2004xm} or (C.9-11) of \cite{NMSSM} show that a clever choice of the corrections included into $Z_H$ can 
make this procedure particularly simple, as $Z_H$ would turn out to be diagonal in the base of gauge-eigenstates. Restricting to neutral
Higgs fields, one has (with $\delta_{S_i,S_j}$ denoting the Kronecker symbol):
\begin{equation}
 \left.Z_{H}\right|_{ij}=Z_{H_u}\left[\delta_{S_i,h_u^0}\delta_{S_j,h_u^0}+\delta_{S_i,a_u^0}\delta_{S_j,a_u^0}\right]
+Z_{H_d}\left[\delta_{S_i,h_d^0}\delta_{S_j,h_d^0}+\delta_{S_i,a_d^0}\delta_{S_j,a_d^0}\right]
+Z_{S}\left[\delta_{S_i,h_s^0}\delta_{S_j,h_s^0}+\delta_{S_i,a_s^0}\delta_{S_j,a_s^0}\right]
\end{equation}
Indeed, considering the contributions of SM-fermions to $Z_H$ ($N_c=3$ is the colour factor; while using the generic notations $u$, $d$, $e$, we will 
be considering only the third generation fermions since we neglect the Yukawa couplings of the two first families), the deviations of the
diagonal scaling factors from unity read\footnote{We use the $\overline{DR}$ scheme. Loop functions are defined in appendix \ref{reffunc}.}:
\begin{align}
 \delta^{\mbox{\tiny SM ferm}}Z_{H_u} &=\frac{1}{16\pi^2}\left\{N_cY_u^2B_0(p,m_u,m_u)\right\}\nonumber\\
 \delta^{\mbox{\tiny SM ferm}}Z_{H_d} &=\frac{1}{16\pi^2}\left\{N_cY_d^2B_0(p,m_d,m_d)+Y_e^2B_0(p,m_e,m_e)\right\}\label{wavefuncsmferm}\\
 \delta^{\mbox{\tiny SM ferm}}Z_{S}\hspace{0.15cm} &=0\nonumber
\end{align}
Similarly, in the approximation where higgsinos and gauginos are simultaneously gauge and mass eigenstates ($\mu$ denotes the doublet higgsino mass;
$m_{\tilde{s}}$, the singlino mass):
\begin{align}
 \delta^{\tilde{h},\tilde{g}}Z_{H_u} &=\frac{1}{16\pi^2}\left\{\frac{g'^2}{2}B_0(p,M_1,\mu)+\frac{3g^2}{2}B_0(p,M_2,\mu)+\lambda^2B_0(p,\mu,m_{\tilde{s}})\right\}\nonumber\\
 \delta^{\tilde{h},\tilde{g}}Z_{H_d} &=\delta^{\tilde{h},\tilde{g}}Z_{H_u}\label{wavefuncino}\\
 \delta^{\tilde{h},\tilde{g}}Z_{S}\hspace{0.15cm} &=\frac{1}{8\pi^2}\left\{\lambda^2B_0(p,\mu,\mu)+\kappa^2B_0(p,m_{\tilde{s}},m_{\tilde{s}})\right\}\nonumber
\end{align}
The last source of corrections to $Z_H$ is the gauge sector -- note that we will be working in the Feynmann gauge. Yet, the corresponding contributions 
are not diagonal in the gauge eigenbase, but rotated by an angle $\beta$ (or $-\beta$, depending on the CP-eigenvalue) in the doublet sector. Noticing 
however that $\tan\beta>1$ in practice, we may keep the $\sin^2\beta$ term in the wave-function scaling while rejecting the remaining $\sin\beta\cos\beta$ and 
$\cos^2\beta$ terms for later treatment as pole-corrections. Then:
\begin{align}
 \delta^{\mbox{\tiny gauge}}Z_{H_u} &=-\frac{\sin^2\beta}{16\pi^2}\left\{g^2B_0(p,M_W,M_W)+\frac{g'^2+g^2}{2}B_0(p,M_Z,M_Z)\right\}\nonumber\\
 \delta^{\mbox{\tiny gauge}}Z_{H_d} &=0\label{wavefuncgauge}\\
 \delta^{\mbox{\tiny gauge}}Z_{S}\hspace{0.15cm} &=0\nonumber
\end{align}
Note that this choice in the gauge sector differs from the default treatment by \verb|NMSSMTools| in the CP-conserving case (see Eq.(C.9-10) of \cite{NMSSM}),
where, moreover, pole corrections from the gauge sector are ignored.

Before setting $Z_{H_u,H_d,S}=1+\left[\delta^{\mbox{\tiny SM ferm}}+\delta^{\tilde{h},\tilde{g}}+\delta^{\mbox{\tiny gauge}}\right]Z_{H_u,H_d,S}$,
one is confronted to the remaining $p^2$-dependence of these coefficients (via the loop functions $B_0$). In the ideal case, $p^2$ would match the 
Higgs squared masses. This, however, is impractical since several mass eigenvalues are present: keeping this $p^2$ dependence, hence working with
$p^2$-dependent fields and mass-matrices, and setting this implicit dependence separately to the corresponding Higgs squared mass after diagonalization of the mass matrix 
would be possible, yet problematic in a numerical evaluation of the mass matrices. The choice of \cite{NMSSM} in the CP-conserving case rested in adding
an artificial dependence of $Z_{H_u,H_d}$ on $\ln(M_A^2/m_t^2)$ -- $M_A$ standing for the mass of the heavy doublet, $m_t$ approximating the SM-like Higgs mass --,
so as to mimic the correct logarithmic dependence after rotation by an angle $-\beta$ (approximating the tree-level diagonalizing rotation in the 
CP-even doublet sector): however an explicit rotation by the angle $-\beta$ shows that this purpose is missed as only the light state receives the proper 
logarithmic factor; in the case of the heavy doublet, the factor is wrong so that the result does not really improve on neglecting the logarithms 
$\ln(M_A^2/m_t^2)$ altogether. Therefore, we settle for the choice which consists in freezing the external momentum to a scale $\mu_H=125$~GeV, allowing for a good precision in the 
characteristics of the SM-like Higgs state -- the most sensitive to radiative corrections. Adequate corrections when the mass is far from this scale 
are rejected to the level of pole-corrections. A final difference with \cite{NMSSM} comes from the implementation of the loop functions: we explicitly 
compute the full relevant $B_0$'s while \cite{NMSSM} only included the leading logarithmic terms in case of large mass hierarchies.

A summary of the wave-function scaling factors is provided in appendix \ref{aploopwave}.

Consistently, the neutral higgs fields are rescaled as:
\begin{equation}
 h_u^0\leftarrow\frac{h_u^0}{\sqrt{Z_{H_u}}}\hspace{0.5cm};\hspace{0.5cm}h_d^0\leftarrow\frac{h_d^0}{\sqrt{Z_{H_d}}}\hspace{0.5cm};\hspace{0.5cm}h_s^0\leftarrow\frac{h_s^0}{\sqrt{Z_{S}}}\hspace{0.5cm};\hspace{0.5cm}a_u^0\leftarrow\frac{a_u^0}{\sqrt{Z_{H_u}}}\hspace{0.5cm};\hspace{0.5cm}a_d^0\leftarrow\frac{a_d^0}{\sqrt{Z_{H_d}}}\hspace{0.5cm};\hspace{0.5cm}a_s^0\leftarrow\frac{a_s^0}{\sqrt{Z_{S}}}\label{neutscaling}
\end{equation}
so that all related quantities (e.g.\ the mass matrices) must be rescaled accordingly. In particular the Higgs v.e.v.'s:
\begin{equation}\label{vevscale}
 v_u(Q)\equiv\frac{v_u}{\sqrt{Z_{H_u}}}\hspace{0.5cm};\hspace{0.5cm}v_d(Q)\equiv\frac{v_d}{\sqrt{Z_{H_d}}}\hspace{0.5cm};\hspace{0.5cm}s(Q)\equiv\frac{s}{\sqrt{Z_{S}}}
\end{equation}
All these operations are carried out in the initialization subroutine \verb|init_CPV.f|.

In the charged-sector, the $p^2$-dependent terms are typically different from those appearing in the neutral case. However, to keep $\beta$ as the 
relevant rotation angle in the charged sector together with the v.e.v. rescaling of Eq.\ref{vevscale}, we will use the same wave-function scaling 
factors $Z_{H_u,H_d}$:
\begin{equation}
 H_u^{\pm}\leftarrow\frac{H_u^{\pm}}{\sqrt{Z_{H_u}}}\hspace{0.5cm};\hspace{0.5cm}H_d^{\pm}\leftarrow\frac{H_d^{\pm}}{\sqrt{Z_{H_d}}}\label{charscaling}
\end{equation}
and restore the appropriate dependence at the level of the pole-corrections.

\subsubsection{Effective potential}
After this discussion relative to the kinetic terms, let us turn to the Higgs potential. At tree-level, it is given by Eq.\ref{tlHiggspot}. Radiative
corrections can be added to this picture by considering diagrams with vanishing external momenta or, equivalently, the Coleman-Weinberg formula for 
one-loop effects. In the $\overline{DR}$ scheme and the Landau gauge, the effective Higgs potential reads:
\begin{equation}
 {\cal V}_{\mbox{\tiny eff.}}(H)={\cal V}^{\mbox{\tiny tree}}_{H}(H)+\delta{\cal V}_{\mbox{\tiny eff.}}(H)\ \ \ \ ;\ \ \ \ \delta{\cal V}_{\mbox{\tiny eff.}}(H)=\frac{1}{64\pi^2}\mbox{Tr}\left\{C_{\Phi}{\cal M}_{\Phi}^4\left[\ln\frac{{\cal M}^2_{\Phi}}{Q^2}-\frac{3}{2}\right]\right\}\label{effpot}
\end{equation}
where the trace applies to all fields $\Phi$ of the model, with $C_{\Phi}$ depending on the Lorentz properties of $\Phi$ -- respectively
$1$, $2$, $-2$, $-4$, $3$ for a real scalar, a complex scalar, a Majorana fermion, a Dirac fermion and a gauge boson -- and ${\cal M}_{\Phi}^2$
is the bilinear (`squared mass') matrix of the fields, where the dependence on Higgs fields has been kept\footnote{In other words, one recovers the 
tree-level squared mass matrix $\left<{\cal M}_{\Phi}^2\right>$ when replacing the Higgs fields by their v.e.v.'s in ${\cal M}_{\Phi}^2$.}. Note
that the gauge or $\mathbb{Z}_3$ symmetries are still explicitly preserved by this potential (but not, in general, by its minimization). On the other
hand, it involves terms of dimension $\geq5$, so that expansions of the potential in the vicinity of its minimum will generically break the symmetries
in an explicit way.

\vspace{0.2cm}
{\em i) Minimization conditions and corrections to the mass matrices}\newline
The Higgs v.e.v.'s $v_u$, $v_d$, $s$ -- of Eq.\ref{vevscale}: remember that we are considering the potential at the scale $Q$ -- are now supposed to 
minimize the full potential of Eq.\ref{effpot}. Consequently, the minimization conditions of Eq.\ref{minpot} (at tree-level) must be modified to account
for the radiative effects. This provides the so-called tadpole equations\footnote{The notation $\left<f\right>$ means that the function $f$ of the 
Higgs fields is evaluated at the Higgs v.e.v.'s.}:
\begin{align}
 &\delta m_{H_u}^2=-\frac{1}{2v_u}\left<\frac{\partial \delta{\cal V}_{\mbox{\tiny eff.}}}{\partial h_u^0/\sqrt{2}}\right>\nonumber\\
 &\delta m_{H_d}^2=-\frac{1}{2v_d}\left<\frac{\partial \delta{\cal V}_{\mbox{\tiny eff.}}}{\partial h_d^0/\sqrt{2}}\right>\label{tadpole}\\
 &\delta m_{S}^2=-\frac{1}{2s}\left<\frac{\partial \delta{\cal V}_{\mbox{\tiny eff.}}}{\partial h_s^0/\sqrt{2}}\right>\nonumber\\
 &\delta(A_{\lambda}\sin\varphi_1)=-\frac{1}{2\lambda sv_d}\left<\frac{\partial \delta{\cal V}_{\mbox{\tiny eff.}}}{\partial a_u^0/\sqrt{2}}\right>=-\frac{1}{2\lambda sv_u}\left<\frac{\partial \delta{\cal V}_{\mbox{\tiny eff.}}}{\partial a_d^0/\sqrt{2}}\right>\nonumber\\
 &\delta(A_{\kappa}\sin\varphi_2)=\frac{1}{2\kappa s^2}\left[\left<\frac{\partial \delta{\cal V}_{\mbox{\tiny eff.}}}{\partial a_s^0/\sqrt{2}}\right>-\frac{v_u}{s}\left<\frac{\partial \delta{\cal V}_{\mbox{\tiny eff.}}}{\partial a_u^0/\sqrt{2}}\right>\right]\nonumber
\end{align}
Given that the parameters $m_{H_u}^2$, $m_{H_d}^2$, $m_{S}^2$, $A_{\lambda}\sin\varphi_1$ and $A_{\kappa}\sin\varphi_2$ have been replaced by their 
tree-level values (Eq.\ref{minpot}) in the tree-level Higgs mass matrices (Eq.\ref{charhiggstreemass} and \ref{neuthiggstreemass}), the shifts of
Eq.\ref{tadpole} must be included into the corrected mass matrices, in addition to the bilinear terms. For the charged Higgs mass matrix, this amounts to:
\begin{align}
 & \delta\left<{\cal M}^2_{H^{\pm}}\right>_{11}=\left<\frac{\partial^2 \delta{\cal V}_{\mbox{\tiny eff.}}}{\partial H_u^-\partial H_u^+}-\frac{1}{2v_u}\frac{\partial \delta{\cal V}_{\mbox{\tiny eff.}}}{\partial h_u^0/\sqrt{2}}\right> & &
\delta\left<{\cal M}^2_{H^{\pm}}\right>_{12}=\left<\frac{\partial^2 \delta{\cal V}_{\mbox{\tiny eff.}}}{\partial H_u^-\partial H_d^+}+\frac{\imath}{2v_d}\frac{\partial \delta{\cal V}_{\mbox{\tiny eff.}}}{\partial a_u^0/\sqrt{2}}\right>\\
 & \delta\left<{\cal M}^2_{H^{\pm}}\right>_{21}=\left<\frac{\partial^2 \delta{\cal V}_{\mbox{\tiny eff.}}}{\partial H_d^-\partial H_u^+}-\frac{\imath}{2v_d}\frac{\partial \delta{\cal V}_{\mbox{\tiny eff.}}}{\partial a_u^0/\sqrt{2}}\right> & &
\delta\left<{\cal M}^2_{H^{\pm}}\right>_{22}=\left<\frac{\partial^2 \delta{\cal V}_{\mbox{\tiny eff.}}}{\partial H_d^-\partial H_d^+}-\frac{1}{2v_d}\frac{\partial \delta{\cal V}_{\mbox{\tiny eff.}}}{\partial h_d^0/\sqrt{2}}\right>\nonumber
\end{align}
and for the neutral Higgs states:
\begin{align}
 & \delta\left<{\cal M}^2_{H^0}\right>_{11}=\frac{1}{2}\left<\frac{\partial^2 \delta{\cal V}_{\mbox{\tiny eff.}}}{(\partial h_u^0/\sqrt{2})^2}-\frac{1}{v_u}\frac{\partial \delta{\cal V}_{\mbox{\tiny eff.}}}{\partial h_u^0/\sqrt{2}}\right> & &
\delta\left<{\cal M}^2_{H^0}\right>_{12}=\frac{1}{2}\left<\frac{\partial^2 \delta{\cal V}_{\mbox{\tiny eff.}}}{\partial h_u^0/\sqrt{2}\partial h_d^0/\sqrt{2}}\right>\nonumber\\
 & \delta\left<{\cal M}^2_{H^0}\right>_{22}=\frac{1}{2}\left<\frac{\partial^2 \delta{\cal V}_{\mbox{\tiny eff.}}}{(\partial h_d^0/\sqrt{2})^2}-\frac{1}{v_d}\frac{\partial \delta{\cal V}_{\mbox{\tiny eff.}}}{\partial h_d^0/\sqrt{2}}\right> & &
\delta\left<{\cal M}^2_{H^0}\right>_{13}=\frac{1}{2}\left<\frac{\partial^2 \delta{\cal V}_{\mbox{\tiny eff.}}}{\partial h_u^0/\sqrt{2}\partial h_s^0/\sqrt{2}}\right>\nonumber\\
 & \delta\left<{\cal M}^2_{H^0}\right>_{23}=\frac{1}{2}\left<\frac{\partial^2 \delta{\cal V}_{\mbox{\tiny eff.}}}{\partial h_d^0/\sqrt{2}\partial h_s^0/\sqrt{2}}\right> & &
\delta\left<{\cal M}^2_{H^0}\right>_{33}=\frac{1}{2}\left<\frac{\partial^2 \delta{\cal V}_{\mbox{\tiny eff.}}}{(\partial h_s^0/\sqrt{2})^2}-\frac{1}{s}\frac{\partial \delta{\cal V}_{\mbox{\tiny eff.}}}{\partial h_s^0/\sqrt{2}}\right>\label{masstadpole}
\end{align}
\begin{align*}
 & \delta\left<{\cal M}^2_{H^0}\right>_{44}=\frac{1}{2}\left<\frac{\partial^2 \delta{\cal V}_{\mbox{\tiny eff.}}}{(\partial a_u^0/\sqrt{2})^2}-\frac{1}{v_u}\frac{\partial \delta{\cal V}_{\mbox{\tiny eff.}}}{\partial h_u^0/\sqrt{2}}\right> & &
\delta\left<{\cal M}^2_{H^0}\right>_{45}=\frac{1}{2}\left<\frac{\partial^2 \delta{\cal V}_{\mbox{\tiny eff.}}}{\partial a_u^0/\sqrt{2}\partial a_d^0/\sqrt{2}}\right>\\
 & \delta\left<{\cal M}^2_{H^0}\right>_{55}=\frac{1}{2}\left<\frac{\partial^2 \delta{\cal V}_{\mbox{\tiny eff.}}}{(\partial a_d^0/\sqrt{2})^2}-\frac{1}{v_d}\frac{\partial \delta{\cal V}_{\mbox{\tiny eff.}}}{\partial h_d^0/\sqrt{2}}\right> & &
\delta\left<{\cal M}^2_{H^0}\right>_{46}=\frac{1}{2}\left<\frac{\partial^2 \delta{\cal V}_{\mbox{\tiny eff.}}}{\partial a_u^0/\sqrt{2}\partial a_s^0/\sqrt{2}}\right>\\
 & \delta\left<{\cal M}^2_{H^0}\right>_{56}=\frac{1}{2}\left<\frac{\partial^2 \delta{\cal V}_{\mbox{\tiny eff.}}}{\partial a_d^0/\sqrt{2}\partial a_s^0/\sqrt{2}}\right> & &
\delta\left<{\cal M}^2_{H^0}\right>_{66}=\frac{1}{2}\left<\frac{\partial^2 \delta{\cal V}_{\mbox{\tiny eff.}}}{(\partial a_s^0/\sqrt{2})^2}-\frac{1}{s}\frac{\partial \delta{\cal V}_{\mbox{\tiny eff.}}}{\partial h_s^0/\sqrt{2}}\right>
\end{align*}
\begin{align*}
 & \delta\left<{\cal M}^2_{H^0}\right>_{14}=\frac{1}{2}\left<\frac{\partial^2 \delta{\cal V}_{\mbox{\tiny eff.}}}{\partial h_u^0/\sqrt{2}\partial a_u^0/\sqrt{2}}\right> & &
\delta\left<{\cal M}^2_{H^0}\right>_{15}=\frac{1}{2}\left<\frac{\partial^2 \delta{\cal V}_{\mbox{\tiny eff.}}}{\partial h_u^0/\sqrt{2}\partial a_d^0/\sqrt{2}}-\frac{1}{v_d}\frac{\partial \delta{\cal V}_{\mbox{\tiny eff.}}}{\partial a_u^0/\sqrt{2}}\right>\\
 & \delta\left<{\cal M}^2_{H^0}\right>_{16}=\frac{1}{2}\left<\frac{\partial^2 \delta{\cal V}_{\mbox{\tiny eff.}}}{\partial h_u^0/\sqrt{2}\partial a_s^0/\sqrt{2}}-\frac{1}{s}\frac{\partial \delta{\cal V}_{\mbox{\tiny eff.}}}{\partial a_u^0/\sqrt{2}}\right> & &
\delta\left<{\cal M}^2_{H^0}\right>_{24}=\frac{1}{2}\left<\frac{\partial^2 \delta{\cal V}_{\mbox{\tiny eff.}}}{\partial h_d^0/\sqrt{2}\partial a_u^0/\sqrt{2}}-\frac{1}{v_d}\frac{\partial \delta{\cal V}_{\mbox{\tiny eff.}}}{\partial a_u^0/\sqrt{2}}\right>\\
 & \delta\left<{\cal M}^2_{H^0}\right>_{25}=\frac{1}{2}\left<\frac{\partial^2 \delta{\cal V}_{\mbox{\tiny eff.}}}{\partial h_d^0/\sqrt{2}\partial a_d^0/\sqrt{2}}\right> & &
\delta\left<{\cal M}^2_{H^0}\right>_{26}=\frac{1}{2}\left<\frac{\partial^2 \delta{\cal V}_{\mbox{\tiny eff.}}}{\partial h_d^0/\sqrt{2}\partial a_s^0/\sqrt{2}}-\frac{v_u}{sv_d}\frac{\partial \delta{\cal V}_{\mbox{\tiny eff.}}}{\partial a_u^0/\sqrt{2}}\right>\\
 & \delta\left<{\cal M}^2_{H^0}\right>_{34}=\frac{1}{2}\left<\frac{\partial^2 \delta{\cal V}_{\mbox{\tiny eff.}}}{\partial h_s^0/\sqrt{2}\partial a_u^0/\sqrt{2}}-\frac{1}{s}\frac{\partial \delta{\cal V}_{\mbox{\tiny eff.}}}{\partial a_u^0/\sqrt{2}}\right> & &
\delta\left<{\cal M}^2_{H^0}\right>_{35}=\frac{1}{2}\left<\frac{\partial^2 \delta{\cal V}_{\mbox{\tiny eff.}}}{\partial h_s^0/\sqrt{2}\partial a_d^0/\sqrt{2}}-\frac{v_u}{sv_d}\frac{\partial \delta{\cal V}_{\mbox{\tiny eff.}}}{\partial a_u^0/\sqrt{2}}\right>
\end{align*}
\begin{align*}
 \delta\left<{\cal M}^2_{H^0}\right>_{36}=\frac{1}{2}\left<\frac{\partial^2 \delta{\cal V}_{\mbox{\tiny eff.}}}{\partial h_s^0/\sqrt{2}\partial a_s^0/\sqrt{2}}-\frac{2}{s}\frac{\partial \delta{\cal V}_{\mbox{\tiny eff.}}}{\partial a_s^0/\sqrt{2}}+\frac{2v_u}{s^2}\frac{\partial \delta{\cal V}_{\mbox{\tiny eff.}}}{\partial a_u^0/\sqrt{2}}\right>
\end{align*}

This concludes the presentation of the general formalism and we may now describe the various contributions to the effective potential which are computed 
within our code.

\vspace{0.2cm}
{\em ii)} \verb|mhiggstree_CPV.f|\newline
This subroutine simply defines the tree-level mass matrices at the scale $Q$ according to Eqs.\ref{charhiggstreemass} and \ref{neuthiggstreemass}.
However, the corrected Higgs masses are not the only information that we want to extract from the effective potential: the Higgs-to-Higgs couplings are
also encoded within this formalism. Therefore, and for reasons that will become clear when we implement the various radiative contributions to the
potential, we wish to match the full effective potential onto the following and simpler one:
\begin{align}
 \tilde{\cal V}_{\mbox{\tiny eff}}= &M_S^2|S|^2+\frac{A_S}{3}\left[e^{\imath\varphi_{A_S}}S^3+h.c.\right]+{\cal V}_0(|S|^2)\label{potential}\\
\null & +(M_u^2+\lambda_P^u|S|^2)|H_u|^2+(M^2_d+\lambda_P^d|S|^2)|H_d|^2+\left[\left(A_{ud} e^{\imath\varphi_{A_{ud}}}S+\lambda_P^Me^{\imath\varphi_M}S^{*2}\right)H_u\cdot H_d+h.c.\right]\nonumber\\
\null & +\frac{\lambda_u}{2}|H_u|^4+\frac{\lambda_d}{2}|H_d|^4+ \lambda_3|H_u|^2|H_d|^2+\lambda_4|H_u\cdot H_d|^2\nonumber\\
\null & +\left[\frac{\lambda_5}{2}e^{\imath\varphi_5}(H_u\cdot H_d)^2+(\lambda_6e^{\imath \varphi_6}|H_u|^2+\lambda_7e^{\imath \varphi_7}|H_d|^2)H_u\cdot H_d+h.c.\right]\nonumber
\end{align}
This is a subset of the most general singlet + two doublet potential which one can write up to dimension 4 terms\footnote{Note however that we do not
specify ${\cal V}_0(|S|^2)$ further. If only terms of dimension $\leq4$ are kept, then the only choice would be ${\cal V}_0(|S|^2)=K^2|S|^4$.}.
The gauge symmetry is observed. However the $\mathbb{Z}_3$-symmetry only holds up to terms quadratic in the doublet fields and is explicitly broken by the terms
in the last line of Eq.\ref{potential}. This potential is meant as an expansion of Eq.\ref{effpot} in the doublet fields and as we mentioned before, 
there is no reason why the $\mathbb{Z}_3$-symmetry should hold in such an expansion. The characteristics of this potential are studied in appendix \ref{apsimppot} 
and matching the tree-level expression of Eq.\ref{tlHiggspot} is straightforward (see appendix \ref{apsimppotmatchtree}).

\vspace{0.2cm}
{\em iii)} \verb|mhiggsloop_sferm_CPV.f|\newline
With this subroutine, we start adding radiative corrections to the effective potential, here those arising from SM-fermion and sfermion loops. These -- 
particularly the contribution associated to the top -- are known to convey the dominant radiative effect and lead e.g.\ to a substancial shift 
of the squared-mass of the SM-like Higgs boson.

The corresponding one-loop effects to the neutral Higgs mass matrix are particularly easy to include in the Coleman-Weinberg formalism of Eq.\ref{effpot}, since the 
bilinear terms provide relatively simple matrices (refer to the appendices \ref{tlsmferm} and \ref{tlsfermmass}). The details of the corrections are 
developed in the appendices \ref{aploopsmferm} and \ref{aploophisferm}. Note that we recover Eqs.(C16-18) of \cite{NMSSM} in the CP-conserving limit.

The situation is slightly more complex for the charged Higgs as well as for the 
Higgs-to-Higgs couplings: we then decide to expand the potential in terms of the doublet fields, up to quartic order $H^4$ and match the corresponding 
expansion onto the simplified potential of Eq.\ref{potential}. This amounts to an expansion in $v/M_{\mbox{\tiny SUSY}}$, where $M_{\mbox{\tiny SUSY}}$
here stands for any sfermion mass. The sfermion contributions to the coefficients of Eq.\ref{potential} are also provided in appendix \ref{aploophisferm}.
Note that this alternative approach allows for a numerical cross-check with the corrections applied to the mass matrix of the neutral Higgs states with the 
method described in the previous paragraph.

In addition to these one-loop effects, the subroutine \verb|mhiggsloop_sferm_CPV.f| also includes two-loop effects of order\footnote{The conventions 
$O(Y_{t,b}^2\alpha_S,Y_{t,b}^4)$ or $O(\alpha_{t,b}\alpha_S,\alpha_{t,b}^2)$ are also used in the literature.} $O(Y_{t,b}^4\alpha_S,Y_{t,b}^6)$
leading to a product of large logarithms in the fermion sector: given that we are working at the average scale of squark masses, the squarks are 
assumed to give subleading contributions. On the other hand, effects associated to SM fermions and gauge bosons will not introduce any additional 
dependence on the new physics phases. The corresponding effects are implemented in the approximations of \cite{Ellwanger:1999ji} -- see also 
Eq.(C.19) of \cite{NMSSM} --, i.e.\ only the contributions to the quartic doublet parameters $\lambda_u$ and $\lambda_d$ of Eq.\ref{potential} are
included -- note that contributions to $M_u^2$ or $M_d^2$ leave the analysis unaffected, while contributions to e.g.\ $A_{ud}$ can be absorbed in
a shift of the tree-level term $A_{\lambda}$, hence only drive a displacement in the parameter space. While these contributions are of two-loop order, 
they may still affect the mass of the SM-like Higgs state by several GeV, which is why we include them. Comparisons to more-elaborate two-loop 
calculations show that this approximation works well numerically (at the GeV level). Two-loop effects beyond this order have been considered in 
\cite{Goodsell:2014pla}.

\vspace{0.2cm}
{\em iv)} \verb|mhiggsloop_inos_CPV.f|\newline
The next subroutine implements the radiative effects associated to charginos and neutralinos. Sticking to the Coleman-Weinberg approach, we
consider the $9\times9$ bilinear term associated with gauginos and higgsinos (refer to appendix \ref{tlchaneumass}). Due to the large rank of
this matrix, we exclusively employ the method which consists in expanding the potential and matching it to the simplified version of Eq.\ref{potential}. 
The corresponding results are collected in appendix \ref{aploophiinos}. Note that they differ from e.g.\ Eq.(C.22-24) of \cite{NMSSM} where additional
simplifying assumptions had been made.

\vspace{0.2cm}
{\em v)} \verb|mhiggsloop_gaugehiggs_CPV.f|\newline
The contributions of the electroweak gauge bosons to the Higgs potential seem easy to include in the Landau gauge: see appendix \ref{aploopsmgauge}.
Yet the drawbacks of the Landau gauge are felt in the Higgs sector, where one then has to handle massless Goldstone bosons. The associated infrared
divergences are of course purely spurious and disappear once confronted to momentum-dependent corrections, as already noted in \cite{Casas:1994us}.
Still it remains a technical issue to manipulate with caution. Moreover, the strategy consisting in diagonalizing the field-dependent bilinear matrices, 
which we have been employing until here, becomes impractical, even in an expansion in terms of the doublet fields, due to the large number of parameters 
and operators involved in the Higgs bilinear terms. Instead, we decide to employ the concurrent strategy in Higgs-mass calculations, which simply consists 
in a direct diagrammatic evaluation of the Higgs self-energies and tadpoles generated by Higgs loops. Nevertheless, disentangling the Higgs and gauge contributions
in this approach proves quite artificial so that we will perform the calculation simultaneously for both types of particles appearing in the loop.

Explicit expressions for the gauge and Higgs one-loop contributions to the Higgs self-energies and tadpoles are summarized e.g.\ in \cite{Pierce:1996zz}
or \cite{Martin} (with different conventions for the loop functions), in the context of the MSSM, and the NMSSM differs only in the definition of the 
couplings and the presence of the singlet fields, hence leads to a formally comparable result. We choose to work in the Feynmann gauge as it is then
possible to set the external momentum to $0$ without generating IR-divergent logarithms. Indeed, we still aim at computing, not only the corrections
to the Higgs masses, but also to the Higgs-to-Higgs couplings. For this, we proceed in the following fashion: after the radiative corrections to the Higgs 
mass matrices are evaluated at zero momentum, we subtract the pure-gauge contribution in the Landau gauge (for which we already know the potential from
appendix \ref{aploopsmgauge}). 
The remaining `Higgs' contributions to the mass matrices can then be identified with those that a $\mathbb{Z}_3$-conserving renormalizable potential 
would produce, allowing for a reconstruction of the corrections to the $\mathbb{Z}_3$-conserving parameters of the potential: this procedure is 
described for the CP-conserving case in \cite{Chalons:2012qe} and is straightforwardly generalized to the CP-violating case. Of course, we then miss 
contributions to the $\mathbb{Z}_3$-violating parameters (the last line of Eq.\ref{potential}) but, as discussed in \cite{Chalons:2012qe}, these are 
subleading in the leading-logarithmic approach\footnote{Note that, while the parameters of the effective potential will then receive contributions which 
are valid only at leading logarithmic order, this is not the case for the contributions to the Higgs masses since they are obtained directly from the 
diagrammatic calculation.}. Further details can be found in appendix \ref{aploophihiggs}.

This completes the list of radiative contributions implemented in the effective potential.

\subsubsection{Pole corrections}
The operations described in the previous lines have provided us with mass matrices for the Higgs states where radiative corrections from the potential
(i.e.\ at zero external momentum) have been included. We will now detail how we account for momentum-dependent corrections. These calculations are
conducted in the subroutine \verb|mhiggsloop_pole_CPV.f|.

First, let us remind the reader that the radiative effects associated with non vanishing external momentum have been partially encoded into the wave-function
scaling factors of paragraph \ref{secwavefunc}. It is necessary to rescale the Higgs mass-matrices in order to account for the re-scaling of the 
Higgs fields:
\begin{equation}
 \left<{\cal M}^2_{H^{\pm}}\right>_{ij}\leftarrow\frac{1}{\sqrt{Z_{H_i}Z_{H_j}}}\left<{\cal M}^2_{H^{\pm}}\right>_{ij}\hspace{1cm};\hspace{1cm}
\left<{\cal M}^2_{H^0}\right>_{ij}\leftarrow\frac{1}{\sqrt{Z_{H_i}Z_{H_j}}}\left<{\cal M}^2_{H^0}\right>_{ij}
\end{equation}
A $\beta$-angle rotation in the pseudoscalar and charged sector allows to rotate away the Goldstone bosons, leaving us with a $5\times5$ symmetric mass 
matrix $\left<\tilde{\cal M}^2_{H^0}\right>_{ij}$ for the neutral sector and a $\overline{DR}$ squared-mass for the charged Higgs 
$m^{2\,\overline{DR}}_{H^{\pm}}$. $\left<\tilde{\cal M}^2_{H^0}\right>$ is now diagonalized according to Eq.\ref{diagneuthiggs}, providing us with
corrected $\overline{DR}$ squared-masses for the neutral Higgs, $m^{2\,\overline{DR}}_{S_i^0}$, and their rotation matrix $X^{H^0}$.

We then apply pole corrections to the $\overline{DR}$ squared-masses in order to evaluate the pole masses:
\begin{align}
& m^{2}_{S_i^0}=m^{2\,\overline{DR}}_{S_i}\Big(1+\sum_{j}\delta Z_{H_j}X^{H^0\,2}_{ij}\Big)-\left[\Pi_{S_i^0S_i^0}(p^2=m^{2}_{S_i^0})-\Pi_{S_i^0S_i^0}(p^2=0)\right]\nonumber\\
& m^{2}_{H^{\pm}}=m^{2\,\overline{DR}}_{H^{\pm}}\left(1+\delta Z_{H_u}\cos^2\beta+\delta Z_{H_d}\sin^2\beta\right)-\left[\Pi_{H^{+}H^{-}}(p^2=m^{2}_{H^{\pm}})-\Pi_{H^{+}H^{-}}(p^2=0)\right]
\end{align}
While ideally the Higgs self energies $\Pi_{S_i^0S_i^0}(p^2)$ and $\Pi_{H^{+}H^{-}}(p^2)$ should be evaluated at the pole masses, we approximate the latter 
by the $\overline{DR}$ masses. The full one-loop pole-corrections are applied. Shifts of the wave-function scaling factors $\delta Z_{H_u,H_d,S}$ are know
from Eqs.\ref{wavefuncsmferm}-\ref{wavefuncgauge}. The shifts in the Higgs self energies are provided in appendix \ref{appolecorr}.

This concludes our evaluation of the masses in the Higgs sector. We now wish to comment briefly on the precision achieved in this calculation. For this, 
it is instructive to consider the impact of the one-loop corrections with respect to the situation at tree-level. For mostly-doublet states, the leading 
effect is driven by the top-quark loop and, respectively to the tree-level mass $m_H$, can be quantified as 
$\sim\frac{N_cY_t^2}{4\pi^2}\frac{m_t^2}{m_H^2}\ln\frac{m_t^2}{Q^2}$. Assuming that $Q=O(\mbox{TeV})$, this amounts to a correction at the percent level
for $m_H=O(\mbox{TeV})$, but reaching a magnitude of $\sim100\%$ for $m_H=O(100~\mbox{GeV})$: this accounts for the well-known sensitivity of the 
SM-like Higgs mass to radiative corrections. Contributions at the two-loop order will involve the strong coupling $g_S$, or the top Yukawa coupling again,
multiplying logarithms of a similar magnitude, so that the typical effect would easily amount to $\sim30\%$ of the one-loop contribution. Now, considering that we
have included the leading double-logarithmic effects in the calculation, we can estimate a reduced uncertainty from higher orders, say at the level of 
$\sim10\%$ of the one-loop corrections. For a Higgs mass at $\sim125$~GeV, this still amounts to an uncertainty of several GeV. For a state at 
$m_H=O(\mbox{TeV})$, this reduces to the permil level. The latter accuracy is treacherous however, as other sources of uncertainty appear e.g.\ in the 
determination of the couplings or neglected electroweak corrections entering the definition of the Higgs v.e.v.'s. In the outcome, one should not expect
a precision under $O(1\%)$ for the masses of the heavy doublet states. Corrections to singlet states are typically smaller, since the associated couplings --
$\lambda$, $\kappa$ -- are of order $\lsim O(0.7)$ and the hierarchies between Higgs bosons and higgsinos may not be as large as those between SM fermions
and sfermions. However, when the singlets mix significantly with doublet states, they will correspondingly acquire part of the larger uncertainties on
doublet masses. 

\subsection{Couplings, decays and constraints}
After the previous subroutines are run, one has a complete set of corrected masses and rotation matrices at one's disposal. The following move
consists in confronting this spectrum to physical processes.

\subsubsection{Supersymmetric and Higgs couplings}
The couplings of supersymmetric particles and Higgs bosons can result in somewhat lengthy expressions. We thus design two subroutines, \verb|susycoup_CPV.f|
and \verb|higgscoup_CPV.f|, in order to evaluate and store them within the code:
\begin{itemize}
 \item The couplings of charginos / neutralinos to sfermions and SM fermions are implemented according to the formulae of appendix \ref{apcoupchasff}
and \ref{apcoupneusff} for the three generations (still neglecting the Yukawa couplings of the two first generations).
 \item The trilinear couplings of the Higgs bosons to sfermions are computed after the results of appendix \ref{apHiSFcoup}.
 \item The couplings of the Higgs bosons to charginos and neutralinos are also included as presented in appendix \ref{aphichacoup}.
 \item Finally, we calculate corrected $\overline{DR}$ Higgs-to-Higgs couplings where radiative effects from sfermions, charginos, neutralinos and Higgs 
bosons are obtained from the simplified effective potential of Eq.\ref{potential}: relevant formulae are provided in appendix \ref{apsimpeffpotcoup}. 
Corrections from fermionic and gauge loops are explicitly incorporated as given in appendix \ref{aploopsmferm} and \ref{aploopsmgauge}. The Yukawa
and gauge couplings employed here have been run to the scale corresponding to the mass of the neutral Higgs state associated with the first index in the coupling.
\end{itemize}
Note that the rescaling of Higgs fields in Eqs.\ref{neutscaling} and \ref{charscaling} is also accounted for when computing the couplings of Higgs bosons.

\subsubsection{Higgs and top decays}
We then adapt the existing \verb|NMSSMTools| subroutines \verb|decay.f| and \verb|tdecay.f| -- respectively computing the Higgs and top two-body decays in 
the CP-conserving NMSSM -- to the CP-violating case.

The subroutine \verb|hidecay_CPV.f| calculates the Higgs widths and the dominant branching ratios. The following decay channels are considered:
\begin{itemize}
 \item decays into a pair of SM fermions: $S_i^0\to\mu^+\mu^-$, $\tau^+\tau^-$, $s\bar{s}$, $c\bar{c}$, $b\bar{b}$, $t\bar{t}$; 
$H^+\to\mu^+\nu_{\mu}$, $\tau^+\nu_{\tau}$, $u\bar{s}$, $u\bar{b}$, $c\bar{s}$, $c\bar{b}$, $t\bar{b}$;
 \item decays into (on- and off-shell) gauge bosons: $S_i^0\to WW$, $ZZ$, $\gamma\gamma$, $Z\gamma$, $gg$;
 \item decays into one Higgs and one gauge boson: $S_i^0\to ZS_j^0$, $W^{\pm}H^{\mp}$; $H^+\to W^+S_j^0$;
 \item Higgs-to-Higgs decays: $S_i^0\to S_j^0S_k^0$, $H^+H^-$;
 \item supersymmetric decays: $S_i^0\to\chi_j^+\chi^-_k$, $\chi_j^0\chi^0_k$, $\tilde{F}_j^*\tilde{F}_k$; $H^+\to\chi_j^+\chi_k^0$,
$\tilde{F}^*_j\tilde{F}'_k$.
\end{itemize}

In the subroutine \verb|tdecay_CPV.f|, we compute the following top decays: $t\to W^+b$, $H^+b$, $\tilde{T}\chi^0$. As in the original
CP-conserving version, leading QCD corrections have been taken into account.

\subsubsection{Phenomenological tests}
We finally propose several tools to confront the CP-violating NMSSM spectrum to experimental constraints.

\verb|checkmin_CPV.f| compares the value of the neutral effective potential at the electroweak symmetry-breaking minimum with that at other points, 
e.g.\ for vanishing v.e.v.'s. Loop effects from the SM fermions and gauge bosons are included explicitely in this evaluation, while other 
radiative effects are encoded within the approximate potential of Eq.\ref{potential}. We also vary the dynamical phases and check whether this
generates a deeper minimum. Finally, the minimization conditions of Eq.\ref{minpot} and \ref{tadpole} are calculated explicitly, which allows e.g.\ to 
test the naturalness of the squared masses $m^2_{H_{u,d}}$ of the potential: they should remain of the order of the SUSY-breaking scale.

In \verb|constsusypart_CPV.f|, we generalize to the CP-violating case LEP limits on superparticle searches that were included in \verb|NMSSMTools|
for the CP-conserving case:
\begin{itemize}
 \item test on chargino, slepton, gluino and squark masses;
 \item limits on $\tilde{T}\to b l\tilde{N}$, $\tilde{T}\to c\chi^0$, $\tilde{B}\to b\chi^0$;
 \item constraint on the invisible $Z$-width and neutralino-pair production.
\end{itemize}

\verb|HBNMSSM_CPV.f| converts our spectrum into input for \verb|HiggsBounds| \cite{HiggsBounds} and \verb|HiggsSignals| \cite{HiggsSignals}. This allows to test the 
Higgs sector in view of LEP, TeVatron and LHC results via a call to the subroutines included within these public tools -- note that \verb|NMSSMTools|, 
\verb|HiggsBounds| and \verb|HiggsSignals| must be interfaced to make use of this subroutine. The chosen input mode is that employing effective couplings (see the 
documentation in \cite{HiggsBounds}). Additional widths and branching ratios are taken from the results in \verb|hidecay_CPV.f| and 
\verb|tdecay_CPV.f|. In the following section, we will be using the current versions \verb|HiggsBounds_4.2.0| and \verb|HiggsSignals_1.3.1|, which
incorporate all the experimental results released till december 2014. The default uncertainty on the Higgs mass precision is set to $3$~GeV and modeled 
as a gaussian distribution. \verb|HiggsBounds| delivers a $95\%$ confidence level cut on the NMSSM parameter
space relative to limits from unsuccessful Higgs boson searches. \verb|HiggsSignals| performs a $\chi^2$-test of the Higgs properties of a given spectrum based on the 
current experimental characteristics of the the signals measured at $\sim125$~GeV. The default setting of version $1.3.1$ includes $81$ test-channels based on the material released by the ATLAS, CMS, CDF and D0 collaborations. 
The output, the $\chi^2$-value, provides a measure of the compatibility of the tested Higgs spectrum with the observed signals. While the implementation 
within \verb|HiggsSignals| accounts for more involved effects, such as correlations among channels or uncertainties, the $\chi^2$ can be grossly understood 
as the sum of squared deviations between theoretical predictions and experimental measurements, normalized for each channel to the sum of squared 
theoretical and experimental uncertainties for this channel: therefore, the smaller the $\chi^2$, the more compatible the Higgs spectrum proves in
view of the measured Higgs signals. Statistical interpretations in terms of P-values are possible, yet depend on the details of the chosen tests or of 
the definition of the statitical ensembles: in our discussion, we will confine to the thumb rule stating that $\chi^2$-values of the order of the number 
of test-channels ($81$ here) are regarded as competitive. The $\chi^2$-value in the SM limit ($\sim78$ in practice) gives another point of reference from which one may 
appreciate the quality of the fit of a particular Higgs spectrum to the observed Higgs signals.

We also include an alternative set of tests for the Higgs sector, based on the original subroutines of \verb|NMSSMTools|. These collect:
\begin{itemize}
 \item \verb|LEP_Higgs_CPV.f|: LEP limits applying on neutral Higgs bosons produced in association with $Z$'s -- $e^+e^-\to Z^*\to S_i^0 Z$ -- or in 
pairs -- $e^+e^-\to Z^*\to S_i^0 S_j^0$ -- \cite{Barate:2003sz}; 
 \item \verb|TeVatron_CHiggs_CPV.f|: TeVatron limits applying on a charged Higgs boson produced in top decays \cite{TeVatron};
 \item \verb|bottomonium_CPV.f|: test for a light mostly CP-odd Higgs intervening in bottomonium decays -- based on \cite{Domingo:2008rr};
 \item \verb|LHC_Higgs_CPV.f|: the inclusion of LHC limits on neutral or charged Higgs searches as well as the confrontation to the signals at
$\sim125$~GeV -- after \cite{Belanger:2013xza} -- are in progress.
\end{itemize}
Note that these routines will not be used in the next section, as we will employ the currently more complete set of tests performed by \verb|HiggsBounds| 
and \verb|HiggsSignals|.

Finally, we design a subroutine \verb|EDM_CPV.f| to estimate the electric dipole moments of the electron, the thallium atom, the neutron and the 
mercury atom. We essentially follow the summary in \cite{Ellis:2008zy} -- in the context of the MSSM; see also \cite{Cheung:2011wn} for previous works 
in the NMSSM. The supersymmetric one-loop effects are mediated by charginos, neutralinos or gluinos and sfermions. Moreover the two-loop 
diagrams of the Bar-Zee type -- involving a fermion or sfermion loop connected to the quark / electron line by a Higgs and a photon propagator -- are 
known to convey a sizable effect: these are particularly sensitive to the phases appearing in the Higgs sector. Other contributions, mediated by 
dimension $6$ operators, are included as well. We estimate the associated uncertainties by adding linearly a $10\%$ error on effects involving no 
coloured particles and a $30\%$ error on contributions involving the coloured sector. Additional uncertainties associated to scale-running or hadronic 
parameters are also incorporated.

\section{A few applications}
We shall now make use of the subroutines which we have just presented and study phenomenological effects associated with the CP-violating NMSSM. This
will be the opportunity to test our tool and compare its predictions with existing results.

\subsection{CP-conserving limit}
Setting all the phases to zero, it is possible to consider the CP-conserving case: in particular this allows to study how our results connect to
the precision calculations implemented within \verb|NMSSMTools|. Given that the input is common, discrepancies directly give an insight on the differences
of treatment and the numerical magnitude of the corresponding effects.

\begin{figure}[!htb]
    \centering
    \includegraphics[width=15cm]{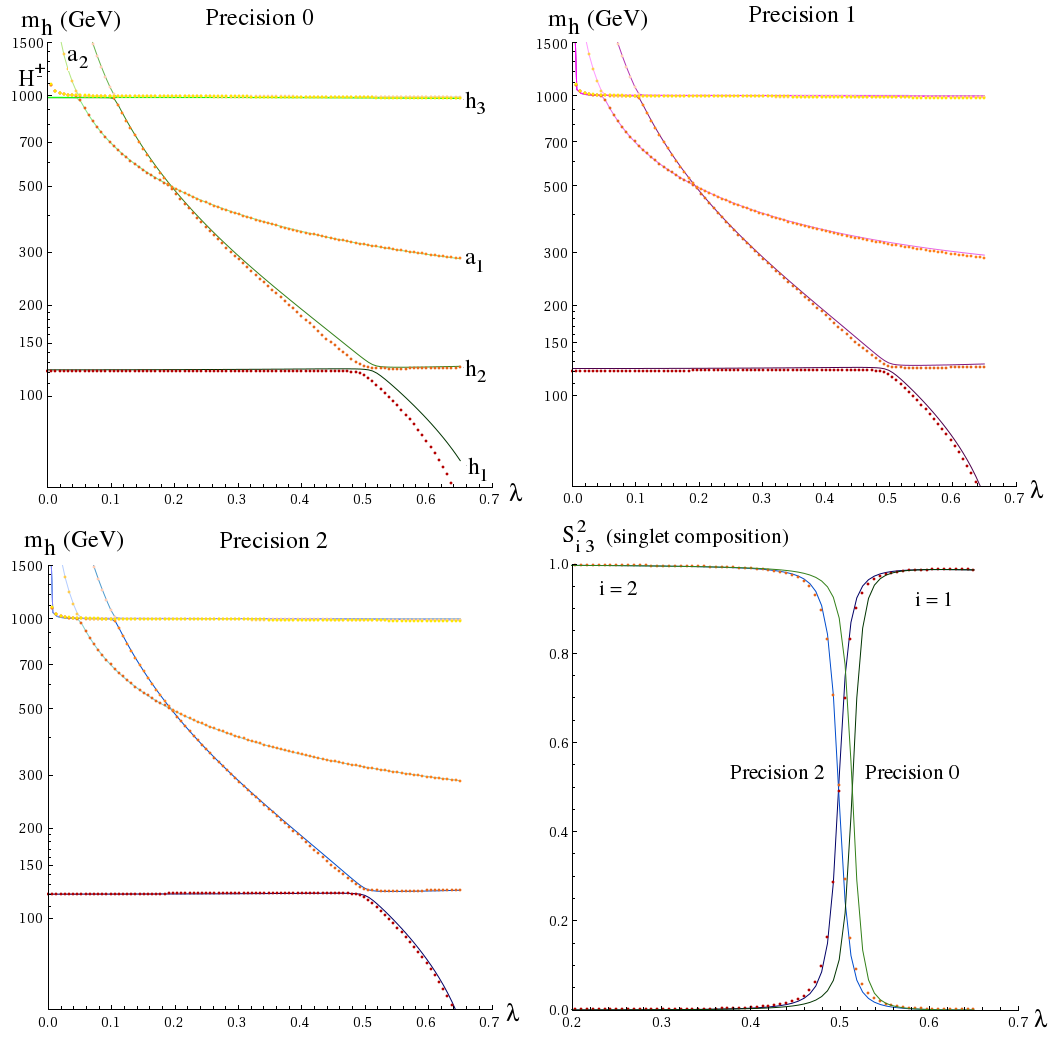}
  \caption{General aspect of the Higgs spectrum in the NMSSMTools procedures with precision $0$ (upper left), precision $1$ (upper right) and
precision $2$ (down left-hand) and in ours (dotted lines). $\lambda\in[0,0.65]$, $\kappa=0.45$, $\tan\beta=8$, $\mu_{\mbox{\tiny eff}}=125$~GeV, $M_A=1$~TeV, 
$A_{\kappa}=-288$~GeV, $m_{\tilde{T},\tilde{B}}=1$~TeV, $m_{\tilde{U},\tilde{D}}=1.5$~TeV, $m_{\tilde{L},\tilde{E}}=200$~GeV, $2M_1=M_2=M_3/3=0.5$~TeV,
$A_t=-2$~TeV, $A_{b,\tau}=-1.5$~TeV. In the down right-hand plot, the singlet composition $S_{i3}^2$ of the two lightest CP-even states is displayed
for precision $0$ (green solid lines), precision $2$ (blue solid lines) and for our calculation (dotted lines).}
  \label{figmH1}
\end{figure}
\vspace{0.2cm}
{\em i) Higgs spectrum}\newline
We shall first consider the Higgs masses. \verb|NMSSMTools| provides three levels of precision in the inclusion of the radiative corrections to the 
$\mathbb{Z}_3$-conserving Higgs sector:
\begin{itemize}
\item `Precision $0$': the default one -- essentially following the procedure described in appendix C of \cite{NMSSM} -- confines to leading logarithmic 
order. Momentum-dependent effects are taken into account only to the extent of wave-function renormalization (where the implementation is slightly 
different from ours: remember the discussion in section \ref{secwavefunc}) and pole-corrections associated with the SM-fermion sector.
\item `Precision $1$': a full one-loop + leading two-loop (to order $O(Y_{t,b}^4\alpha_S)$) implementation without momentum-dependent effects.
\item `Precision $2$': a full one-loop + leading two-loop (to order $O(Y_{t,b}^4\alpha_S)$) implementation including momentum-dependent effects. It 
follows the work of \cite{Degrassi:2009yq}.
\end{itemize}
Formally, our implementation -- full one-loop including momentum-dependent corrections + leading two-loop double logarithms of order $O(Y_{t,b}^4\alpha_S,
Y_{t,b}^6)$ -- should fall somewhere between these three procedures in terms of precision.

\begin{figure}[!htb]
    \centering
    \includegraphics[width=15cm]{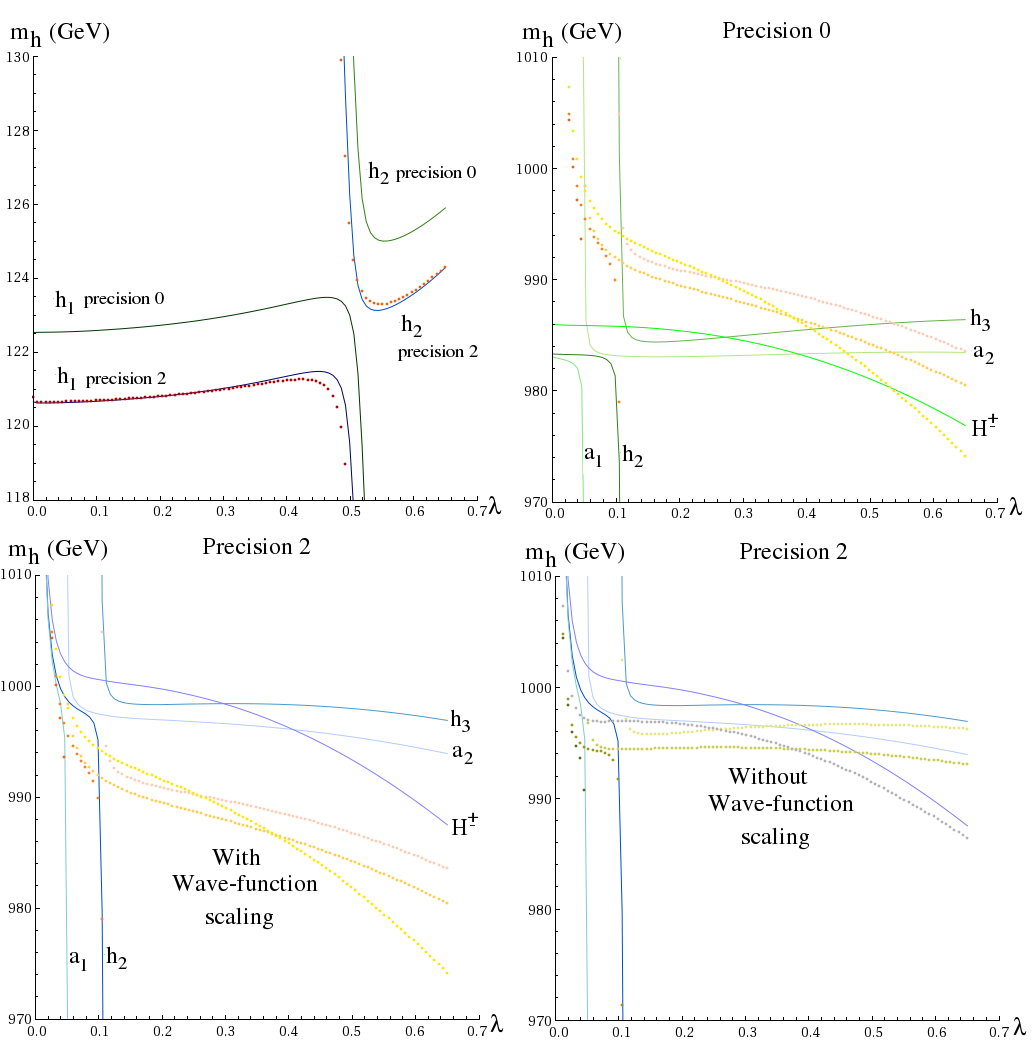}
  \caption{Details on the Higgs masses of Fig.\ref{figmH1}. On the upper left-hand quadrant, we show the Higgs masses close to $\sim125$~GeV for 
precision $0$ (green lines), precision $2$ (blue lines) and for our implementations (red dots). The plot on the upper right-hand side compares our 
results (dots) for the `heavy' masses with those of precision $0$ (green lines). The same exercise is carried out in the lower left-hand corner for
precision $2$ (blue lines). In the lower right-hand quadrant, we alter our implementation of the Higgs mass corrections so that all $p^2$-dependent 
terms are taken into account as pole-corrections only (so that the wave-scaling factors are set to $1$): the results are displayed as khaki dots and 
compared to the masses obtained in the procedure of precision $2$.}
  \label{figmH1bis}
\end{figure}

We first test our results in a region of the parameter space where $\kappa=0.45$, $\tan\beta=8$, $\mu_{\mbox{\tiny eff}}=125$~GeV, $M_A=1$~TeV, 
$A_{\kappa}=-288$~GeV, $m_{\tilde{T},\tilde{B}}=1$~TeV, $m_{\tilde{U},\tilde{D}}=1.5$~TeV, $m_{\tilde{L},\tilde{E}}=200$~GeV, $2M_1=M_2=M_3/3=0.5$~TeV,
$A_t=-2$~TeV, $A_{b,\tau}=-1.5$~TeV and we scan over $\lambda\in[0,0.65]$. The Higgs masses are displayed in Fig.\ref{figmH1} and \ref{figmH1bis}: the 
results of \verb|NMSSMTools| for precision `$0$' (greenish colors), `$1$' (pink colors) and `$2$' (bluish colors) are shown as solid lines while our 
calculation corresponds to the dots (yellow to red tones). We 
observe a significant variation of the masses corresponding to the mostly-singlet states while the doublet masses are grossly constant with varying 
$\lambda$. A typical NMSSM effect develops when singlet masses are close to doublet masses, as significant mixing may appear. In particular, when the
singlet state is slightly lighter than the doublet one, the mixing tends to uplift the mass of the mostly-doublet Higgs. This is what occurs in this
example for the CP-even sector in the upper range of $\lambda$. 
In Fig.\ref{figmH1}, we see that our results fit quite closely the predictions of the procedure with precision $2$, while larger 
discrepancies appear with respect to precision $0$, especially at large $\lambda$.

Fig.\ref{figmH1bis} allows for a closer comparison among Higgs masses. For the Higgs states with mass close to $\sim125$~GeV (upper left-hand quadrant), 
we note a remarkable agreement between our calculation and the masses obtained with the precision setting $2$, while the results obtained with precision 
$0$ are about $2$~GeV off: this fact should not make us forget that the uncertainties affecting the Higgs mass computations (also in the setting of 
precision $2$) are of the order of several GeV. However, it justifies the observation that the leading two-loop effects are captured by the simpler 
inclusion of double logarithmic terms.

\begin{figure}[!htb]
    \centering
    \includegraphics[width=15cm]{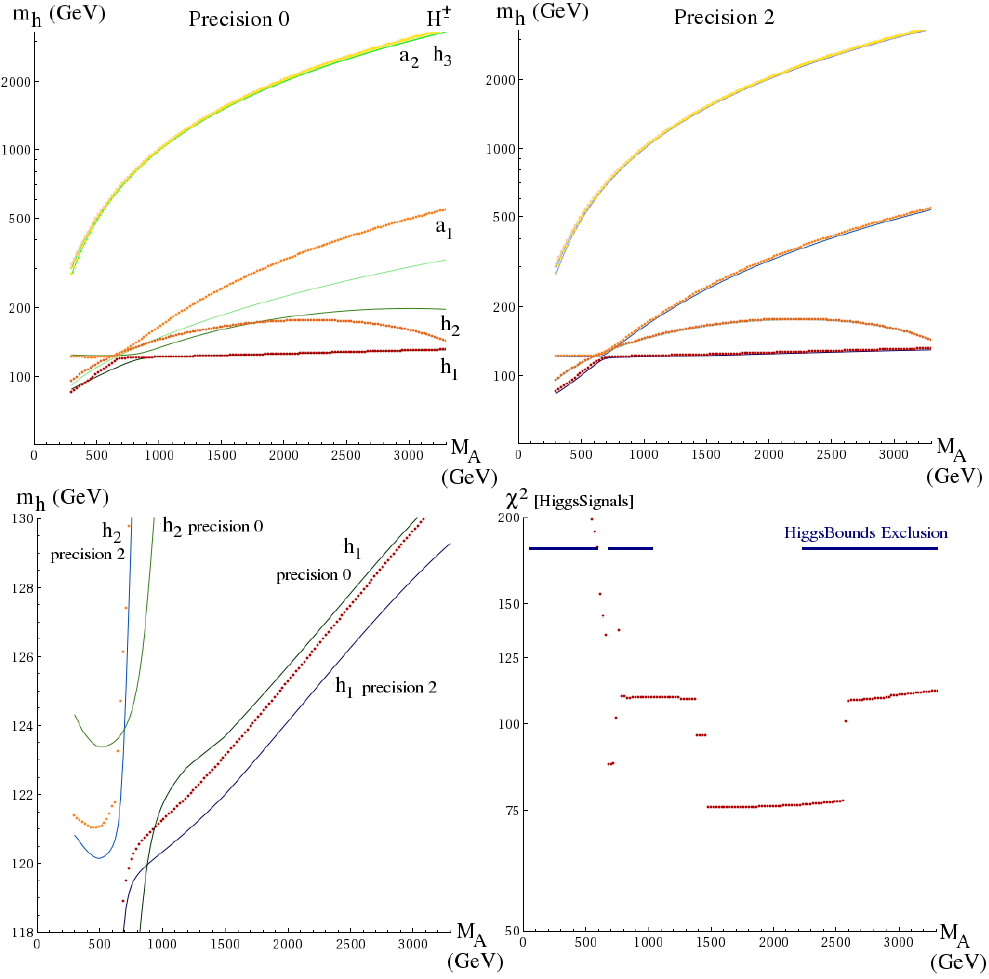}
  \caption{Higgs spectrum for $\lambda=0.7$, $\kappa=0.1$, $\tan\beta=2$, $M_A\simeq 2.33\,\mu_{\mbox{\tiny eff}}+20.45$~GeV, $M_A\in[0.3,3]$~TeV, 
$A_{\kappa}=-50$~GeV, $m_{\tilde{T},\tilde{B}}=0.5$~TeV, $m_{\tilde{U},\tilde{D}}=1.5$~TeV, $m_{\tilde{L},\tilde{E}}=110$~GeV, $2M_1=M_2=150$~GeV, $M_3=1.5$~TeV
$A_{t,b,\tau}=-0.1$~TeV. Comparison of our results (dots) with precision $0$ (upper left-hand plot), precision $2$ (upper right-hand plot). Focus on
the masses close to $125$~GeV (bottom left-hand plot). Finally, results obtained with HiggsBounds and HiggsSignals.}
  \label{figmH2}
\end{figure}
Concerning the heavy mass states, we observe in Fig.\ref{figmH1bis} -- in the upper-right and lower-left hand quadrants -- that our results are 
intermediary between the calculations with precision $0$ and precision $2$. However, we note that the 
leading difference with precision $2$ originates in the implementation of the wave-function scaling factors. Indeed, if we set the `$Z$-factors' to $1$ 
and modify the pole-corrections accordingly, we observe that our result -- corresponding to the khaki dots in the lower-right-hand corner of Fig.\ref{figmH1bis} -- 
then matches that with precision $2$ somewhat more closely (at the permil level). It is quite easy to see how the discrepancy develops between these two procedures.
For this, let us focus on the CP-odd doublet state, where we will neglect the mixing with the singlet. In the case where the wave-function scaling factors
are set to $1$, the squared mass of this state is -- schematically: the effect of potential and pole corrections are encoded as $\delta^{\mbox{\tiny pot},
\mbox{\tiny pole}}$ -- obtained as (all the $p^2$-dependent terms are treated as pole corrections): 
\begin{equation}
\left.m_A^{2}\right|_{\mbox{\small no }Z}=\lambda s\left[A_{\lambda}\cos\varphi_1+\kappa s\cos(\varphi_{\lambda}-\varphi_{\kappa})+\delta^{\mbox{\tiny pot}}\right]\frac{v_u^2+v_d^2}{v_uv_d}\Big(1+\delta^{\mbox{\tiny pole}}\Big) \label{mAwoZ}
\end{equation}
In the approach where the wave-function scaling factors are taken into account at the level of the kinetic terms, the $Z$-factors intervene in the calculation
at several steps: first, for the scaling of the v.e.v.'s, which transforms the tree-level mass-matrix in the CP-odd doublet sector to: 
\begin{equation}
\frac{\lambda s}{\sqrt{Z_S}}\left[A_{\lambda}\cos\varphi_1+\frac{\kappa s}{\sqrt{Z_S}}\cos(\varphi_{\lambda}-\varphi_{\kappa})\right]\begin{pmatrix}\sqrt{\frac{Z_{H_u}}{Z_{H_d}}}\frac{v_d}{v_u}&1\\1&\sqrt{\frac{Z_{H_d}}{Z_{H_u}}}\frac{v_u}{v_d}\end{pmatrix} 
\end{equation}
The scaling effect on the potential corrections can be neglected as being of higher order. Then comes the scaling of the mass-matrix:
\begin{equation}
\frac{\lambda s}{\sqrt{Z_SZ_{H_u}Z_{H_d}}}\left[A_{\lambda}\cos\varphi_1+\frac{\kappa s}{\sqrt{Z_S}}\cos(\varphi_{\lambda}-\varphi_{\kappa})+\delta^{\mbox{\tiny pot}}\right]\begin{pmatrix}\frac{v_d}{v_u}&1\\1&\frac{v_u}{v_d}\end{pmatrix} 
\end{equation}
so that we can extract the $\overline{DR}$ squared mass for the physical state via a $\beta$-angle rotation. Finally, the $Z$-factors have to be 
subtracted from the pole corrections, since they have been accounted for elsewhere. This provides:
\begin{equation}
\left.m_A^{2}\right|_{Z}=\frac{\lambda s}{\sqrt{Z_SZ_{H_u}Z_{H_d}}}\left[A_{\lambda}\cos\varphi_1+\frac{\kappa s}{\sqrt{Z_S}}\cos(\varphi_{\lambda}-\varphi_{\kappa})+\delta^{\mbox{\tiny pot}}\right]{\textstyle\frac{v_u^2+v_d^2}{v_uv_d}}\Big[1+\delta^{\mbox{\tiny pole}}+\cos^2\beta (Z_{H_u}-1)+\sin^2\beta (Z_{H_d}-1)\Big]\label{mAwZ} 
\end{equation}
Expanding the $Z$-factors as $Z_{\cdot}=1+\delta Z_{\cdot}$, we see that Eqs.\ref{mAwoZ} and \ref{mAwZ} differ by a factor $1-\frac{\delta Z_S}{2}+\frac{\cos{2\beta}}{2}(\delta Z_{H_u}-\delta Z_{H_d})$.
This explains the mismatch, reaching the order of one-loop effects, that is $O(1\%)$ here. In particular, the steeper apparent slope with varying $\lambda$, 
in Fig.\ref{figmH1bis} is largely driven by the $Z_S$ factor. In principle, the approach including the wave-function scaling factors is the most refined
among the two methods, hence should be prefered. On the other hand, our choice of setting the $Z$-factors at a low-value of the external momentum,
$\mu_H=125$~GeV, is not optimized for heavy states. In any case, a $1\%$ effect should not be taken too seriously in view of the various additional
sources of uncertainty (parametric errors, running, etc.).

We then consider a second example with $\lambda=0.7$, $\kappa=0.1$, $\tan\beta=2$, $\mu_{\mbox{\tiny eff}}\simeq 2.33\,M_A+20.45$~GeV, $M_A\in[0.3,3]$~TeV, 
$A_{\kappa}=-50$~GeV, $m_{\tilde{T},\tilde{B}}=0.5$~TeV, $m_{\tilde{U},\tilde{D}}=1.5$~TeV, $m_{\tilde{L},\tilde{E}}=110$~GeV, $2M_1=M_2=150$~GeV, $M_3=1.5$~TeV
$A_{t,b,\tau}=-0.1$~TeV. The results are displayed in Fig.\ref{figmH2}. This region of the parameter space highlights another effect in the NMSSM Higgs 
sector, namely the large contribution of F-terms to the mass of the SM-like state for large $\lambda$ and low $\tan\beta$. Indeed, the low value of 
$\tan\beta$, the low mass of the squarks of third generation and the moderate trilinear soft terms would result in a Higgs mass below $M_Z$ in the MSSM,
making this regime incompatible with LEP limits and the LHC measurement. In the NMSSM however, we observe that the mass of the 
SM-like state remains above $120$~GeV: this is a consequence of the specific tree-level contributions to the Higgs mass matrices, associated with $\lambda$.
Comparison of our results with the masses obtained with \verb|NMSSMTools| for precision settings
$0$ and $2$ again show that our calculation is typically closer to precision $2$, although the differences are larger than in the previous scan (about
$1$~GeV for the two light CP-even states, as can be observed on the plot on the lower left-hand corner). We also display the output of \verb|HiggsBounds| and 
HiggsSignals for our results (plot on the lower right-hand side): \verb|HiggsBounds| exclusions apply e.g.\ in the presence of very light Higgs-states with 
non-vanishing doublet composition. The $\chi^2$ test of \verb|HiggsSignals| provides values down to $\sim75$ -- for comparison, we obtain $\sim78$ in the SM 
limit -- when a light doublet is present close to $\sim125$~GeV.

\vspace{0.2cm}
{\em ii) Higgsino and gaugino masses}\newline
\begin{figure}[!htb]
    \centering
    \includegraphics[width=11cm]{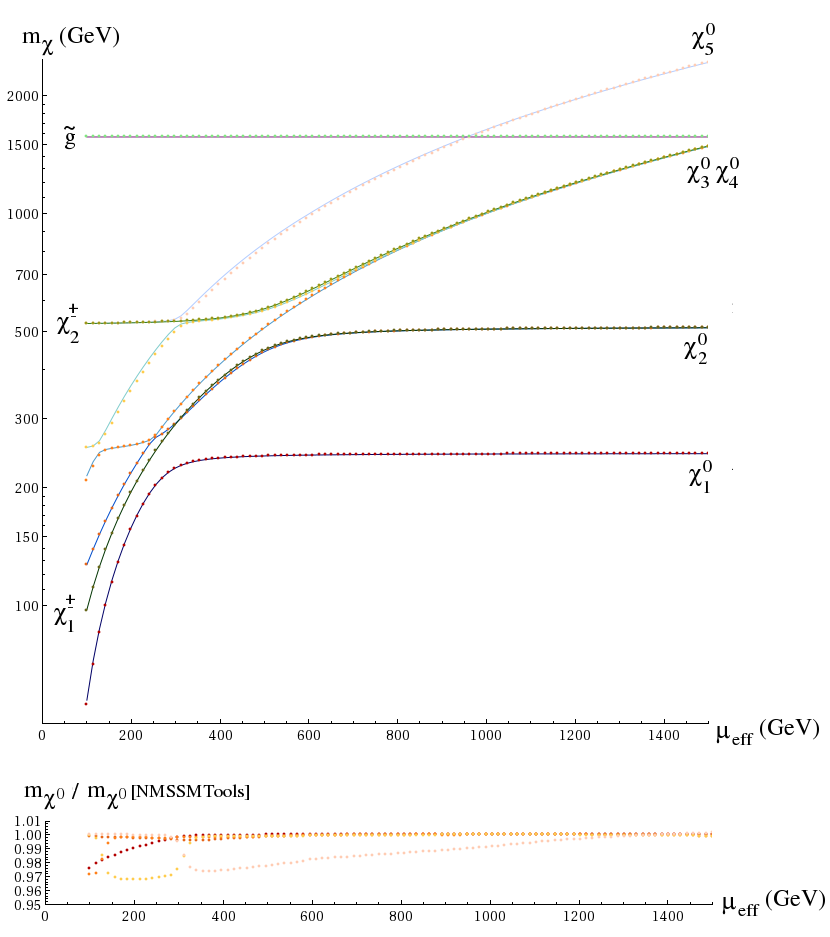}
  \caption{Higgsino / gaugino spectrum for $\lambda=0.55$, $\kappa=0.45$, $\tan\beta=12$, $\mu_{\mbox{\tiny eff}}\in[100,1500]$~GeV, $M_A=1$~TeV, 
$A_{\kappa}=-300$~GeV, $m_{\tilde{T},\tilde{B}}=1$~TeV, $m_{\tilde{U},\tilde{D}}=1.5$~TeV, $m_{\tilde{L},\tilde{E}}=200$~GeV, $2M_1=M_2=M_3/3=0.5$~TeV,
$A_{t,b,\tau}=-1.5$~TeV. Comparison of our results (dots) with the implementation within NMSSMTools (solid lines; note that we actually display the absolute
values of the masses). The plot below shows the ratio between our results for neutralino masses and those delivered by NMSSMTools.}
  \label{figchi}
\end{figure}
Our implementation of the chargino, neutralino and gluino masses should prove very similar to the original subroutines within \verb|NMSSMTools| in the 
CP-conserving limit. Nevertheless, small technical differences should be noted:
\begin{itemize}
 \item we take into account the Higgs-higgsino-singlino couplings which had been neglected in \verb|NMSSMTools|: this results in additional corrections 
to the higgsino and singlino masses; 
 \item similarly, bino and winos are not assumed degenerate in the calculation of loop corrections to the higgsino masses;
 \item all masses are chosen real and positive: this is possible since the diagonalizing matrices are complex. The convention in \verb|NMSSMTools| consisted 
in keeping these matrices real, so that some masses could take negative values.
\end{itemize}

We consider the following region in the NMSSM parameter space: $\lambda=0.55$, $\kappa=0.45$, $\tan\beta=12$, $\mu_{\mbox{\tiny eff}}\in[100,1500]$~GeV, $M_A=1$~TeV, 
$A_{\kappa}=-300$~GeV, $m_{\tilde{T},\tilde{B}}=1$~TeV, $m_{\tilde{U},\tilde{D}}=1.5$~TeV, $m_{\tilde{L},\tilde{E}}=200$~GeV, $2M_1=M_2=M_3/3=0.5$~TeV,
$A_{t,b,\tau}=-1.5$~TeV. The masses of the higgsinos and gauginos are shown in Fig.\ref{figchi}. The scan over $\mu_{\mbox{\tiny eff}}$ drives a 
significant variation of the higgsino masses, while the gaugino masses remain essentially constant. Once again, the masses obtained with the 
original routine of \verb|NMSSMTools| are depicted with a solid line, whereas our results appear as dots: the general features are identical. More 
quantitatively, the main deviation reaches $\sim3\%$ at the level of the neutralino masses: it originates from the corrections to the singlino mass, 
which were neglected in \verb|NMSSMTools|.

For the rest of the spectrum, e.g.\ the sfermion masses, our calculation reduces, in the CP-conserving limit, to the original implementation within \verb|NMSSMTools|. 
Therefore, we will not push the comparison in this limit any further.

\subsection{CP-violating case}
CP-violation could induce several phenomenological effects at colliders. The most immediate one would be the measurement of EDM's. The absence of 
any hint in corresponding searches thus places stringent limits on new-physics phases. Note however that, at one loop order, these effects are essentially driven by 
the gaugino phases. In other words, new-physics phases associated to the Higgs sector or the third generation sfermions enter the EDM's at the two-loop
level only and are thus more loosely constrained. CP-violation could also intervene in rare flavour decays and oscillations, which are consistent so far 
with the SM-interpretation (where only the CKM phase is present): such effects have not been included in our study yet and we will not discuss them here.

\vspace{0.2cm}
{\em i) CP-violating effects in the NMSSM Higgs spectrum}\newline
CP-violation could enter the Higgs sector at tree-level, via a non-vanishing phase $\varphi_{\lambda}-\varphi_{\kappa}$, or at the loop-level, e.g.\ via 
the phases associated to the sfermions of third generation. As a first consequence, the neutral Higgs states would become scalar / pseudoscalar
admixtures, which affects their couplings to SM particles: for doublet states, the pseudoscalar component does not couple to a $ZZ$ or $W^+W^-$ pair, 
so that the corresponding decay channels, as compared to the fermionic decays, are suppressed / enhanced with respect to the case of pure CP-even / 
CP-odd eigenstates. Other effects can be measured in the fermionic channels, provided, however, that the fermion masses are sufficiently large. 
Therefore the presence of CP-violation in the Higgs sector could be tested in precision analyses of the Higgs properties -- for the 
observed or hypothetical new states. Note however that doublet Higgs states are typically shielded from CP-violating mixing -- consider e.g.\ the 
zero-entries in the tree-level mass-matrix of Eq.\ref{neuthiggstreemass} --, so that only a very high degree of precision in the measurement of the 
branching ratios would be likely to detect the tiny -- radiatively-generated -- pseudoscalar component of a mostly CP-even state. Moreover, the current
limits on Higgs searches tend to favour a sizable mass-hierarchy between the SM-like Higgs state and the approximately degenerate `heavy-doublet' states.
This makes the presence of a pseudoscalar doublet component within the observed Higgs state unlikely, as the mixing of this state with the 
`heavy-doublet' pseudoscalar would be suppressed in proportion to the large mass gap. Another test would involve the two
`heavy-doublet' neutral states, which are generically close in mass, so that their mixing could be significant. Yet, the detection of CP-violation there 
will still require high-precision experiments (and the discovery of these states), due to a typically reduced production cross-section -- with respect 
to a SM Higgs boson at the same mass; this is related to the mostly $H_d$-nature of these states -- as well as the opening of many less-controlled
decays (e.g.\ towards new-physics states).
\begin{figure}[!htb]
    \centering
    \includegraphics[width=17.2cm]{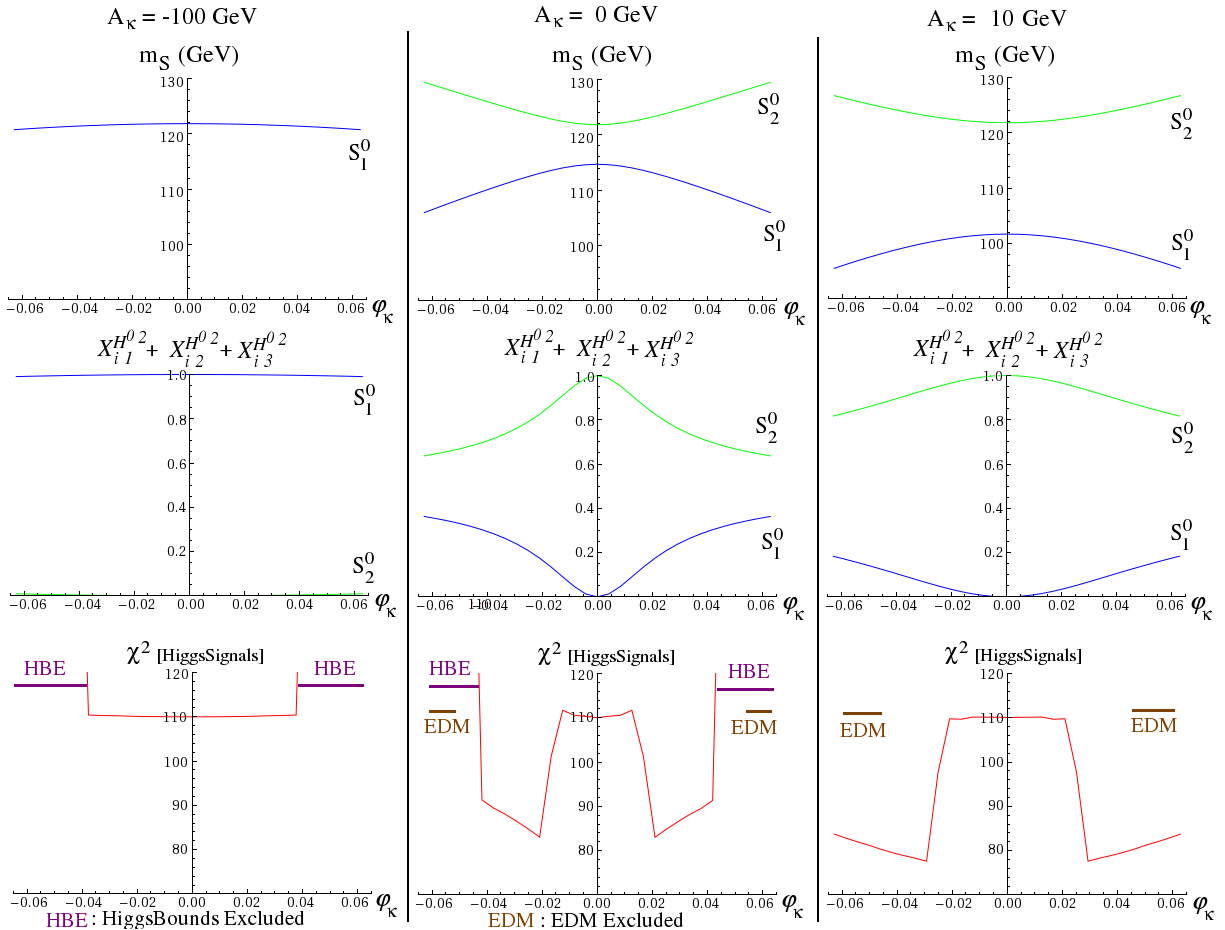}
  \caption{Characteristics of the light Higgs states for $\lambda=0.68$, $\kappa=0.1$, $\tan\beta=2$, $\mu_{\mbox{\tiny eff}}=635$~GeV, $M_A=1.5$~TeV, 
$A_{\kappa}=-100\,,\,0\,,\,10$~GeV, $m_{\tilde{T},\tilde{B},\tilde{\tau}}=0.5$~TeV, $m_{\tilde{U},\tilde{D},\tilde{E}}=1.5$~TeV, $2M_1=M_2=M_3/3=0.5$~TeV,
$A_{t,b,\tau}=-0.1$~TeV. The plots on the first line show how the masses of the Higgs states close to $\sim125$~GeV vary with $\varphi_{\kappa}$; the 
second line displays the magnitude of the scalar components of these states. Characteristics of the lightest state are shown in blue and those for the 
second lightest are shown in green. Finally, the lower series of plots shows how the points compare to phenomenological limits: the violet mark indicates 
that the points are excluded by the test in HiggsBounds while the brown mark stands for tensions with the EDM's. The $\chi^2$-test of the Higgs data is
obtained with HiggsSignals and corresponds to the red curve.}
  \label{fig6}
\end{figure}

In the NMSSM, another type of CP-violating mixing is allowed: a mostly CP-odd singlet may mix with the doublet CP-even states -- provided $\lambda$ and 
$\kappa$ are large and $\varphi_{\lambda}-\varphi_{\kappa}$ is non-vanishing -- and this effect could be fairly important if these states are close in 
mass. In the following, we focus on the SM-like Higgs state at $\sim125$~GeV. Such a scenario is studied in Fig.\ref{fig6} for $\lambda=0.68$, 
$\kappa=0.1$, $\tan\beta=2$, $\mu_{\mbox{\tiny eff}}=635$~GeV, $M_A=1.5$~TeV, $m_{\tilde{T},\tilde{B},\tilde{\tau}}=0.5$~TeV, 
$m_{\tilde{U},\tilde{D},\tilde{E}}=1.5$~TeV, $2M_1=M_2=M_3/3=0.5$~TeV, $A_{t,b,\tau}=-0.1$~TeV. CP-violation is induced through variations of 
$\varphi_{\kappa}$: note that this strategy is the safest in view of the EDM's, as non-vanishing $\varphi_{\lambda}$ produces direct CP-violation in the
doublet higgsino sector (as well as in the sfermion sector). In the first column of Fig.\ref{fig6}, $A_{\kappa}=-100$~GeV, and the mostly CP-odd state 
is relatively far in mass ($\sim250$~GeV): correspondingly,
the mixing with the SM-like state does not reach $1\%$. The latter state has a somewhat low mass of $\sim121$~GeV which translates into a mediocre fit
to the LHC-observed signals, hence a high $\chi^2$-value with \verb|HiggsSignals|. In the second column, we take $A_{\kappa}=0$~GeV, so that the CP-odd
singlet is close in mass to the SM-like state: at $\varphi_{\kappa}=0$, the singlet has a mass of about $\sim115$~GeV. Consequently, a significant
mixing develops between the two light states as soon as $\varphi_{\kappa}\neq0$, the effect reaching the level of $30$ to $40\%$. A consequence is the 
uplift in mass of the heavier SM-like state so that the associated signal gives an improved fit with the LHC data. The column on the right is obtained
for $A_{\kappa}=10$~GeV: the CP-odd singlet is then somewhat lighter ($\sim100$~GeV), so that the mixing effect at non-vanishing $\varphi_{\kappa}$
remains milder than in the previous case, yet generates an uplift of the mass of the SM-like state as well. It is to be noted that the mostly CP-odd
singlet acquires a CP-even doublet component which reaches $O(10\%)$ (at the level of the squared mixing angles): the latter would generate a signal at the $O(10\%)$-level as compared to a SM-like 
state at the same mass -- indeed, the production cross-section at colliders is essentially mediated by the doublet components. For a state with mass $\sim100$~GeV,
the corresponding signal could be consistent with the LEP $\sim2.3\,\sigma$ excess in Higgs searches with a $b\bar{b}$ final state \cite{Barate:2003sz},
even though the state is dominantly CP-odd.

\begin{figure}[!htb]
    \centering
    \includegraphics[width=17.2cm]{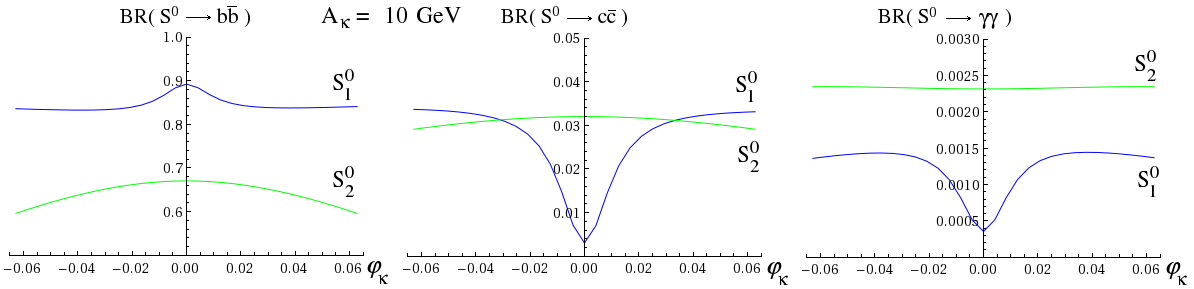}
  \caption{Branching ratios of the light Higgs states for $A_{\kappa}=10$~GeV in the scan of Fig.\ref{fig6}.}
  \label{fig6bis}
\end{figure}
Note that the two effects that we highlighted -- uplift of the mass of the SM-like state via its mixing with the singlet and presence of a `miniature' 
Higgs boson under $125$~GeV -- are well-known in the CP-conserving NMSSM \cite{natsusy}, provided the auxiliary singlet is CP-even. CP-violation extends
this possibility to CP-odd singlets. Further consequences appear on Fig.\ref{fig6bis} at the level of the branching fractions of the Higgs states -- we
display their values for the $b\bar{b}$, $c\bar{c}$ and $\gamma\gamma$ final states --: similarly to the case where the SM-like Higgs boson mixes with a CP-even singlet,
the proportions among doublet components $h_u^0$ and $h_d^0$ may fluctuate, displacing the branching ratios. However, the main effect in Fig.\ref{fig6bis}
concerns the rates of the lighter singlet state which become dominated by CP-even-like channels -- for fermionic final states, rates differ at the radiative
level depending on the CP property\footnote{There is also some difference at tree-level, but the corresponding effect is very small for light fermions.} --, 
while the fluctuations of the branching fractions of the mostly CP-even doublet are dominated by the variations of the associated Higgs mass.

Disentangling this scenario -- where a light mostly CP-odd singlet mixes with the SM-like Higgs boson -- from the CP-conserving one -- where the light
singlet-like state is genuinely CP-even -- is likely to prove very difficult. The reason rests with the observation that the singlets do not lend specific
properties to the SM-like Higgs state -- they simply reduce its total width and might alter its branching ratios at the percent level. Moreover their 
decays are essentially mediated by the doublet component which they acquire in the mixing, i.e.\ a CP-even one in both cases. Typical singlet decays -- 
towards hypothetically lighter singlet states or singlinos -- would not necessarily help to discriminate among CP-even and CP-odd mixing and would be 
problematic in terms of compatibility with the measured Higgs signals. Indeed, the standard rates would then be suppressed in proportion of the magnitude of 
the unconventional decays. While deviations of the rates of the observed Higgs state from the standard ones might be interpreted via such a mixing effect 
-- should such deviations be detected at the LHC or a future linear collider --, it is questionable anyway whether the light singlet could be detected 
-- possibly in Higgs-pair production: see e.g.\ \cite{pairprod} in the CP-even case.

At the outcome of this discussion, we see that, while the CP-violating effects involving singlets in the Higgs sector may be larger than in the pure 
doublet case, they are also more difficult to trace and could be mistaken for CP-conserving phenomena. For this reason, it is essential that CP-violation
be tested in processes where the CP-properties are well-controlled, which brings us back to EDM's or rare flavour transitions. Spectral effects in the 
Higgs sector are unlikely to allow for discrimination with the CP-conserving case.

\vspace{0.2cm}
{\em ii) Comparison of the Higgs mass predictions with the existing literature}\newline
We will now compare some of our results with existing analyses in the literature, where CP-violation has been considered. Note that, contrarily to the 
comparison with the calculations in the CP-conserving \verb|NMSSMTools|, one should not expect much more than a qualitative agreement. Indeed, the choice 
of disparate procedures in different tools, e.g.\ concerning the definition of the input -- such as the choice of running Yukawa couplings or that of $A_{\kappa}$ 
versus $A_{\kappa}\cos\varphi_2$ --, are known to lead
to sizable deviations, already in the CP-conserving case. The level of precision in radiative corrections is also to be considered.

\verb|NMSSMCALC| \cite{NMSSMCALC} is a public tool computing the Higgs spectrum and decays in the $\mathbb{Z}_3$-conserving but possibly CP-violating
NMSSM. The chosen approach is that of a diagrammatic calculation. The level of precision has recently been extended to include the dominant two-loop 
corrections \cite{Muhlleitner:2014vsa}.

\begin{figure}[!htb]
    \centering
    \includegraphics[width=15cm]{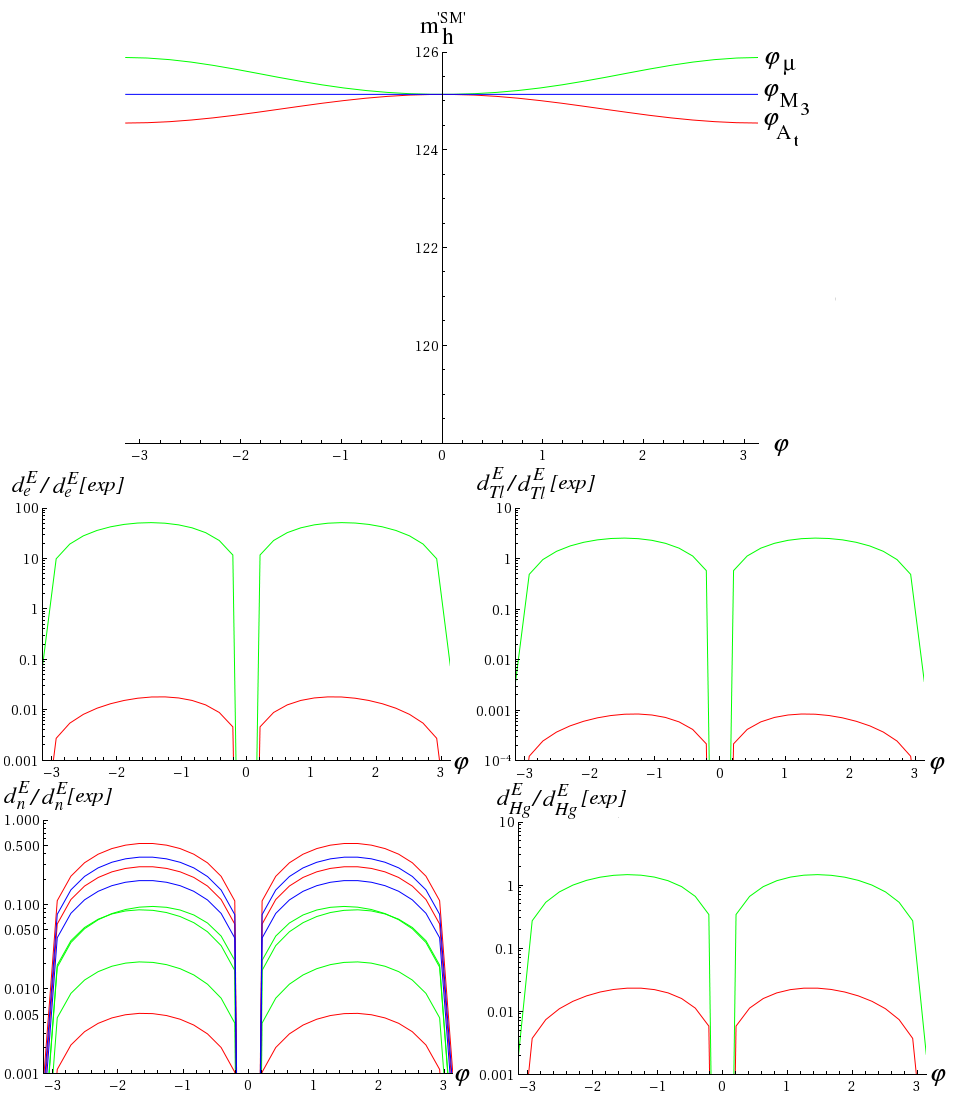}
  \caption{Top: phase-dependence of the mostly $h_u^0$ Higgs state as obtained from section 4.1 of \cite{Muhlleitner:2014vsa}. Below: estimates of the 
EDM's (normalized to their experimental upper bounds: see text). The color coding follows that of Fig.6 of \cite{Muhlleitner:2014vsa}: green for a scan
over $\varphi_{\mu}\equiv\varphi_{\lambda}=\varphi_{\kappa}$, red for a scan over $\varphi_{A_t}$ and blue for a scan over $\phi_{M_3}$. When the blue curve does not appear, the reason is that 
associated values are negligibly small. Note that several estimates are employed for the neutron EDM.}
  \label{mhphase}
\end{figure}

First, we focus on the results of \cite{Muhlleitner:2014vsa} dealing with CP-violating effects, i.e.\ essentially Fig.6 and the surrounding text in that
paper. If we blindly input the parameters given in section 4.1 of this reference into our framework\footnote{Note that this addresses the 
$\overline{DR}$-parameters in the reference, since the parameters within NMSSMTools are regarded as $\overline{DR}$.}, the spectrum -- not unexpectedly and
already with the CP-conserving \verb|NMSSSMTools| -- turns out to be 
slightly different from the quoted one: in particular, the mostly CP-even and mostly CP-odd singlet states appear with masses $\sim108$~GeV and
$\sim113$~GeV respectively. Yet, this discrepancy can be absorbed within a small shift of $A_{\kappa}\cos\varphi_2$: using the value $203$~GeV, we then 
recover states at $\sim 103$ and $\sim128$~GeV so that the Higgs spectrum then largely coincides with the one provided in \cite{Muhlleitner:2014vsa}.

In any case,
this manipulation has little effect on the properties of the mostly $h_u^0$-state, close to $125$~GeV. Scanning over the phases $\varphi_{A_t}$, 
$\varphi_{\mu}$ -- a scan over $\varphi_{\mu}$ in the notations of \cite{Muhlleitner:2014vsa} would correspond to a scan over $\varphi_{\lambda}$, 
keeping $\varphi_{\kappa}=\varphi_{\lambda}$, in ours,
so that CP-violation does not enter the Higgs sector at tree-level --
and $\phi_{M_3}$, we obtain the plots of Fig.\ref{mhphase}. On the upper part, we observe that the general dependence of the `SM-like' Higgs mass on 
$\varphi_{A_t}$ and $\varphi_{\mu}$ is largely reminiscent in shape and magnitude of that observed in Fig.6 of \cite{Muhlleitner:2014vsa}. In these two
cases, CP-violation enters the Higgs sector via radiative corrections, where the leading effect is generated by the sfermion corrections. On the other 
hand, the mass obtained with our code is independent from $\phi_{M_3}$, while such a dependence already appears at one-loop in \cite{Muhlleitner:2014vsa}. 
Note that one does not expect gluino corrections to the Higgs mass at one-loop order and it is thus not surprising that our implementation does not show 
any variation with $\phi_{M_3}$. The corresponding effect in \cite{Muhlleitner:2014vsa} is explained there as an artifact of the top-Yukawa $\overline{DR}$
counterterm-fixing of higher order. Note also that the corresponding fluctuations, at the GeV level, are small compared to the uncertainty that one 
naively expects for the Higgs mass (a few GeV).

In addition, we show the values of the EDM's that we obtain in these scans. These have been normalized to the experimental upper bounds: $\sim1\cdot10^{-28}$~$e$~cm
for the electron \cite{Baron:2013eja} -- estimate from thorium monoxide experiment --, $\sim1.3\cdot10^{-24}$~$e$~cm for the Thallium atom \cite{Regan:2002ta},
$\sim3.1\cdot10^{-29}$~$e$~cm for the Mercury EDM \cite{Griffith:2009zz} and $\sim3\cdot10^{-26}$~$e$~cm for the neutron \cite{Baker:2006ts}. Note that only the central 
values are displayed, without error bands. The color code is the same as in Fig.6 of \cite{Muhlleitner:2014vsa}, i.e.\ green for the scan on $\varphi_{\mu}$,
red for that on $\varphi_{A_t}$ and blue for the one over $\phi_{M_3}$ (when the curve does not appear in the plot, this is because the corresponding 
values are negligibly small). We see that the scan over $\varphi_{\mu}$ may generate tensions with the EDM's -- mostly the electron EDM -- when 
$\varphi_{\mu}$ is not trivial.

\begin{figure}[!htb]
    \centering
    \includegraphics[width=15cm]{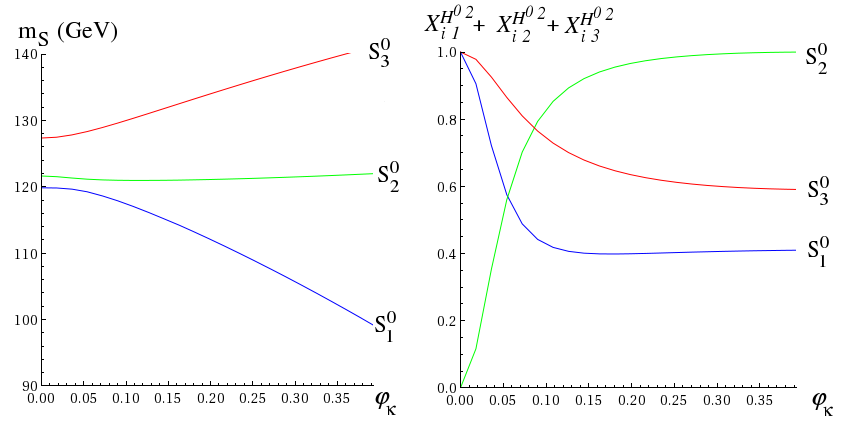}
  \caption{Masses and CP-even composition ($X^{H^0\,2}_{i1}+X^{H^0\,2}_{i2}+X^{H^0\,2}_{i3}$) for the three lightest Higgs states in the scenario of
section 4.1.1 of \cite{Graf:2012hh}.}
  \label{Figmhgraf1}
\end{figure}
\begin{figure}[!htb]
    \centering
    \includegraphics[width=15cm]{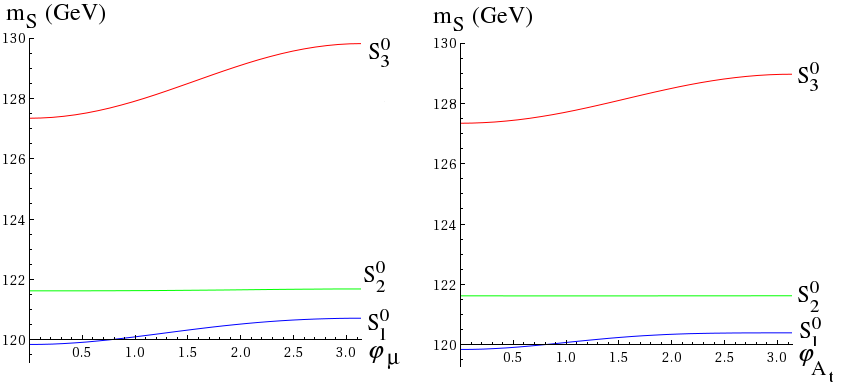}
  \caption{Masses in the scenarios of section 4.1.2 (scan over $\varphi_{\mu}\equiv\varphi_{\lambda}=\varphi_{\kappa}$) and section 4.1.3 (scan over 
$\varphi_{A_t}$) of \cite{Graf:2012hh}.}
  \label{Figmhgraf2}
\end{figure}
We now turn to the one-loop analysis proposed in \cite{Graf:2012hh}. We first consider the scenario presented in section 4.1.1 of this reference,
where CP-violation intervenes in the Higgs sector at tree-level via the phase $\varphi_{\kappa}$. Again, a qualitatively close spectrum can be recovered
with little alteration of the input proposed in the reference and our results are displayed in Fig.\ref{Figmhgraf1}: while small differences appear,
both the Higgs masses and the composition of the states agree reasonably well with those of \cite{Graf:2012hh}. The major source of deviation is 
associated to the use of different input -- $A_{\kappa}$ in \verb|NMSSMCALC| instead of $A_{\kappa}\cos\varphi_2$ in our case --, so that the comparison
makes limited sense when $\varphi_2$ becomes large (i.e.\ for $\varphi_{\kappa}\sim\pi/8$). In the regime considered here, the CP-even and 
CP-odd singlet states are close in mass to the SM-like Higgs boson, so that the non-vanishing $\varphi_{\kappa}$ generates a substantial mixing of these 
three states.

\cite{Graf:2012hh} then considers the case where CP-violation is absent in the tree-level Higgs sector, but radiatively generated via phases in the 
supersymmetric spectrum. In the first case (section 4.1.2), the `active' phase is $\varphi_{\lambda}$ but the condition $\varphi_{\kappa}=
\varphi_{\lambda}$ ensures that no CP-violation enters the Higgs potential at tree-level -- we will recycle the previous notation $\varphi_{\mu}$ for 
this scenario. In the second case, only the phase $\varphi_{A_t}$ is non-trivial. We display our results in Fig.\ref{Figmhgraf2} and observe that 
they capture the effects depicted in Fig.5 and Fig.7 of \cite{Graf:2012hh}.

Our code is thus able to reproduce the main qualitative features that were observed in the CP-violating case by \verb|NMSSMCALC| analyses. We stress that a more
quantitative study would have limited interest, as the divergent treatment of the input already generates discrepancies between the CP-conserving
\verb|NMSSMTools| and \verb|NMSSMCALC|.


\section{Conclusions}

We have presented a series of Fortran tools extending \verb|NMSSMTools| to the CP-violating case. Radiative corrections to the supersymmetric and Higgs masses
are computed at one-loop order. Dominant two-loop effects to the Higgs masses are also included in the double-log approximation. Additionally,
Higgs couplings and decays, as well as top two-body decays and EDM's are implemented and allow for phenomenological tests of the spectra. We have shown
that our code compares competitively with existing results, both in the CP-conserving and CP-violating cases. The new tools
will be made publicly available on the \verb|NMSSMTools| website \cite{NMSSMTools} in the near future.

We also highlighted a scenario made possible by CP-violation, where the SM-like Higgs would mix with a mostly CP-odd singlet state. The consequences on
the Higgs phenomenology are similar to the CP-conserving mixing with a light CP-even singlet so that both scenarii should prove difficult to 
discriminate, unless genuine CP-violating effects -- e.g.\ in EDM's or flavour physics -- are discovered simultaneously.

Finally, we would like to close this discussion with some details concerning the future developments which we plan to consider. First, an extension of our 
tools including $\mathbb{Z}_3$-violating terms should raise little difficulty. Then, flavour constraints are relevant in the CP-violating NMSSM and we 
intend to design phenomenological tests accordingly. Finally, the dominant two-loop corrections to the Higgs masses will be calculated in a more 
quantitative way. 

\section*{Acknowledgements}
The author thanks U.~Ellwanger for constructive comments. This work has been supported by the Collaborative Research Center SFB676 of the DFG, 
``Particles, Strings, and the Early Universe''.

\newpage
\appendix

\section{Reference functions}\label{reffunc}

{\begin{align*}
 {\cal F}_0(m^2)&=m^2\left[\ln\frac{m^2}{Q^2}-1\right]\\
 {\cal F}_1(m^2,M^2)&=\frac{1}{m^2-M^2}\left[m^2\left(\ln\frac{m^2}{Q^2}-1\right)-M^2\left(\ln\frac{M^2}{Q^2}-1\right)\right]\\
 {\cal F}_3(m^2,M^2)&=\frac{m^2\left[\ln\frac{m^2}{M^2}-1\right]+M^2}{(m^2-M^2)^2}\\
 {\cal F}_5(m_1^2,m_2^2,m_3^2)&=\frac{m_1^2m_2^2\ln\frac{m_1^2}{m_2^2}-m_1^2m_3^2\ln\frac{m_1^2}{m_3^2}+m_2^2m_3^2\ln\frac{m_2^2}{m_3^2}}{(m_1^2-m_2^2)(m_2^2-m_3^2)(m_1^2-m_3^2)}\\
 {\cal F}_6(m_1^2,m_2^2,m_3^2)&=\frac{m_1^2\ln\frac{m_1^2}{m_3^2}}{(m_1^2-m_2^2)(m_1^2-m_3^2)^2}+\frac{m_2^2\ln\frac{m_2^2}{m_3^2}}{(m_2^2-m_1^2)(m_2^2-m_3^2)^2}+\frac{1}{(m_1^2-m_3^2)(m_2^2-m_3^2)}\\
 {\cal F}_7(m^2,M^2)&=-\frac{m^2\left[\ln\frac{m^2}{M^2}-2\right]-M^2\left[\ln\frac{M^2}{m^2}-2\right]}{(m^2-M^2)^3}
\end{align*}}

We only consider the finite part of the loop integrals:
{\begin{align*}
A_0(m)&=-{\cal F}_0(m^2)\\
B_0(p,m,M)&=-\int_0^1{\ln\frac{xm^2+(1-x)(M^2-p^2)+(1-x)^2p^2}{Q^2}dx}\\
B_1(p,m,M)&=\int_0^1{(1-x)\ln\frac{xm^2+(1-x)(M^2-p^2)+(1-x)^2p^2}{Q^2}dx}
\end{align*}}

Finally, we borrow some of the notations of \cite{Martin}:
\begin{align*}
 B_{FF}(p,m,M)&=(p^2-m^2-M^2)B_0(p,m,M)-A_0(m)-A_0(M)\\
 B_{SV}(p,m,M)&=-(2p^2+2m^2-M^2)B_0(p,m,M)+A_0(m)-2A_0(M)\hspace{1cm}\mbox{(Feynmann gauge)}
\end{align*}

\section{The tree-level masses and couplings}\label{sectl}
This appendix provides the reader with a detailed presentation of the tree-level spectrum and couplings of the CP-violating, minimal-flavour-violating,
R-parity and $\mathbb{Z}_3$ conserving NMSSM.

\subsection{Tree-level masses}
Here we derive the tree-level bilinear terms of the lagrangian. For a later application to the Higgs couplings as well as to the loop-corrections in
the Coleman-Weinberg effective potential, we will try to keep a full dependence in the Higgs scalar fields $S$, $H_u=(H_u^+,H_u^0)^T$ and 
$H_d=(H_d^0,H_d^-)^T$. To evaluate the masses, one of course simply needs to replace these fields by their v.e.v.'s.

\subsubsection{SM fermions}\label{tlsmferm}
The Higgs-fermion potential reads:
\begin{equation}
 {\cal V}_f=-(H_u^+d_LV_{CKM}^+-H_u^0u_L)Y_uu_R^c+(H_d^0d_L-H_d^-u_LV_{CKM})Y_dd_R^c+(H_d^0e_L-H_d^-\nu_L)Y_ee_R^c+h.c.
\end{equation}
Focussing on the third generation (and neglecting off-diagonal CKM elements), we may cast under matrix form:
\begin{multline}
 {\cal V}_{f^3}=(\bar{u}_L,u_R^c,\bar{d}_L,d_R^c){\cal M}_{q^3}\begin{pmatrix} u_L\\ \bar{u}_R^c\\ d_L \\ \bar{d}_R^c \end{pmatrix}+(\bar{\nu}_L,\bar{e}_L,e_R^c){\cal M}_{l^3}\begin{pmatrix} \nu_L\\ e_L \\ \bar{e}_R^c \end{pmatrix}\\{\cal M}_{q^3}\equiv\begin{bmatrix}
0 & Y_tH_u^{0\,*} & 0 & -Y_b H_d^+ \\ Y_t H_u^0 & 0 & -Y_tH_u^+ & 0 \\ 0 & -Y_tH_u^- & 0 & Y_bH_d^0 \\ -Y_b H_d^- & 0 & Y_bH_d^{0\,*} & 0
\end{bmatrix}\ \ \ \ ;\ \ \ \ {\cal M}_{l^3}\equiv\begin{bmatrix}
0 &  0 & -Y_{\tau} H_d^+ \\ 0 & 0 & Y_{\tau}H_d^0 \\ -Y_{\tau} H_d^- & Y_{\tau}H_d^{0\,*} & 0
\end{bmatrix}
\end{multline}
from which we derive the squared-mass matrices:
\begin{multline}
{\cal M}_{q^3}^2\equiv{\cal M}_{q^3}^{\dag}{\cal M}_{q^3}=\\\begin{bmatrix}
 Y_t^2|H_u^0|^2+Y_b^2H_d^+H_d^- & 0 & -Y_t^2H_u^{0\,*}H_u^+-Y_b^2H_d^{0}H_d^+ & 0\\
 0 & Y_t^2(|H_u^0|^2+H_u^+H_u^-) & 0 & -Y_tY_b(H_u^0H_d^++H_d^0H_u^+)\\
 -Y_t^2H_u^{0}H_u^--Y_b^2H_d^{0\,*}H_d^- & 0 & Y_b^2|H_d^0|^2+Y_t^2H_u^+H_u^- & 0\\
0 & -Y_tY_b(H_u^{0\,*}H_d^-+H_d^{0\,*}H_u^-) & 0 & Y_b^2(|H_d^0|^2+H_d^+H_d^-)
                            \end{bmatrix}\label{sqmquarks}
\end{multline}
\begin{equation}
{\cal M}_{l^3}^2\equiv{\cal M}_{l^3}^{\dag}{\cal M}_{l^3}=\begin{bmatrix}
 Y_{\tau}^2H_d^+H_d^- & -Y_{\tau}^2H_d^{0}H_d^+ & 0\\
 -Y_{\tau}^2H_d^{0\,*}H_d^- & Y_{\tau}^2|H_d^0|^2& 0\\
0 & 0 & Y_{\tau}^2(|H_d^0|^2+H_d^+H_d^-)
                            \end{bmatrix}\label{sqmleptons}
\end{equation}
Replacing the Higgs fields by their v.e.v.'s, one obtains diagonal matrices $\left<{\cal M}_{q^3}^2\right>$ and $\left<{\cal M}_{l^3}^2\right>$, with 
the usual relations: $m_t^2=Y_t^2v_u^2$, $m_b^2=Y_b^2v_d^2$, $m_{\nu}^2=0$, $m_{\tau}^2=Y_{\tau}^2v_d^2$.

\subsubsection{Electroweak gauge bosons}
From the Higgs kinetic terms, one obtains the Higgs-gauge potential (where we omit the derivative Higgs couplings):
\begin{equation}
 {\cal V}_G=\frac{1}{4}\left[g'^2B_{\mu}B^{\mu}+g^2\overrightarrow{W}_{\mu}\cdot\overrightarrow{W}^{\mu}\right]\left(|H_u|^2+|H_d|^2\right)+\frac{gg'}{2}B_{\mu}\overrightarrow{W}^{\mu}\cdot\left(H_u^{\dag}\overrightarrow{\sigma}H_u-H_d^{\dag}\overrightarrow{\sigma}H_d\right)
\end{equation}
After the fields are rotated to the mass-states, 
\begin{equation}
 A_{\mu}\equiv\frac{gB_{\mu}+g'W^3_{\mu}}{\sqrt{g'^2+g^2}}\ \ \ ;\ \ \ Z_{\mu}\equiv\frac{-g'B_{\mu}+gW^3_{\mu}}{\sqrt{g'^2+g^2}}\ \ \ ;\ \ \ W^{\pm}_{\mu}=\frac{W^1_{\mu}\mp\imath W^2_{\mu}}{\sqrt{2}}
\end{equation}
we derive:
\begin{multline}
 {\cal V}_G=\frac{1}{2}(A_{\mu},Z_{\mu})\left[\begin{pmatrix}0 & 0\\ 0 & \frac{g^2+g'^2}{2}\end{pmatrix}(|H_u^0|^2+|H_d^0|^2)+\begin{pmatrix}\frac{2g^2g'^2}{g^2+g'^2} &\frac{gg'(g^2-g'^2)}{g^2+g'^2} \\ 
\frac{gg'(g^2-g'^2)}{g^2+g'^2} & \frac{(g^2-g'^2)^2}{2(g^2+g'^2)}\end{pmatrix}(H_u^+H_u^-+H_d^+H_d^-)\right]\begin{pmatrix}A^{\mu}\\Z^{\mu}\end{pmatrix}\\
+\frac{g^2}{2}(|H_u^0|^2+|H_d^0|^2+H_u^+H_u^-+H_d^+H_d^-)W_{\mu}^-W^{+\,\mu}\\+\frac{gg'}{\sqrt{2}\sqrt{g^2+g'^2}}(gA^{\mu}-g'Z^{\mu})\left[W^+_{\mu}(H_u^-H_u^0-H_d^{0\,*}H_d^-)+W^-_{\mu}(H_u^+H_u^{0\,*}-H_d^0H_d^+)\right]\label{massgauge}
\end{multline}
This leads to the usual gauge-boson masses: $M_{\gamma}^2=0$, $M_{W}^2=\frac{g^2}{2}(v_u^2+v_d^2)$, $M_{Z}^2=\frac{g'^2+g^2}{2}(v_u^2+v_d^2)$.

\subsubsection{Sfermions}\label{tlsfermmass}
The Higgs-sfermion potential originates from soft, $F$ and $D$ terms:
\begin{align}
 {\cal V}_{\tilde{F}}=&m_Q^2Q_L^{\dag}Q_L+m_U^2U_R^{c\,\dag}U_R^c+m_D^2D_R^{c\,\dag}D_R^c+m_L^2L_L^{\dag}L_L+m_E^2E_R^{c\,\dag}E_R^c\\
& +(H_u^0U_L-H_u^+D_LV^{\dag}_{CKM})Y_uA_ue^{\imath\varphi_{A_u}}U_R^c+(H_d^0D_L-H_d^-U_LV_{CKM})Y_dA_de^{\imath\varphi_{A_d}}D_R^c\nonumber\\
&\null\hspace{8cm}+(H_d^0E_L-H_d^-N_L)Y_eA_ee^{\imath\varphi_{A_e}}E_R^c+h.c.\nonumber\\
& +\left|(H_u^+D_LV_{CKM}^{\dag}-H_u^0U_L)Y_u\right|^2+\left|Y_uV_{CKM}H_u^+U_R^{c}-Y_dH_d^0D_R^c\right|^2+\left|Y_uH_u^0U_R^{c}-Y_dH_d^-V_{CKM}^{\dag}D_R^c\right|^2\nonumber\\
& +\left|(H_d^0D_L-H_d^-U_LV_{CKM})Y_d\right|^2+\left|(H_d^0E_L-H_d^-N_L)Y_e\right|^2+\left|Y_eH_d^-E_R^c\right|^2+\left|Y_eH_d^0E_R^c\right|^2\nonumber\\
& +\left|\lambda e^{\imath\varphi_{\lambda}}SH_u^+-U_LV_{CKM}Y_dD_R^c-N_LY_eE_R^c\right|^2+\left|\lambda e^{\imath\varphi_{\lambda}}SH_u^0-D_LY_dD_R^c-E_LY_eE_R^c\right|^2\nonumber\\
& +\left|\lambda e^{\imath\varphi_{\lambda}}SH_d^0-U_LY_uU_R^c\right|^2+\left|\lambda e^{\imath\varphi_{\lambda}}SH_d^--D_LV_{CKM}^{\dag}Y_uU_R^c\right|^2\nonumber\\
& +\frac{g'^2}{8}\left|H_u^{\dag}H_u-H_d^{\dag}H_d+\frac{1}{3}Q_L^{\dag}Q_L-\frac{4}{3}U_R^{c\,\dag}U_R^c+\frac{2}{3}D_R^{c\,\dag}D_R^c-L_L^{\dag}L_L+2E_R^{c\,\dag}E_R^c\right|^2\nonumber\\
& +\frac{g^2}{8}\left|H_u^{\dag}\overrightarrow{\sigma}H_u+H_d^{\dag}\overrightarrow{\sigma}H_d+Q_L^{\dag}\overrightarrow{\sigma}Q_L+L_L^{\dag}\overrightarrow{\sigma}L_L\right|^2\nonumber
\end{align}
The bilinear sfermion terms can be cast under matrix form:
\begin{equation}
 {\cal V}_{\tilde{F}}\ni(U_L^{\dag},U_R^c,D_L^{\dag},D_R^c)\begin{pmatrix}{\cal M}^2_{U} & {\cal M}^{2\,\dag}_{D^{\dag}U}\\{\cal M}^2_{D^{\dag}U}&{\cal M}^2_{D}\end{pmatrix}\begin{pmatrix}U_L\\U_R^{c\,\dag}\\D_L\\D_R^{c\,\dag}\end{pmatrix}+(N_L^{\dag},E_L^{\dag},E_R^c)\begin{pmatrix}{\cal M}^2_{N} & {\cal M}^{2\,\dag}_{E^{\dag}N}\\{\cal M}^2_{E^{\dag}N}&{\cal M}^2_{E}\end{pmatrix}\begin{pmatrix}N_L\\E_L\\E_R^{c\,\dag}\end{pmatrix}
\end{equation}
with the matrix blocks:
\begin{align}
{\cal M}^2_{U}= &\begin{pmatrix}m_{Q}^2+Y_u^2|H_u^0|^2+\frac{1}{4}\left(\frac{g'^2}{3}-g^2\right)(|H_u^0|^2-|H_d^0|^2)&Y_u\left[A_ue^{-\imath\varphi_{A_u}}H_u^{0\,*}-\lambda e^{\imath\varphi_{\lambda}}SH_d^0\right]\label{massup}\\
Y_u\left[A_ue^{\imath\varphi_{A_u}}H_u^{0}-\lambda e^{-\imath\varphi_{\lambda}}S^*H_d^{0\,*}\right]&m_U^2+Y_u^2|H_u^0|^2-\frac{g'^2}{3}(|H_u^0|^2-|H_d^0|^2)\end{pmatrix}\\
 &+\begin{pmatrix}V_{CKM}Y_d^2V_{CKM}^{\dag}H_d^+H_d^-+\frac{1}{4}\left(\frac{g'^2}{3}+g^2\right)(H_u^+H_u^--H_d^+H_d^-)&0\\
0&Y_u^2H_u^+H_u^--\frac{g'^2}{3}(H_u^+H_u^--H_d^+H_d^-)\end{pmatrix}\nonumber
\end{align}
\begin{align}
{\cal M}^2_{D}= &\begin{pmatrix}m_{Q}^2+Y_d^2|H_d^0|^2+\frac{1}{4}\left(\frac{g'^2}{3}+g^2\right)(|H_u^0|^2-|H_d^0|^2)&Y_d\left[A_de^{-\imath\varphi_{A_d}}H_d^{0\,*}-\lambda e^{\imath\varphi_{\lambda}}SH_u^0\right]\label{massdown}\\
Y_d\left[A_de^{\imath\varphi_{A_d}}H_d^{0}-\lambda e^{-\imath\varphi_{\lambda}}S^*H_u^{0\,*}\right]&m_D^2+Y_d^2|H_d^0|^2+\frac{g'^2}{6}(|H_u^0|^2-|H_d^0|^2)\end{pmatrix}\\
 &+\begin{pmatrix}V_{CKM}^{\dag}Y_u^2V_{CKM}H_u^+H_u^-+\frac{1}{4}\left(\frac{g'^2}{3}-g^2\right)(H_u^+H_u^--H_d^+H_d^-)&0\\
0&Y_d^2H_d^+H_d^-+\frac{g'^2}{6}(H_u^+H_u^--H_d^+H_d^-)\end{pmatrix}\nonumber
\end{align}
\begin{equation}
 {\cal M}^2_{D^{\dag}U}= \begin{pmatrix}-V^{\dag}_{CKM}Y_u^2H_u^0H_u^--Y_d^2V_{CKM}^{\dag}H_d^{0\,*}H_d^-+\frac{g^2}{2}V_{CKM}^{\dag}(H_u^0H_u^-+H_d^{0\,*}H_d^-)&-V_{CKM}^{\dag}Y_u\left[A_ue^{-\imath\varphi_{A_u}}H_u^-+\lambda e^{\imath\varphi_{\lambda}}SH_d^-\right]\label{massqmix}\\
-Y_d\left[A_de^{\imath\varphi_{A_d}}H_d^-+\lambda e^{-\imath\varphi_{\lambda}}S^*H_u^-\right]V_{CKM}^{\dag}&-Y_dV_{CKM}^{\dag}Y_u(H_d^0H_u^-+H_u^{0\,*}H_d^-)\end{pmatrix}
\end{equation}
\begin{equation}
 {\cal M}^2_{N}=m_L^2-\frac{g'^2+g^2}{4}(|H_u^0|^2-|H_d^0|^2)+Y_e^2H_d^+H_d^-+\frac{-g'^2+g^2}{4}(H_u^+H_u^--H_d^+H_d^-)\label{massneut}
\end{equation}
\begin{align}
{\cal M}^2_{E}= &\begin{pmatrix}m_{L}^2+Y_e^2|H_d^0|^2+\frac{-g'^2+g^2}{4}(|H_u^0|^2-|H_d^0|^2)&Y_e\left[A_ee^{-\imath\varphi_{A_e}}H_d^{0\,*}-\lambda e^{\imath\varphi_{\lambda}}SH_u^0\right]\label{massel}\\
Y_e\left[A_ee^{\imath\varphi_{A_e}}H_d^{0}-\lambda e^{-\imath\varphi_{\lambda}}S^*H_u^{0\,*}\right]&m_E^2+Y_e^2|H_d^0|^2+\frac{g'^2}{2}(|H_u^0|^2-|H_d^0|^2)\end{pmatrix}\\
 &+\begin{pmatrix}-\frac{g'^2+g^2}{4}(H_u^+H_u^--H_d^+H_d^-)&0\\
0&Y_e^2H_d^+H_d^-+\frac{g'^2}{2}(H_u^+H_u^--H_d^+H_d^-)\end{pmatrix}\nonumber
\end{align}
\begin{equation}
 {\cal M}^2_{E^{\dag}N}= \begin{pmatrix}-\left(Y_e^2-\frac{g^2}{2}\right)H_d^{0\,*}H_d^-+\frac{g^2}{2}H_u^{0}H_u^-&-Y_e\left[A_ee^{\imath\varphi_{A_e}}H_d^-+\lambda e^{-\imath\varphi_{\lambda}}S^*H_u^-\right]\end{pmatrix}\label{massemix}
\end{equation}
Moving to the v.e.v.'s, the matrices become block diagonal -- each block being associated to a given electric charge of the sfermion fields. Under
our Minimal Flavour Violation hypothesis the various generations also decouple so that we are left with $2\times2$ (hermitian) mass-matrices 
$\left<{\cal M}^2_F\right>$. Those can be diagonalized via unitary matrices $X^F$, according to:
\begin{multline}
 \left<{\cal M}^2_F\right>=X^{F\,\dag}\mbox{diag}(m^2_{F_1},m^2_{F_2})X^F\ \ \ \ \ \ ;\ \ \ \ \ \ X^F\equiv\begin{pmatrix}\cos\theta_F&-\sin\theta_Fe^{\imath\varphi_F}\\
\sin\theta_Fe^{-\imath\varphi_F}&\cos\theta_F\end{pmatrix}\\
\begin{cases}
 m^2_{F_1}=\frac{1}{2}\left[\left<{\cal M}^2_F\right>_{11}+\left<{\cal M}^2_F\right>_{22}-\sqrt{\left(\left<{\cal M}^2_F\right>_{11}-\left<{\cal M}^2_F\right>_{22}\right)^2+4\left<{\cal M}^2_F\right>_{12}^2}\right]\\
 m^2_{F_2}=\frac{1}{2}\left[\left<{\cal M}^2_F\right>_{11}+\left<{\cal M}^2_F\right>_{22}+\sqrt{\left(\left<{\cal M}^2_F\right>_{11}-\left<{\cal M}^2_F\right>_{22}\right)^2+4\left<{\cal M}^2_F\right>_{12}^2}\right]
\end{cases}\\
\begin{cases}
 \theta_F=\arctan\left[\frac{\left<{\cal M}^2_F\right>_{11}-\left<{\cal M}^2_F\right>_{22}+\sqrt{\left(\left<{\cal M}^2_F\right>_{11}-\left<{\cal M}^2_F\right>_{22}\right)^2+4\left<{\cal M}^2_F\right>_{12}^2}}{2|\left<{\cal M}^2_F\right>_{12}|}\right]\\
 \varphi_F=\mbox{arg}[\left<{\cal M}^2_F\right>_{12}]
\end{cases}
\end{multline}
The mass-states are given by $F_i=X^F_{iL}F_L+X^F_{iR}F_R^{c\,*}$ (where, in our notation $1\leftrightarrow L$ and $2\leftrightarrow R$).

\subsubsection{Charginos and neutralinos}\label{tlchaneumass}
The gaugino-higgsino-Higgs potential may also be cast under matrix form:
\begin{displaymath}
{\cal V}_{\chi}=\frac{1}{2}\chi^T\begin{pmatrix}
 0 & {\cal M}_{\chi^{-+}} & {\cal M}_{\chi^{-0}}\\{\cal M}_{\chi^{+-}} & 0 & {\cal M}_{\chi^{+0}}\\{\cal M}_{\chi^{0-}}&{\cal M}_{\chi^{0+}}&{\cal M}_{\chi^{0}}
\end{pmatrix}\chi+h.c.\ \ \ \ ;\ \ \ \ \chi^T\equiv(-\imath \tilde{w}^-,\tilde{h}_d^-,-\imath \tilde{w}^+,\tilde{h}_u^+,-\imath \tilde{b},-\imath \tilde{w}^3,\tilde{h}_u^0,\tilde{h}_d^0,\tilde{h}_s^0)
\end{displaymath}
\begin{equation}
 {\cal M}_{\chi^{-+}}=\begin{pmatrix}M_2e^{\imath\phi_{M_2}}&gH_u^{0\,*}\\gH_d^{0\,*}&\lambda e^{\imath\varphi_{\lambda}}S\end{pmatrix}={\cal M}_{\chi^{+-}}^T\label{masschaneu}
\end{equation}
\begin{displaymath}
 {\cal M}_{\chi^{-0}}=\begin{pmatrix} 0 & 0 & 0 & gH_d^+ & 0\\ -\frac{g'}{\sqrt{2}}H_d^+ & -\frac{g}{\sqrt{2}}H_d^+ & 0 & 0 & \lambda e^{\imath \varphi_{\lambda}}H_u^+ \end{pmatrix}={\cal M}_{\chi^{0-}}^T
\end{displaymath}
\begin{displaymath}
{\cal M}_{\chi^{+0}}=\begin{pmatrix} 0 & 0 & gH_u^- & 0 & 0\\ \frac{g'}{\sqrt{2}}H_u^- & \frac{g}{\sqrt{2}}H_u^- & 0 & 0 & \lambda e^{\imath \varphi_{\lambda}}H_d^- \end{pmatrix}={\cal M}_{\chi^{0+}}^T
\end{displaymath}
\begin{displaymath}
{\cal M}_{\chi^{0}}=\begin{pmatrix} M_1 e^{\imath\phi_{M_1}} & 0 & \frac{g'}{\sqrt{2}}H_u^{0\,*} & -\frac{g'}{\sqrt{2}}H_d^{0\,*} & 0\\
 0 & M_2 e^{\imath\phi_{M_2}} & -\frac{g}{\sqrt{2}}H_u^{0\,*} & \frac{g}{\sqrt{2}}H_d^{0\,*} & 0 \\
 \frac{g'}{\sqrt{2}}H_u^{0\,*} & -\frac{g}{\sqrt{2}}H_u^{0\,*} & 0 & -\lambda e^{\imath\varphi_{\lambda}}S & -\lambda e^{\imath\varphi_{\lambda}}H_d^0\\
 -\frac{g'}{\sqrt{2}}H_d^{0\,*} & \frac{g}{\sqrt{2}}H_d^{0\,*} & -\lambda e^{\imath\varphi_{\lambda}}S & 0 & -\lambda e^{\imath\varphi_{\lambda}}H_u^0\\
 0 & 0 & -\lambda e^{\imath\varphi_{\lambda}}H_d^0 & -\lambda e^{\imath\varphi_{\lambda}}H_u^0 & 2\kappa e^{\imath\varphi_{\kappa}}S
\end{pmatrix}={\cal M}_{\chi^{0}}^T
\end{displaymath}

{\em i) Charginos}

The $2\times2$ chargino mass-matrix may be diagonalized via two unitary matrices $U$ and $V$:\newline 
$\left<{\cal M}_{\chi^{-+}}\right>=U^T\mbox{diag}(m_{\chi^{\pm}_1},m_{\chi^{\pm}_2})V$.
To determine $m_{\chi^{\pm}_1}$, $m_{\chi^{\pm}_2}$, $U$ and $V$, we consider the hermitian matrices:
\begin{align*}
 &\left<{\cal M}_{\chi^{+}}^2\right>\equiv\left<{\cal M}_{\chi^{-+}}\right>^{\dag}\left<{\cal M}_{\chi^{-+}}\right>=V^{\dag}\mbox{diag}(m^2_{\chi^{\pm}_1},m^2_{\chi^{\pm}_2})V\\
 &\left<{\cal M}_{\chi^{-}}^2\right>\equiv\left<{\cal M}_{\chi^{-+}}\right>\left<{\cal M}_{\chi^{-+}}\right>^{\dag}=U^{T}\mbox{diag}(m^2_{\chi^{\pm}_1},m^2_{\chi^{\pm}_2})U^*
\end{align*}
which provide:
\begin{align*}
&\begin{cases}
 m^2_{\chi^{\pm}_1}=\frac{1}{2}\left[\left<{\cal M}_{\chi^{+}}^2\right>_{11}+\left<{\cal M}_{\chi^{+}}^2\right>_{22}-\sqrt{\left(\left<{\cal M}_{\chi^{+}}^2\right>_{11}-\left<{\cal M}_{\chi^{+}}^2\right>_{22}\right)^2+4\left<{\cal M}_{\chi^{+}}^2\right>_{12}^2}\right]\\
 m^2_{\chi^{\pm}_2}=\frac{1}{2}\left[\left<{\cal M}_{\chi^{+}}^2\right>_{11}+\left<{\cal M}_{\chi^{+}}^2\right>_{22}+\sqrt{\left(\left<{\cal M}_{\chi^{+}}^2\right>_{11}-\left<{\cal M}_{\chi^{+}}^2\right>_{22}\right)^2+4\left<{\cal M}_{\chi^{+}}^2\right>_{12}^2}\right]
\end{cases}\\
&\begin{cases}
 \theta_V=\arctan\left[\frac{\left<{\cal M}_{\chi^{+}}^2\right>_{11}-\left<{\cal M}_{\chi^{+}}^2\right>_{22}+\sqrt{\left(\left<{\cal M}_{\chi^{+}}^2\right>_{11}-\left<{\cal M}_{\chi^{+}}^2\right>_{22}\right)^2+4\left<{\cal M}_{\chi^{+}}^2\right>_{12}^2}}{2|\left<{\cal M}_{\chi^{+}}^2\right>_{12}|}\right]\\
 \varphi_V=\mbox{arg}[\left<{\cal M}_{\chi^{+}}^2\right>_{12}]
\end{cases}\\
&\begin{cases}
 \theta_U=\arctan\left[\frac{\left<{\cal M}_{\chi^{-}}^2\right>_{11}-\left<{\cal M}_{\chi^{-}}^2\right>_{22}+\sqrt{\left(\left<{\cal M}_{\chi^{-}}^2\right>_{11}-\left<{\cal M}_{\chi^{-}}^2\right>_{22}\right)^2+4\left<{\cal M}_{\chi^{-}}^2\right>_{12}^2}}{2|\left<{\cal M}_{\chi^{-}}^2\right>_{12}|}\right]\\
 \varphi_U=-\mbox{arg}[\left<{\cal M}_{\chi^{-}}^2\right>_{12}]
\end{cases}\\
&U\equiv\begin{pmatrix}e^{\imath\hat{\varphi}_U} & 0\\ 0 & 1\end{pmatrix}
\begin{pmatrix}\cos\theta_U&-\sin\theta_Ue^{\imath\varphi_U}\\
\sin\theta_Ue^{-\imath\varphi_U}&\cos\theta_U\end{pmatrix}\ \ ;\ \ 
V\equiv\begin{pmatrix}1 & 0\\ 0 & e^{\imath\hat{\varphi}_V}\end{pmatrix}
\begin{pmatrix}\cos\theta_V&-\sin\theta_Ve^{\imath\varphi_V}\\
\sin\theta_Ve^{-\imath\varphi_V}&\cos\theta_V\end{pmatrix}
\end{align*}
The choice of phases $\hat{\varphi}_U$, $\hat{\varphi}_V$ is a priori arbitrary. We decide to determine them by the requirement that
$m_{\chi^{\pm}_1}$ and $m_{\chi^{\pm}_2}$, obtained in the matrix product $U^*\left<{\cal M}_{\chi^{-+}}\right>V^{\dag}$, are real and positive.
The associated mass-states are then:
\begin{displaymath}
 \chi^+=V_{i1}(-\imath\tilde{w}^+)+V_{i2}\tilde{h}_u^+\equiv V_{iw}(-\imath\tilde{w}^+)+V_{iu}\tilde{h}_u^+\ \ \ \ ;\ \ \ \ \chi^-=U_{i1}(-\imath\tilde{w}^-)+U_{i2}\tilde{h}_d^-\equiv U_{iw}(-\imath\tilde{w}^-)+U_{id}\tilde{h}_d^-
\end{displaymath}

{\em ii) Neutralinos}

The $5\times5$ neutralino mass-matrix is symmetric, hence is diagonalizable via a single unitary matrix $N$: \newline
$\left<{\cal M}_{\chi^{0}}\right>=N^T\mbox{diag}(m_{\chi^{0}_i},i=1,\ldots,5)N$. As before, we first consider the hermitian matrix
\begin{displaymath}
\left<{\cal M}_{\chi^{0}}^2\right>\equiv\left<{\cal M}_{\chi^{0}}\right>^{\dag}\left<{\cal M}_{\chi^{0}}\right>=N^{\dag}\mbox{diag}(m^2_{\chi^{0}_i},i=1,\ldots,5)N
\end{displaymath}
This hermitian matrix -- or equivalently the $10\times10$ symmetric matrix $\begin{pmatrix} \mbox{Re}&\mbox{Im}\\-\mbox{Im} &\mbox{Re}\end{pmatrix}\left<{\cal M}_{\chi^{0}}^2\right>$ -- may be diagonalized numerically,
providing us with $m^2_{\chi^{0}_i},i=1,\ldots,5$ and a diagonalization matrix $N_0$. We define $N=\mbox{diag}(e^{\imath\varphi_{\chi^{0}_i}},i=1,\ldots,5)N_0$,
where the phases $\varphi_{\chi^{0}_i},i=1,\ldots,5$ are determined by the requirement that the masses $m_{\chi^{0}_i},i=1,\ldots,5$ obtained
from the matrix product $N^*\left<{\cal M}_{\chi^{0}}\right>N^{\dag}$ are real and positive. The neutralino mass-states are then defined as:
\begin{displaymath}
 \chi^0=N_{i1}(-\imath \tilde{b})+N_{i2}(-\imath \tilde{w}^3)+N_{i3}\tilde{h}_u^0+N_{i4}\tilde{h}_d^0+N_{i5}\tilde{h}_s^0\equiv N_{ib}(-\imath \tilde{b})+N_{iw}(-\imath \tilde{w}^3)+N_{iu}\tilde{h}_u^0+N_{id}\tilde{h}_d^0+N_{is}\tilde{h}_s^0
\end{displaymath}

\subsubsection{Gluinos}
The gluons of course remain massless. Concerning their supersymmetric partners, the gluino bilinear terms read:
\begin{equation}
 {\cal V}_{\tilde{g}}=-M_3 e^{\imath\phi_{M_3}}\tilde{g}_{a}\tilde{g}_{a}+h.c.
\end{equation}
so that we define the mass states $\tilde{\cal G}_a\equiv-\imath e^{\frac{\imath}{2}\phi_{M_3}}\tilde{g}_{a}$, with mass $M_3$.

\subsubsection{Higgs sector\label{Higgsmass}}
The tree-level Higgs potential is given in Eq.\ref{tlHiggspot}.

\vspace{0.2cm}
{\em i) Minimization Conditions}

First derivatives of the potential must vanish at the minimum, which provides:
\begin{align*}
 &\frac{1}{2}\left<\frac{\partial {\cal V}_{H^0}}{\partial h_u^0/\sqrt{2}}\right>=0=\left[m_{H_u}^2+\lambda^2(s^2+v_d^2)+\frac{g'^2+g^2}{4}(v_u^2-v_d^2)\right]v_u-\lambda s\left[A_{\lambda}\cos\varphi_1+\kappa s\cos(\varphi_{\lambda}-\varphi_{\kappa})\right]v_d\\
 &\frac{1}{2}\left<\frac{\partial {\cal V}_{H^0}}{\partial h_d^0/\sqrt{2}}\right>=0=\left[m_{H_d}^2+\lambda^2(s^2+v_u^2)-\frac{g'^2+g^2}{4}(v_u^2-v_d^2)\right]v_d-\lambda s\left[A_{\lambda}\cos\varphi_1+\kappa s\cos(\varphi_{\lambda}-\varphi_{\kappa})\right]v_u\\
 &\frac{1}{2}\left<\frac{\partial {\cal V}_{H^0}}{\partial h_s^0/\sqrt{2}}\right>=0=\left[m_{S}^2+\kappa s(A_{\kappa}\cos\varphi_2)+2\kappa s+\lambda^2(v_u^2+v_d^2)\right]s-\lambda\left[A_{\lambda}\cos\varphi_1+2\kappa s\cos(\varphi_{\lambda}-\varphi_{\kappa})\right]v_uv_d\\
 &\frac{1}{2}\left<\frac{\partial {\cal V}_{H^0}}{\partial a_u^0/\sqrt{2}}\right>=0=\lambda s\left[A_{\lambda}\sin\varphi_1+\kappa s\sin(\varphi_{\lambda}-\varphi_{\kappa})\right]v_d\\
 &\frac{1}{2}\left<\frac{\partial {\cal V}_{H^0}}{\partial a_d^0/\sqrt{2}}\right>=0=\lambda s\left[A_{\lambda}\sin\varphi_1+\kappa s\sin(\varphi_{\lambda}-\varphi_{\kappa})\right]v_u\\
 &\frac{1}{2}\left<\frac{\partial {\cal V}_{H^0}}{\partial a_s^0/\sqrt{2}}\right>=0=\lambda \left[A_{\lambda}\sin\varphi_1-2\kappa s\sin(\varphi_{\lambda}-\varphi_{\kappa})\right]v_uv_d-\kappa s^2A_{\kappa}\sin\varphi_2
\end{align*}
So that one can express certain parameters in terms of the v.e.v.'s:
\begin{align*}
 &m_{H_u}^2=\lambda s\left[A_{\lambda}\cos\varphi_1+\kappa s\cos(\varphi_{\lambda}-\varphi_{\kappa})\right]\frac{v_d}{v_u}-\lambda^2(s^2+v_d^2)-\frac{g'^2+g^2}{4}(v_u^2-v_d^2)\\
 &m_{H_d}^2=\lambda s\left[A_{\lambda}\cos\varphi_1+\kappa s\cos(\varphi_{\lambda}-\varphi_{\kappa})\right]\frac{v_u}{v_d}-\lambda^2(s^2+v_u^2)+\frac{g'^2+g^2}{4}(v_u^2-v_d^2)\\
 &m_{S}^2=\lambda\left[A_{\lambda}\cos\varphi_1+2\kappa s\cos(\varphi_{\lambda}-\varphi_{\kappa})\right]\frac{v_uv_d}{s}-\kappa s(A_{\kappa}\cos\varphi_2)+2\kappa s-\lambda^2(v_u^2+v_d^2)\\
 &A_{\lambda}\sin\varphi_1=-\kappa s\sin(\varphi_{\lambda}-\varphi_{\kappa})\\
 &A_{\kappa}\sin\varphi_2=\frac{\lambda}{\kappa}\left[A_{\lambda}\sin\varphi_1-2\kappa s\sin(\varphi_{\lambda}-\varphi_{\kappa})\right]\frac{v_uv_d}{s^2}=-3\lambda\frac{v_uv_d}{s}\sin(\varphi_{\lambda}-\varphi_{\kappa})
\end{align*}

\vspace{0.2cm}
{\em ii) Charged Higgs}

The $2\times2$ charged-Higgs bilinear terms read:
\begin{equation}
 {V}_{H^{\pm}}=(H_u^-,H_d^-){\cal M}^2_{H^{\pm}}\begin{pmatrix}H_u^+\\H_d^+\end{pmatrix}
\end{equation}
\begin{align*}
 \left.{\cal M}^2_{H^{\pm}}\right|_{11}& =m_{H_u}^2+\lambda^2|S|^2+\frac{g'^2+g^2}{4}\left(|H_u^0|^2-|H_d^0|^2\right)+\frac{g^2}{2}|H_d^0|^2+\lambda^2H_d^+H_d^-+\frac{g'^2+g^2}{4}\left[2H_u^+H_u^--H_d^+H_d^-\right]\\
 & =\lambda s\left[A_{\lambda}\cos\varphi_1+\kappa s\cos(\varphi_{\lambda}-\varphi_{\kappa})\right]\frac{v_d}{v_u}-\left(\lambda^2-\frac{g^2}{2}\right)v_d^2+\lambda^2H_d^+H_d^-+\frac{g'^2+g^2}{4}\left[2H_u^+H_u^--H_d^+H_d^-\right]\\
 &\null\hspace{0.1cm}+\lambda^2(|S|^2-s^2)+\frac{g'^2+g^2}{4}\left(|H_u^0|^2-|H_d^0|^2-v_u^2+v_d^2\right)+\frac{g^2}{2}(|H_d^0|^2-v_d^2)
\end{align*}
\begin{align*}
 \left.{\cal M}^2_{H^{\pm}}\right|_{22}& =m_{H_d}^2+\lambda^2|S|^2-\frac{g'^2+g^2}{4}\left(|H_u^0|^2-|H_d^0|^2\right)+\frac{g^2}{2}|H_u^0|^2+\lambda^2H_u^+H_u^-+\frac{g'^2+g^2}{4}\left[2H_d^+H_d^--H_u^+H_u^-\right]\\
 & =\lambda s\left[A_{\lambda}\cos\varphi_1+\kappa s\cos(\varphi_{\lambda}-\varphi_{\kappa})\right]\frac{v_u}{v_d}-\left(\lambda^2-\frac{g^2}{2}\right)v_u^2+\lambda^2H_u^+H_u^-+\frac{g'^2+g^2}{4}\left[2H_d^+H_d^--H_u^+H_u^-\right]\\
 &\null\hspace{0.1cm}+\lambda^2(|S|^2-s^2)-\frac{g'^2+g^2}{4}\left(|H_u^0|^2-|H_d^0|^2-v_u^2+v_d^2\right)+\frac{g^2}{2}(|H_u^0|^2-v_u^2)
\end{align*}
\begin{align*}
 \left.{\cal M}^2_{H^{\pm}}\right|_{12}& =\lambda \left[A_{\lambda}e^{-\imath\varphi_1}S^*+\kappa e^{-\imath(\varphi_{\lambda}-\varphi_{\kappa})}S^2\right]-(\lambda^2-\frac{g^2}{2})H_u^0H_d^0+(\lambda^2-\frac{g'^2+g^2}{4})H_u^+H_d^-=\left.{\cal M}^2_{H^{\pm}}\right|_{21}^*\\
 & =\lambda s\left[A_{\lambda}\cos\varphi_1+\kappa s\cos(\varphi_{\lambda}-\varphi_{\kappa})\right]-\left(\lambda^2-\frac{g^2}{2}\right)v_uv_d+(\lambda^2-\frac{g'^2+g^2}{4})H_u^+H_d^-\\
 &\null\hspace{0cm}+\lambda \left[A_{\lambda}\cos\varphi_1(S^*-s)+\kappa \cos(\varphi_{\lambda}-\varphi_{\kappa})(S^2-s^2)-\imath\kappa \sin(\varphi_{\lambda}-\varphi_{\kappa})(S^2-sS^*)\right]-(\lambda^2-\frac{g^2}{2})(H_u^0H_d^0-v_uv_d)
\end{align*}

Obviously,
\begin{multline}
 \left<{\cal M}^2_{H^{\pm}}\right>=\left\{\lambda s\left[A_{\lambda}\cos\varphi_1+\kappa s\cos(\varphi_{\lambda}-\varphi_{\kappa})\right]-\left(\lambda^2-\frac{g^2}{2}\right)v_uv_d\right\}\begin{pmatrix}\frac{v_d}{v_u}&1\\1&\frac{v_u}{v_d}\end{pmatrix}\\
=\begin{pmatrix}-\sin\beta&\cos\beta\\\cos\beta&\sin\beta\end{pmatrix}\begin{pmatrix}0&0\\0&m_{H^{\pm}}^2\end{pmatrix}\begin{pmatrix}-\sin\beta&\cos\beta\\\cos\beta&\sin\beta\end{pmatrix}\\
m_{H^{\pm}}^2\equiv\left\{\frac{\lambda s}{v_uv_d}\left[A_{\lambda}\cos\varphi_1+\kappa s\cos(\varphi_{\lambda}-\varphi_{\kappa})\right]-\left(\lambda^2-\frac{g^2}{2}\right)\right\}(v_u^2+v_d^2)\ \ \ \ ;\ \ \ \ \tan\beta\equiv\frac{v_u}{v_d}
\end{multline}
with the Goldstone boson $G^{\pm}=-\sin\beta H_u^{\pm}+\cos\beta H_d^{\pm}$ and the charged Higgs state $H^{\pm}=\cos\beta H_u^{\pm}+\sin\beta H_d^{\pm}$.
We will denote the corresponding rotation matrix as follows:
\begin{displaymath}
 X^C\equiv\begin{pmatrix}-\sin\beta&\cos\beta\\ \cos\beta&\sin\beta\end{pmatrix}\ \ \ \ ;\ \ \ \ H_1^{\pm}\equiv G^{\pm}\ \ \ ;\ \ \ H_2^{\pm}\equiv H^{\pm}
\end{displaymath}

\vspace{0.2cm}
{\em iii) Neutral Higgs}

The symmetric $6\times6$ bilinear Higgs matrix ${\cal M}^2_{H^{0}}\equiv\Big[\frac{1}{2}\frac{\partial^2{\cal V}_{H^0}}{\partial S_i/\sqrt{2}\partial S_j/\sqrt{2}}, S_{i,j}=h_u^0,h_d^0,h_s^0,a_u^0,a_d^0,a_s^0\Big]$
includes the following elements:
\begin{align*}
 \left.{\cal M}^2_{H^{0}}\right|_{11}& =m_{H_u}^2+\lambda^2(|S|^2+|H_d^0|^2)+\frac{g'^2+g^2}{4}\left(2\mbox{Re}(H_u^0)^2+|H_u^0|^2-|H_d^0|^2\right)+\frac{g'^2+g^2}{4}H_u^+H_u^--\frac{g'^2-g^2}{4}H_d^+H_d^-\\
 & =\lambda s\left[A_{\lambda}\cos\varphi_1+\kappa s\cos(\varphi_{\lambda}-\varphi_{\kappa})\right]\frac{v_d}{v_u}+\frac{g'^2+g^2}{2}v_u^2+\frac{g'^2+g^2}{4}H_u^+H_u^--\frac{g'^2-g^2}{4}H_d^+H_d^-\\
 &\null\hspace{3cm}+\lambda^2(|S|^2-s^2+|H_d^0|^2-v_d^2)+\frac{g'^2+g^2}{4}\left(2\mbox{Re}(H_u^0)^2+|H_u^0|^2-|H_d^0|^2-3v_u^2+v_d^2\right)
\end{align*}
\begin{align*}
 \left.{\cal M}^2_{H^{0}}\right|_{12}& =-\lambda \mbox{Re}\left[A_{\lambda}e^{\imath\varphi_1}S+\kappa e^{\imath(\varphi_{\lambda}-\varphi_{\kappa})}S^{*\,2}\right]+2\left(\lambda^2-\frac{g'^2+g^2}{4}\right)\mbox{Re}(H_u^0)\mbox{Re}(H_d^0)\\
 & =-\lambda s\left[A_{\lambda}\cos\varphi_1+\kappa s\cos(\varphi_{\lambda}-\varphi_{\kappa})\right]+2\left(\lambda^2-\frac{g'^2+g^2}{4}\right)v_uv_d-(\lambda^2-\frac{g^2}{2})\mbox{Re}(H_u^+H_d^-)\\
 &\null\hspace{1cm}-\lambda \left[A_{\lambda}\cos\varphi_1\mbox{Re}(S-s)+\kappa \cos(\varphi_{\lambda}-\varphi_{\kappa})\mbox{Re}(S^2-s^2)+\kappa \sin(\varphi_{\lambda}-\varphi_{\kappa})\mbox{Im}(S)\mbox{Re}(2S+s)\right]\\
 &\null\hspace{5cm}+2\left(\lambda^2-\frac{g'^2+g^2}{4}\right)(\mbox{Re}(H_u^0)\mbox{Re}(H_d^0)-v_uv_d)-(\lambda^2-\frac{g^2}{2})\mbox{Re}(H_u^+H_d^-)
\end{align*}
\begin{align*}
 \left.{\cal M}^2_{H^{0}}\right|_{22}& =m_{H_d}^2+\lambda^2(|S|^2+|H_u^0|^2)+\frac{g'^2+g^2}{4}\left(2\mbox{Re}(H_d^0)^2-|H_u^0|^2+|H_d^0|^2\right)+\frac{g'^2+g^2}{4}H_d^+H_d^--\frac{g'^2-g^2}{4}H_u^+H_u^-\\
 & =\lambda s\left[A_{\lambda}\cos\varphi_1+\kappa s\cos(\varphi_{\lambda}-\varphi_{\kappa})\right]\frac{v_u}{v_d}+\frac{g'^2+g^2}{2}v_d^2+\frac{g'^2+g^2}{4}H_d^+H_d^--\frac{g'^2-g^2}{4}H_u^+H_u^-\\
 &\null\hspace{3cm}+\lambda^2(|S|^2-s^2+|H_u^0|^2-v_u^2)+\frac{g'^2+g^2}{4}\left(2\mbox{Re}(H_d^0)^2+|H_d^0|^2-|H_u^0|^2-3v_d^2+v_u^2\right)
\end{align*}
\begin{align*}
 \left.{\cal M}^2_{H^{0}}\right|_{13}& =-\lambda \mbox{Re}\left[(A_{\lambda}e^{\imath\varphi_1}+2\kappa e^{\imath(\varphi_{\lambda}-\varphi_{\kappa})}S^{*})H_d^0\right]+2\lambda^2\mbox{Re}(S)\mbox{Re}(H_u^0)\\
 & =-\lambda v_d\left[A_{\lambda}\cos\varphi_1+2\kappa s\cos(\varphi_{\lambda}-\varphi_{\kappa})\right]+2\lambda^2sv_u\\
 &\null\hspace{1cm}-\lambda \left[A_{\lambda}\cos\varphi_1\mbox{Re}(H_d^0-v_d)+2\kappa \cos(\varphi_{\lambda}-\varphi_{\kappa})\mbox{Re}(S^*H_d^0-sv_d)\right.\\
 &\null\hspace{1.5cm}+\left.\kappa \sin(\varphi_{\lambda}-\varphi_{\kappa})\left(\mbox{Im}(H_d^0)\mbox{Re}(s-2S)+2\mbox{Im}(S)\mbox{Re}(H_d^0)\right)\right]+2\lambda^2\left(\mbox{Re}(S)\mbox{Re}(H_u^0)-sv_u\right)
\end{align*}
\begin{align*}
 \left.{\cal M}^2_{H^{0}}\right|_{23}& =-\lambda \mbox{Re}\left[(A_{\lambda}e^{\imath\varphi_1}+2\kappa e^{\imath(\varphi_{\lambda}-\varphi_{\kappa})}S^{*})H_u^0\right]+2\lambda^2\mbox{Re}(S)\mbox{Re}(H_d^0)\\
 & =-\lambda v_u\left[A_{\lambda}\cos\varphi_1+2\kappa s\cos(\varphi_{\lambda}-\varphi_{\kappa})\right]+2\lambda^2sv_d\\
 &\null\hspace{1cm}-\lambda \left[A_{\lambda}\cos\varphi_1\mbox{Re}(H_u^0-v_u)+2\kappa \cos(\varphi_{\lambda}-\varphi_{\kappa})\mbox{Re}(S^*H_u^0-sv_u)\right.\\
 &\null\hspace{1.5cm}+\left.\kappa \sin(\varphi_{\lambda}-\varphi_{\kappa})\left(\mbox{Im}(H_u^0)\mbox{Re}(s-2S)+2\mbox{Im}(S)\mbox{Re}(H_u^0)\right)\right]+2\lambda^2\left(\mbox{Re}(S)\mbox{Re}(H_d^0)-sv_d\right)
\end{align*}
\begin{align*}
 \left.{\cal M}^2_{H^{0}}\right|_{33}& =m_S^2+2\kappa A_{\kappa}\mbox{Re}(e^{\imath\varphi_2}S)+2\kappa^2\left[2\mbox{Re}(S)^2+|S|^2\right]+\lambda^2(|H_u^0|^2+|H_d^0|^2)-2\lambda\kappa\mbox{Re}\left[e^{\imath(\varphi_{\lambda}-\varphi_{\kappa})}H_u^0H_d^0\right]\\
 &\null\hspace{5cm}+\lambda^2\left[H_u^+H_u^-+H_d^+H_d^-\right]+2\lambda\kappa\mbox{Re}\left[e^{\imath(\varphi_{\lambda}-\varphi_{\kappa})}H_u^+H_d^-\right]\\
 & =\kappa s\left[A_{\kappa}\cos\varphi_2+4\kappa s\right]+\lambda A_{\lambda}\cos\varphi_1\frac{v_uv_d}{s}+\lambda^2\left[H_u^+H_u^-+H_d^+H_d^-\right]+2\lambda\kappa\mbox{Re}\left[e^{\imath(\varphi_{\lambda}-\varphi_{\kappa})}H_u^+H_d^-\right]\\
 &\null\hspace{0.5cm}+2\kappa\left[A_{\kappa}\cos\varphi_2\mbox{Re}(S-s)+\kappa\left(2\mbox{Re}(S)^2+|S|^2-3s^2\right)\right]+\lambda^2\left[|H_u^0|^2+|H_d^0|^2-v_u^2-v_d^2\right]\\
 &\null\hspace{0.5cm}-2\lambda\kappa\left[\cos{(\varphi_{\lambda}-\varphi_{\kappa})}\mbox{Re}(H_u^0H_d^0-v_uv_d)-\sin{(\varphi_{\lambda}-\varphi_{\kappa})}\left(\mbox{Re}(H_u^0)\mbox{Im}(H_d^0)+\mbox{Im}(H_u^0)\mbox{Re}(H_d^0)+3\frac{v_uv_d}{s}\mbox{Im}(S)\right)\right]
\end{align*}
\begin{displaymath}
 \left.{\cal M}^2_{H^{0}}\right|_{14}=\frac{g'^2+g^2}{2}\mbox{Re}(H_u^0)\mbox{Im}(H_u^0)\hspace{11cm}\null
\end{displaymath}
\begin{align*}
 \left.{\cal M}^2_{H^{0}}\right|_{24}& =-\lambda \mbox{Re}\left[\imath\left(A_{\lambda}e^{\imath\varphi_1}S+\kappa e^{\imath(\varphi_{\lambda}-\varphi_{\kappa})}S^{*\,2}\right)\right]+2\left(\lambda^2-\frac{g'^2+g^2}{4}\right)\mbox{Im}(H_u^0)\mbox{Re}(H_d^0)-(\lambda^2-\frac{g^2}{2})\mbox{Im}(H_u^+H_d^-)\\
 & =\lambda \left[\left(A_{\lambda}\cos\varphi_1-2\kappa \cos(\varphi_{\lambda}-\varphi_{\kappa})\mbox{Re}(S)\right)\mbox{Im}(S)+\kappa \sin(\varphi_{\lambda}-\varphi_{\kappa})\mbox{Re}(S^2-sS)\right]\\
 &\null\hspace{5cm}+2\left(\lambda^2-\frac{g'^2+g^2}{4}\right)\mbox{Im}(H_u^0)\mbox{Re}(H_d^0)-(\lambda^2-\frac{g^2}{2})\mbox{Im}(H_u^+H_d^-)
\end{align*}
\begin{align*}
 \left.{\cal M}^2_{H^{0}}\right|_{34}& =-\lambda \mbox{Re}\left[\imath\left(A_{\lambda}e^{\imath\varphi_1}+2\kappa e^{\imath(\varphi_{\lambda}-\varphi_{\kappa})}S^{*}\right)H_d^0\right]+2\lambda^2\mbox{Re}(S)\mbox{Im}(H_u^0)\\
 & =\lambda\kappa sv_d\sin(\varphi_{\lambda}-\varphi_{\kappa})\hspace{0.3cm}+\lambda \left[A_{\lambda}\cos\varphi_1\mbox{Im}(H_d^0)+2\kappa \cos(\varphi_{\lambda}-\varphi_{\kappa})\left(\mbox{Re}(S)\mbox{Im}(H_d^0)-\mbox{Im}(S)\mbox{Re}(H_d^0)\right)\right.\\
 &\null\hspace{5cm}\left.+\kappa \sin(\varphi_{\lambda}-\varphi_{\kappa})\mbox{Re}((2S^*-s)H_d^0-sv_d)\right]+2\lambda^2\mbox{Im}(H_u^0)\mbox{Re}(S)
\end{align*}
\begin{align*}
 \left.{\cal M}^2_{H^{0}}\right|_{44}& =m_{H_u}^2+\lambda^2(|S|^2+|H_d^0|^2)+\frac{g'^2+g^2}{4}\left(2\mbox{Im}(H_u^0)^2+|H_u^0|^2-|H_d^0|^2\right)+\frac{g'^2+g^2}{4}H_u^+H_u^--\frac{g'^2-g^2}{4}H_d^+H_d^-\\
 & =\lambda s\left[A_{\lambda}\cos\varphi_1+\kappa s\cos(\varphi_{\lambda}-\varphi_{\kappa})\right]\frac{v_d}{v_u}+\frac{g'^2+g^2}{4}H_u^+H_u^--\frac{g'^2-g^2}{4}H_d^+H_d^-\\
 &\null\hspace{3cm}+\lambda^2(|S|^2-s^2+|H_d^0|^2-v_d^2)+\frac{g'^2+g^2}{4}\left(2\mbox{Im}(H_u^0)^2+|H_u^0|^2-|H_d^0|^2-v_u^2+v_d^2\right)
\end{align*}
\begin{align*}
 \left.{\cal M}^2_{H^{0}}\right|_{15}& =-\lambda \mbox{Re}\left[\imath\left(A_{\lambda}e^{\imath\varphi_1}S+\kappa e^{\imath(\varphi_{\lambda}-\varphi_{\kappa})}S^{*\,2}\right)\right]+2\left(\lambda^2-\frac{g'^2+g^2}{4}\right)\mbox{Re}(H_u^0)\mbox{Im}(H_d^0)-(\lambda^2-\frac{g^2}{2})\mbox{Im}(H_u^+H_d^-)\\
 & =\lambda \left[\left(A_{\lambda}\cos\varphi_1-2\kappa \cos(\varphi_{\lambda}-\varphi_{\kappa})\mbox{Re}(S)\right)\mbox{Im}(S)+\kappa \sin(\varphi_{\lambda}-\varphi_{\kappa})\mbox{Re}(S^2-sS)\right]\\
 &\null\hspace{5cm}+2\left(\lambda^2-\frac{g'^2+g^2}{4}\right)\mbox{Re}(H_u^0)\mbox{Im}(H_d^0)-(\lambda^2-\frac{g^2}{2})\mbox{Im}(H_u^+H_d^-)
\end{align*}
\begin{displaymath}
 \left.{\cal M}^2_{H^{0}}\right|_{25}=\frac{g'^2+g^2}{2}\mbox{Re}(H_d^0)\mbox{Im}(H_d^0)\hspace{11cm}\null
\end{displaymath}
\begin{align*}
 \left.{\cal M}^2_{H^{0}}\right|_{35}& =-\lambda \mbox{Re}\left[\imath\left(A_{\lambda}e^{\imath\varphi_1}+2\kappa e^{\imath(\varphi_{\lambda}-\varphi_{\kappa})}S^{*}\right)H_u^0\right]+2\lambda^2\mbox{Re}(S)\mbox{Im}(H_d^0)\\
 & =\lambda\kappa sv_u\sin(\varphi_{\lambda}-\varphi_{\kappa})\hspace{0.3cm}+\lambda \left[A_{\lambda}\cos\varphi_1\mbox{Im}(H_u^0)+2\kappa \cos(\varphi_{\lambda}-\varphi_{\kappa})\left(\mbox{Re}(S)\mbox{Im}(H_u^0)-\mbox{Im}(S)\mbox{Re}(H_u^0)\right)\right.\\
 &\null\hspace{5cm}\left.+\kappa \sin(\varphi_{\lambda}-\varphi_{\kappa})\mbox{Re}((2S^*-s)H_u^0-sv_u)\right]+2\lambda^2\mbox{Im}(H_d^0)\mbox{Re}(S)
\end{align*}
\begin{align*}
 \left.{\cal M}^2_{H^{0}}\right|_{45}& =\lambda \mbox{Re}\left[A_{\lambda}e^{\imath\varphi_1}S+\kappa e^{\imath(\varphi_{\lambda}-\varphi_{\kappa})}S^{*\,2}\right]+2\left(\lambda^2-\frac{g'^2+g^2}{4}\right)\mbox{Im}(H_u^0)\mbox{Im}(H_d^0)+(\lambda^2-\frac{g^2}{2})\mbox{Re}(H_u^+H_d^-)\\
 & =\lambda s\left[A_{\lambda}\cos\varphi_1+\kappa s\cos(\varphi_{\lambda}-\varphi_{\kappa})\right]\\
 &\null\hspace{1cm}+\lambda \left[A_{\lambda}\cos\varphi_1\mbox{Re}(S-s)+\kappa \cos(\varphi_{\lambda}-\varphi_{\kappa})\mbox{Re}(S^2-s^2)+\kappa \sin(\varphi_{\lambda}-\varphi_{\kappa})\mbox{Im}(S)\mbox{Re}(2S+s)\right]\\
 &\null\hspace{0.1cm}+2\left(\lambda^2-\frac{g'^2+g^2}{4}\right)\mbox{Im}(H_u^0)\mbox{Im}(H_d^0)-(\lambda^2-\frac{g^2}{2})\mbox{Re}(H_u^+H_d^-)
\end{align*}
\begin{align*}
 \left.{\cal M}^2_{H^{0}}\right|_{55}& =m_{H_d}^2+\lambda^2(|S|^2+|H_u^0|^2)+\frac{g'^2+g^2}{4}\left(2\mbox{Im}(H_d^0)^2-|H_u^0|^2+|H_d^0|^2\right)+\frac{g'^2+g^2}{4}H_d^+H_d^--\frac{g'^2-g^2}{4}H_u^+H_u^-\\
 & =\lambda s\left[A_{\lambda}\cos\varphi_1+\kappa s\cos(\varphi_{\lambda}-\varphi_{\kappa})\right]\frac{v_u}{v_d}+\frac{g'^2+g^2}{4}H_d^+H_d^--\frac{g'^2-g^2}{4}H_u^+H_u^-\\
 &\null\hspace{3cm}+\lambda^2(|S|^2-s^2+|H_u^0|^2-v_u^2)+\frac{g'^2+g^2}{4}\left(2\mbox{Im}(H_d^0)^2+|H_d^0|^2-|H_u^0|^2-3v_d^2+v_u^2\right)
\end{align*}
\begin{align*}
 \left.{\cal M}^2_{H^{0}}\right|_{16}& =-\lambda \mbox{Re}\left[\imath\left(A_{\lambda}e^{\imath\varphi_1}-2\kappa e^{\imath(\varphi_{\lambda}-\varphi_{\kappa})}S^{*}\right)H_d^0\right]+2\lambda^2\mbox{Im}(S)\mbox{Re}(H_u^0)\\
 & =-3\lambda\kappa sv_d\sin(\varphi_{\lambda}-\varphi_{\kappa})\hspace{0.3cm}+\lambda \left[A_{\lambda}\cos\varphi_1\mbox{Im}(H_d^0)+2\kappa \cos(\varphi_{\lambda}-\varphi_{\kappa})\left(\mbox{Im}(S)\mbox{Re}(H_d^0)-\mbox{Re}(S)\mbox{Im}(H_d^0)\right)\right.\\
 &\null\hspace{5cm}\left.-\kappa \sin(\varphi_{\lambda}-\varphi_{\kappa})\mbox{Re}((2S^*+s)H_d^0-3sv_d)\right]+2\lambda^2\mbox{Im}(S)\mbox{Re}(H_u^0)
\end{align*}
\begin{align*}
 \left.{\cal M}^2_{H^{0}}\right|_{26}& =-\lambda \mbox{Re}\left[\imath\left(A_{\lambda}e^{\imath\varphi_1}-2\kappa e^{\imath(\varphi_{\lambda}-\varphi_{\kappa})}S^{*}\right)H_u^0\right]+2\lambda^2\mbox{Im}(S)\mbox{Re}(H_d^0)\\
 & =-3\lambda\kappa sv_u\sin(\varphi_{\lambda}-\varphi_{\kappa})\hspace{0.3cm}+\lambda \left[A_{\lambda}\cos\varphi_1\mbox{Im}(H_u^0)+2\kappa \cos(\varphi_{\lambda}-\varphi_{\kappa})\left(\mbox{Im}(S)\mbox{Re}(H_u^0)-\mbox{Re}(S)\mbox{Im}(H_u^0)\right)\right.\\
 &\null\hspace{5cm}\left.-\kappa \sin(\varphi_{\lambda}-\varphi_{\kappa})\mbox{Re}((2S^*+s)H_u^0-3sv_u)\right]+2\lambda^2\mbox{Im}(S)\mbox{Re}(H_d^0)
\end{align*}
\begin{align*}
 \left.{\cal M}^2_{H^{0}}\right|_{36}& =2\kappa A_{\kappa} \mbox{Re}\left[\imath e^{\imath\varphi_2}S\right]+4\kappa^2\mbox{Re}(S)\mbox{Im}(S)+2\lambda\kappa\mbox{Re}\left[\imath e^{\imath(\varphi_{\lambda}-\varphi_{\kappa})}H_u^0H_d^0\right]-2\lambda\kappa\mbox{Re}\left[\imath e^{\imath(\varphi_{\lambda}-\varphi_{\kappa})}H_u^+H_d^-\right]\\
 & =4\lambda\kappa v_uv_d\sin(\varphi_{\lambda}-\varphi_{\kappa})\hspace{0.3cm}+2\lambda\kappa \sin(\varphi_{\lambda}-\varphi_{\kappa})\left[3\frac{\mbox{\small Re}(S)}{s}v_uv_d-\mbox{Re}(H_u^0H_d^0)-2v_uv_d\right]\\
 &\null\hspace{1.5cm}-2\kappa\left[A_{\kappa}\cos\varphi_2-2\kappa\mbox{Re}(S)\right]\mbox{Im}(S)-2\lambda\kappa \cos(\varphi_{\lambda}-\varphi_{\kappa})\left[\mbox{Im}(H_u^0)\mbox{Re}(H_d^0)+\mbox{Re}(H_u^0)\mbox{Im}(H_d^0)\right]\\
 &\null\hspace{3cm}-2\lambda\kappa\mbox{Re}\left[\imath e^{\imath(\varphi_{\lambda}-\varphi_{\kappa})}H_u^+H_d^-\right]
\end{align*}
\begin{align*}
 \left.{\cal M}^2_{H^{0}}\right|_{46}& =\lambda \mbox{Re}\left[(A_{\lambda}e^{\imath\varphi_1}-2\kappa e^{\imath(\varphi_{\lambda}-\varphi_{\kappa})}S^{*})H_d^0\right]+2\lambda^2\mbox{Im}(S)\mbox{Im}(H_u^0)\\
 & =\lambda v_d\left[A_{\lambda}\cos\varphi_1-2\kappa s\cos(\varphi_{\lambda}-\varphi_{\kappa})\right]\\
 &\null\hspace{1cm}+\lambda \left[A_{\lambda}\cos\varphi_1\mbox{Re}(H_d^0-v_d)-2\kappa \cos(\varphi_{\lambda}-\varphi_{\kappa})\mbox{Re}(S^*H_d^0-sv_d)\right.\\
 &\null\hspace{1.5cm}+\left.\kappa \sin(\varphi_{\lambda}-\varphi_{\kappa})\left(\mbox{Im}(H_d^0)\mbox{Re}(2S+s)-2\mbox{Im}(S)\mbox{Re}(H_d^0)\right)\right]+2\lambda^2\mbox{Im}(S)\mbox{Im}(H_u^0)
\end{align*}
\begin{align*}
 \left.{\cal M}^2_{H^{0}}\right|_{56}& =-\lambda \mbox{Re}\left[(A_{\lambda}e^{\imath\varphi_1}-2\kappa e^{\imath(\varphi_{\lambda}-\varphi_{\kappa})}S^{*})H_u^0\right]+2\lambda^2\mbox{Im}(S)\mbox{Im}(H_d^0)\\
 & =\lambda v_u\left[A_{\lambda}\cos\varphi_1-2\kappa s\cos(\varphi_{\lambda}-\varphi_{\kappa})\right]\\
 &\null\hspace{1cm}+\lambda \left[A_{\lambda}\cos\varphi_1\mbox{Re}(H_u^0-v_u)-2\kappa \cos(\varphi_{\lambda}-\varphi_{\kappa})\mbox{Re}(S^*H_u^0-sv_u)\right.\\
 &\null\hspace{1.5cm}+\left.\kappa \sin(\varphi_{\lambda}-\varphi_{\kappa})\left(\mbox{Im}(H_u^0)\mbox{Re}(2S+s)-2\mbox{Im}(S)\mbox{Re}(H_u^0)\right)\right]+2\lambda^2\mbox{Im}(S)\mbox{Im}(H_d^0)
\end{align*}
\begin{align*}
 \left.{\cal M}^2_{H^{0}}\right|_{66}& =m_S^2-2\kappa A_{\kappa}\mbox{Re}(e^{\imath\varphi_2}S)-2\kappa^2\left[2\mbox{Im}(S)^2+|S|^2\right]+\lambda^2(|H_u^0|^2+|H_d^0|^2)+2\lambda\kappa\mbox{Re}\left[e^{\imath(\varphi_{\lambda}-\varphi_{\kappa})}H_u^0H_d^0\right]\\
 &\null\hspace{3cm}+\lambda^2\left[H_u^+H_u^-+H_d^+H_d^-\right]-2\lambda\kappa\mbox{Re}\left[e^{\imath(\varphi_{\lambda}-\varphi_{\kappa})}H_u^+H_d^-\right]\\
 & =-3\kappa sA_{\kappa}\cos\varphi_2+\lambda\frac{v_uv_d}{s} \left[A_{\lambda}\cos\varphi_1+4\kappa s\cos(\varphi_{\lambda}-\varphi_{\kappa})\right]+\lambda^2\left[H_u^+H_u^-+H_d^+H_d^-\right]-2\lambda\kappa\mbox{Re}\left[e^{\imath(\varphi_{\lambda}-\varphi_{\kappa})}H_u^+H_d^-\right]\\
 &\null\hspace{0.5cm}-2\kappa A_{\kappa}\cos\varphi_2\mbox{Re}(S-s)+2\kappa^2\left(2\mbox{Im}(S)^2+|S|^2-s^2\right)+\lambda^2\left[|H_u^0|^2+|H_d^0|^2-v_u^2-v_d^2\right]\\
 &\null\hspace{0.5cm}+2\lambda\kappa\left[\cos{(\varphi_{\lambda}-\varphi_{\kappa})}\mbox{Re}(H_u^0H_d^0-v_uv_d)-\sin{(\varphi_{\lambda}-\varphi_{\kappa})}\left(\mbox{Re}(H_u^0)\mbox{Im}(H_d^0)+\mbox{Im}(H_u^0)\mbox{Re}(H_d^0)+3\frac{v_uv_d}{s}\mbox{Im}(S)\right)\right]
\end{align*}

As for the case of the charged Higgs, the Goldstone boson $G^0\equiv-\sin\beta a_u^0+\cos\beta a_d^0$ can be separated from the doublet CP-odd
state $a^0\equiv\cos\beta a_u^0+\sin\beta a_d^0$ by a rotation of angle $\beta$. The remaining (symmetric) $5\times5$ sub-matrix of massive states
$\left<\tilde{\cal M}^2_{H^{0}}\right>$ may be diagonalized (numerically) through an orthogonal matrix $X^{H^0}$:
\begin{equation}
 \left<\tilde{\cal M}^2_{H^{0}}\right>=X^{H^0\,T}\mbox{diag}(m_{S_i^0}^2,i=1,\ldots,5)X^{H^0}
\end{equation}
The corresponding mass-states are then:
\begin{displaymath}
 S_i^0=X^{H^0}_{i1}h_u^0+X^{H^0}_{i2}h_d^0+X^{H^0}_{i3}h_s^0+X^{H^0}_{i4}a^0+X^{H^0}_{i5}a_s^0\equiv X^{R}_{iu}h_u^0+X^{R}_{id}h_d^0+X^{R}_{is}h_s^0+X^{I}_{ia}a^0+X^{I}_{is}a_s^0
\end{displaymath}

\vspace{0.2cm}
{\em iv) Charged-Neutral Higgs terms}

For completeness we indicate the bilinear terms mixing charged and neutral Higgs states (note that $\left.{\cal M}^2\right|_{S^-S^0}=\left.{\cal M}^2\right|_{S^+S^0}^*$):
\begin{align*}
 \left.{\cal M}^2\right|_{H_u^+h_u^0}& =\frac{1}{\sqrt{2}}\left[-(\lambda^2-\frac{g^2}{2})H_d^-H_d^{0\,*}+\frac{g'^2+g^2}{2}H_u^-\mbox{Re}(H_u^0)\right]\\
 \left.{\cal M}^2\right|_{H_u^+h_d^0}& =\frac{1}{\sqrt{2}}\left[-(\lambda^2-\frac{g^2}{2})H_d^-H_u^{0\,*}-\frac{g'^2-g^2}{2}H_u^-\mbox{Re}(H_d^0)\right]\\
 \left.{\cal M}^2\right|_{H_u^+h_s^0}& =\frac{1}{\sqrt{2}}\left[\lambda\left(A_{\lambda}e^{\imath\varphi_1}+2\kappa e^{\imath(\varphi_{\lambda}-\varphi_{\kappa})}S^*\right)H_d^-+2\lambda^2\mbox{Re}(S)H_u^-\right]\\
 \left.{\cal M}^2\right|_{H_d^+h_u^0}& =\frac{1}{\sqrt{2}}\left[-(\lambda^2-\frac{g^2}{2})H_u^-H_d^{0}-\frac{g'^2-g^2}{2}H_d^-\mbox{Re}(H_u^0)\right]\\
 \left.{\cal M}^2\right|_{H_d^+h_d^0}& =\frac{1}{\sqrt{2}}\left[-(\lambda^2-\frac{g^2}{2})H_u^-H_u^{0}+\frac{g'^2+g^2}{2}H_d^-\mbox{Re}(H_d^0)\right]\\
 \left.{\cal M}^2\right|_{H_d^+h_s^0}& =\frac{1}{\sqrt{2}}\left[\lambda\left(A_{\lambda}e^{-\imath\varphi_1}+2\kappa e^{-\imath(\varphi_{\lambda}-\varphi_{\kappa})}S\right)H_u^-+2\lambda^2\mbox{Re}(S)H_d^-\right]
\end{align*}
\begin{align*}
 \left.{\cal M}^2\right|_{H_u^+a_u^0}& =\frac{1}{\sqrt{2}}\left[\imath(\lambda^2-\frac{g^2}{2})H_d^-H_d^{0\,*}+\frac{g'^2+g^2}{2}H_u^-\mbox{Im}(H_u^0)\right]\\
 \left.{\cal M}^2\right|_{H_u^+a_d^0}& =\frac{1}{\sqrt{2}}\left[\imath(\lambda^2-\frac{g^2}{2})H_d^-H_u^{0\,*}-\frac{g'^2-g^2}{2}H_u^-\mbox{Im}(H_d^0)\right]\\
 \left.{\cal M}^2\right|_{H_u^+a_s^0}& =\frac{1}{\sqrt{2}}\left[\imath\lambda\left(A_{\lambda}e^{\imath\varphi_1}-2\kappa e^{\imath(\varphi_{\lambda}-\varphi_{\kappa})}S^*\right)H_d^-+2\lambda^2\mbox{Im}(S)H_u^-\right]\\
 \left.{\cal M}^2\right|_{H_d^+a_u^0}& =\frac{1}{\sqrt{2}}\left[-\imath(\lambda^2-\frac{g^2}{2})H_u^-H_d^{0}-\frac{g'^2-g^2}{2}H_d^-\mbox{Im}(H_u^0)\right]\\
 \left.{\cal M}^2\right|_{H_d^+a_d^0}& =\frac{1}{\sqrt{2}}\left[-\imath(\lambda^2-\frac{g^2}{2})H_u^-H_u^{0}+\frac{g'^2+g^2}{2}H_d^-\mbox{Im}(H_d^0)\right]\\
 \left.{\cal M}^2\right|_{H_d^+a_s^0}& =\frac{1}{\sqrt{2}}\left[-\imath\lambda\left(A_{\lambda}e^{-\imath\varphi_1}+2\kappa e^{-\imath(\varphi_{\lambda}-\varphi_{\kappa})}S\right)H_u^-+2\lambda^2\mbox{Im}(S)H_d^-\right]
\end{align*}

\subsection{Tree-level Higgs couplings}
Having presented the spectrum and our conventions, we may now turn to the Higgs couplings.

\subsubsection{Higgs-SM fermions}
Employing the Dirac-fermion notation, the Higgs couplings to SM fermions may be cast in the following form (with the usual left- and right-handed projectors $P_{L,R}$):
\begin{equation}
 {\cal V}_f\ni\bar{f}\left[g_L^{S\bar{f}f'}P_L+g_R^{S\bar{f}f'}P_R\right]f'S\ \ \ \ ;\ \ \ \ g_R^{S\bar{f}f'}=\left(g_L^{S^*\bar{f}'f}\right)^*\label{scalfermccoup}
\end{equation}
with the (non-vanishing) values of $g_L^{S\bar{f}f'}$:
\begin{align*}
 &g_L^{S_i^0\bar{u}u}=\frac{Y_u}{\sqrt{2}}\left[X^R_{iu}+\imath X^I_{iu}\right] &g_L^{H^+\bar{u}d}=-Y_u\cos\beta\\
 &g_L^{S_i^0\bar{d}d}=\frac{Y_d}{\sqrt{2}}\left[X^R_{id}+\imath X^I_{id}\right] &g_L^{H^-\bar{d}u}=-Y_b\sin\beta\\
 &g_L^{S_i^0\bar{e}e}=\frac{Y_e}{\sqrt{2}}\left[X^R_{id}+\imath X^I_{id}\right] &g_L^{H^-\bar{e}\nu}=-Y_e\sin\beta
\end{align*}

\subsubsection{Higgs-gauge}
The situation is unchanged with respect to the CP-conserving case:
\begin{equation}
 {\cal V}_G\ni g^{SVV'}_{\mu\nu}SV^{\mu}V'^{\nu}+g^{SS'VV'}_{\mu\nu}SS'V^{\mu}V'^{\nu}+g^{SS'V}\imath\left(S\partial_{\mu}S'-S'\partial_{\mu}S\right)V^{\mu}\label{Higgsgauge}
\end{equation}
\begin{eqnarray*}
2g^{S_i^0ZZ}_{\mu\nu}=g_{\mu\nu}\frac{g'^2+g^2}{\sqrt{2}}(v_d X_{id}^R+v_u X_{iu}^R) & & 2g^{S_i^0S_j^0Z}=\imath\frac{\sqrt{g'^2+g^2}}{2}\left[X_{id}^R X_{jd}^I-X_{iu}^R X_{ju}^I-X_{id}^I X_{jd}^R+X_{iu}^I X_{ju}^R\right]\\
g^{S_i^0W^+W^-}_{\mu\nu}=g_{\mu\nu}\frac{g^2}{\sqrt{2}}(v_d X_{id}^R+v_u X_{iu}^R)\hspace{0.3cm}\null & &g^{S_i^0H_j^+W^-}=\frac{g}{\sqrt{2}}\left[(X_{id}^R-\imath X_{id}^I)X^C_{jd} -(X_{iu}^R+\imath X_{iu}^I)X^C_{ju}\right]\\
2g^{H_i^+H_j^-\gamma}=\frac{gg'}{\sqrt{g'^2+g^2}}\delta_{ij}\hspace{1.9cm}\null & &2g^{H^+_i\gamma W^-}=\frac{g^2g'}{\sqrt{2(g^2+g'^2)}}\left[v_uX^C_{iu}-v_dX^C_{id}\right]\\
2g^{H^+_iH^-_jZ}=-\frac{g'^2-g^2}{2\sqrt{g'^2+g^2}}\delta_{ij}\hspace{1.4cm}\null & & 2g^{H^+_iZ W^-}=-\frac{gg'^2}{\sqrt{2(g^2+g'^2)}}\left[v_uX^C_{iu}-v_dX^C_{id}\right] \\
2g^{H_i^+H_j^-\gamma\gamma}_{\mu\nu}=g_{\mu\nu}\frac{2g^2g'^2}{g'^2+g^2}\delta_{ij}\hspace{1.6cm}\null & &2g^{H^+_iH^-_jZ\gamma}_{\mu\nu}=g_{\mu\nu}\frac{gg'(g'^2-g^2)}{g'^2+g^2}\delta_{ij}\\
2g^{H^+_iH^-_jZZ}_{\mu\nu}=g_{\mu\nu}\frac{(g'^2-g^2)^2}{2(g'^2+g^2)}\delta_{ij}\hspace{1cm}\null & &4g^{S_i^0S_j^0ZZ}_{\mu\nu}=g_{\mu\nu}\frac{g'^2+g^2}{2}(X_{id}^R X_{jd}^R+X_{iu}^R X_{ju}^R+X_{id}^I X_{jd}^I+X_{iu}^I X_{ju}^I)\\
g^{H^+_iH^-_jW^+W^-}_{\mu\nu}=g_{\mu\nu}\frac{g^2}{2}\delta_{ij}\hspace{1.9cm}\null & &2g^{S_i^0S_j^0W^+W^-}_{\mu\nu}=g_{\mu\nu}\frac{g^2}{2}(X_{id}^R X_{jd}^R+X_{iu}^R X_{ju}^R+X_{id}^I X_{jd}^I+X_{iu}^I X_{ju}^I)\\
 & &2g^{S_{i}^0H^+_jW^-\gamma}_{\mu\nu}=\frac{g_{\mu\nu}g^2g'}{\sqrt{2(g'^2+g^2)}}\left[(X_{iu}^R+\imath X_{iu}^I)X^C_{ju}-(X_{id}^R-\imath X_{id}^I)X^C_{jd}\right]\\
 & &2g^{S_{i}^0H^+_jW^-Z}_{\mu\nu}=-\frac{g_{\mu\nu}g'^2g}{\sqrt{2(g'^2+g^2)}}\left[(X_{iu}^R+\imath X_{iu}^I)X^C_{ju}-(X_{id}^R-\imath X_{id}^I)X^C_{jd}\right]
\end{eqnarray*}

\subsubsection{Higgs-sfermions}\label{apHiSFcoup}
The Higgs-sfermion vertices read:
\begin{equation}
 {\cal V}_F\ni g^{SF^*F'}SF^*F'+g^{SS'F^*F'}SS'F^*F'
\end{equation}
with:
\begin{align*}
g^{S_i^0U_k^*U_l}= &\sqrt{2}\left[Y_u^2v_uX^R_{iu}+\frac{1}{4}\left(\frac{g'^2}{3}-g^2\right)(v_uX^R_{iu}-v_dX^R_{id})\right]X^U_{kL}\left(X^U_{lL}\right)^*\hspace{6.cm}\\
 &+\sqrt{2}\left[Y_u^2v_uX^R_{iu}-\frac{g'^2}{3}(v_uX^R_{iu}-v_dX^R_{id})\right]X^U_{kR}\left(X^U_{lR}\right)^*\\
 &+\frac{Y_u}{\sqrt{2}}\left[A_u e^{\imath\varphi_{A_u}}(X_{iu}^R+\imath X^I_{iu})-\lambda e^{-\imath\varphi_{\lambda}}\left(s(X_{id}^R-\imath X^I_{id})+v_d(X_{is}^R-\imath X^I_{is})\right)\right]X^U_{kR}\left(X^U_{lL}\right)^*\\
 &+\frac{Y_u}{\sqrt{2}}\left[A_u e^{-\imath\varphi_{A_u}}(X_{iu}^R-\imath X^I_{iu})-\lambda e^{\imath\varphi_{\lambda}}\left(s(X_{id}^R+\imath X^I_{id})+v_d(X_{is}^R+\imath X^I_{is})\right)\right]X^U_{kL}\left(X^U_{lR}\right)^*
\end{align*}
\begin{align*}
g^{S_i^0D_k^*D_l}= &\sqrt{2}\left[Y_d^2v_dX^R_{id}+\frac{1}{4}\left(\frac{g'^2}{3}+g^2\right)(v_uX^R_{iu}-v_dX^R_{id})\right]X^D_{kL}\left(X^D_{lL}\right)^*\hspace{6.cm}\\
 &+\sqrt{2}\left[Y_d^2v_dX^R_{id}+\frac{g'^2}{6}(v_uX^R_{iu}-v_dX^R_{id})\right]X^D_{kR}\left(X^D_{lR}\right)^*\\
 &+\frac{Y_d}{\sqrt{2}}\left[A_d e^{\imath\varphi_{A_d}}(X_{id}^R+\imath X^I_{id})-\lambda e^{-\imath\varphi_{\lambda}}\left(s(X_{iu}^R-\imath X^I_{iu})+v_u(X_{is}^R-\imath X^I_{is})\right)\right]X^D_{kR}\left(X^D_{lL}\right)^*\\
 &+\frac{Y_d}{\sqrt{2}}\left[A_d e^{-\imath\varphi_{A_d}}(X_{id}^R-\imath X^I_{id})-\lambda e^{\imath\varphi_{\lambda}}\left(s(X_{iu}^R+\imath X^I_{iu})+v_u(X_{is}^R+\imath X^I_{is})\right)\right]X^D_{kL}\left(X^D_{lR}\right)^*
\end{align*}
\begin{align*}
g^{S_i^0E_k^*E_l}= &\sqrt{2}\left[Y_e^2v_dX^R_{id}+\frac{-g'^2+g^2}{4}(v_uX^R_{iu}-v_dX^R_{id})\right]X^E_{kL}\left(X^E_{lL}\right)^*\hspace{6.7cm}\\
 &+\sqrt{2}\left[Y_e^2v_dX^R_{id}+\frac{g'^2}{2}(v_uX^R_{iu}-v_dX^R_{id})\right]X^E_{kR}\left(X^E_{lR}\right)^*\\
 &+\frac{Y_e}{\sqrt{2}}\left[A_e e^{\imath\varphi_{A_e}}(X_{id}^R+\imath X^I_{id})-\lambda e^{-\imath\varphi_{\lambda}}\left(s(X_{iu}^R-\imath X^I_{iu})+v_u(X_{is}^R-\imath X^I_{is})\right)\right]X^E_{kR}\left(X^E_{lL}\right)^*\\
 &+\frac{Y_e}{\sqrt{2}}\left[A_e e^{-\imath\varphi_{A_e}}(X_{id}^R-\imath X^I_{id})-\lambda e^{\imath\varphi_{\lambda}}\left(s(X_{iu}^R+\imath X^I_{iu})+v_u(X_{is}^R+\imath X^I_{is})\right)\right]X^E_{kL}\left(X^E_{lR}\right)^*
\end{align*}
\begin{displaymath}
g^{S_i^0N_L^*N_L}= -\frac{\sqrt{2}}{4}\left(g'^2+g^2\right)\left[v_uX^R_{iu}-v_dX^R_{id}\right]\hspace{10.1cm}\null
\end{displaymath}
\begin{align*}
g^{H^+_iU_k^*D_l}= &-V^{ud}_{CKM}\left\{\left[\left(Y_u^2-\frac{g^2}{2}\right)v_uX^C_{iu}+\left(Y_d^2-\frac{g^2}{2}\right)v_dX^C_{id}\right] X^U_{kL}\left(X^D_{lL}\right)^*+Y_uY_d\left[v_uX^C_{id}+v_dX^C_{iu}\right]X^U_{kR}\left(X^D_{lR}\right)^*\right.\\
 &\left.+Y_u\left[A_u e^{\imath\varphi_{A_u}}X^C_{iu}+\lambda s e^{-\imath\varphi_{\lambda}}X^C_{id}\right]X^U_{kR}\left(X^D_{lL}\right)^*
+Y_d\left[A_d e^{-\imath\varphi_{A_d}}X^C_{id}+\lambda s e^{\imath\varphi_{\lambda}}X^C_{iu}\right]X^U_{kL}\left(X^D_{lR}\right)^*\right\}\\
 &\null\hspace{-0.4cm}=\left(g^{H^-D_l^*U_k}\right)^*
\end{align*}
\begin{align*}
g^{H^+_iN_L^*E_l}= &-\left\{\left[\left(Y_e^2-\frac{g^2}{2}\right)v_dX^C_{id}-\frac{g^2}{2}v_uX^C_{iu}\right]\left(X^E_{lL}\right)^*
+Y_e\left[A_e e^{-\imath\varphi_{A_e}}X^C_{id}+\lambda s e^{\imath\varphi_{\lambda}}X^C_{iu}\right]\left(X^E_{lR}\right)^*\right\}\hspace{1.8cm}\null\\
 &\null\hspace{-0.4cm}=\left(g^{H^-E_l^*N_L}\right)^*
\end{align*}
\begin{align*}
g^{S_i^0S_j^0U_k^*U_l}= &\frac{1}{2}\left[Y_u^2\left(X^R_{iu}X^R_{ju}+X^I_{iu}X^I_{ju}\right)+\frac{1}{4}\left(\frac{g'^2}{3}-g^2\right)(X^R_{iu}X^R_{ju}+X^I_{iu}X^I_{ju}-X^R_{id}X^R_{jd}-X^I_{id}X^I_{jd})\right]X^U_{kL}\left(X^U_{lL}\right)^*\\
 &+\frac{1}{2}\left[Y_u^2\left(X^R_{iu}X^R_{ju}+X^I_{iu}X^I_{ju}\right)-\frac{g'^2}{3}(X^R_{iu}X^R_{ju}+X^I_{iu}X^I_{ju}-X^R_{id}X^R_{jd}-X^I_{id}X^I_{jd})\right]X^U_{kR}\left(X^U_{lR}\right)^*\\
 &+\frac{Y_u\lambda}{4} e^{-\imath\varphi_{\lambda}}\left[(X_{is}^R-\imath X^I_{is})(X_{jd}^R-\imath X^I_{jd})+(X_{js}^R-\imath X^I_{js})(X_{id}^R-\imath X^I_{id})\right]X^U_{kR}\left(X^U_{lL}\right)^*\\
 &+\frac{Y_u\lambda}{4} e^{\imath\varphi_{\lambda}}\left[(X_{is}^R+\imath X^I_{is})(X_{jd}^R+\imath X^I_{jd})+(X_{js}^R+\imath X^I_{js})(X_{id}^R+\imath X^I_{id})\right]X^U_{kL}\left(X^U_{lR}\right)^*
\end{align*}
\begin{align*}
g^{S_i^0S_j^0D_k^*D_l}= &\frac{1}{2}\left[Y_d^2\left(X^R_{id}X^R_{jd}+X^I_{id}X^I_{jd}\right)+\frac{1}{4}\left(\frac{g'^2}{3}+g^2\right)(X^R_{iu}X^R_{ju}+X^I_{iu}X^I_{ju}-X^R_{id}X^R_{jd}-X^I_{id}X^I_{jd})\right]X^D_{kL}\left(X^D_{lL}\right)^*\\
 &+\frac{1}{2}\left[Y_d^2\left(X^R_{id}X^R_{jd}+X^I_{id}X^I_{jd}\right)+\frac{g'^2}{6}(X^R_{iu}X^R_{ju}+X^I_{iu}X^I_{ju}-X^R_{id}X^R_{jd}-X^I_{id}X^I_{jd})\right]X^D_{kR}\left(X^D_{lR}\right)^*\\
 &+\frac{Y_d\lambda}{4} e^{-\imath\varphi_{\lambda}}\left[(X_{is}^R-\imath X^I_{is})(X_{ju}^R-\imath X^I_{ju})+(X_{js}^R-\imath X^I_{js})(X_{iu}^R-\imath X^I_{iu})\right]X^D_{kR}\left(X^D_{lL}\right)^*\\
 &+\frac{Y_d\lambda}{4} e^{\imath\varphi_{\lambda}}\left[(X_{is}^R+\imath X^I_{is})(X_{ju}^R+\imath X^I_{ju})+(X_{js}^R+\imath X^I_{js})(X_{iu}^R+\imath X^I_{iu})\right]X^D_{kL}\left(X^D_{lR}\right)^*
\end{align*}
\begin{align*}
g^{S_i^0S_j^0E_k^*E_l}= &\frac{1}{2}\left[Y_e^2\left(X^R_{id}X^R_{jd}+X^I_{id}X^I_{jd}\right)+\frac{-g'^2+g^2}{4}(X^R_{iu}X^R_{ju}+X^I_{iu}X^I_{ju}-X^R_{id}X^R_{jd}-X^I_{id}X^I_{jd})\right]X^E_{kL}\left(X^E_{lL}\right)^*\\
 &+\frac{1}{2}\left[Y_e^2\left(X^R_{id}X^R_{jd}+X^I_{id}X^I_{jd}\right)+\frac{g'^2}{2}(X^R_{iu}X^R_{ju}+X^I_{iu}X^I_{ju}-X^R_{id}X^R_{jd}-X^I_{id}X^I_{jd})\right]X^E_{kR}\left(X^E_{lR}\right)^*\\
 &+\frac{Y_e\lambda}{4} e^{-\imath\varphi_{\lambda}}\left[(X_{is}^R-\imath X^I_{is})(X_{ju}^R-\imath X^I_{ju})+(X_{js}^R-\imath X^I_{js})(X_{iu}^R-\imath X^I_{iu})\right]X^E_{kR}\left(X^E_{lL}\right)^*\\
 &+\frac{Y_e\lambda}{4} e^{\imath\varphi_{\lambda}}\left[(X_{is}^R+\imath X^I_{is})(X_{ju}^R+\imath X^I_{ju})+(X_{js}^R+\imath X^I_{js})(X_{iu}^R+\imath X^I_{iu})\right]X^E_{kL}\left(X^E_{lR}\right)^*
\end{align*}
\begin{displaymath}
g^{S_i^0S_j^0N_L^*N_L}= -\frac{1}{8}\left(g'^2+g^2\right)\left[X^R_{iu}X^R_{ju}-X^R_{id}X^R_{jd}\right]\hspace{8.5cm}\null
\end{displaymath}
\begin{align*}
g^{S_i^0H^+_jU_k^*D_l}= &-\frac{V^{ud}_{CKM}}{\sqrt{2}}\left\{\left[(Y_u^2-\frac{g^2}{2})X^C_{ju}(X^R_{iu}-\imath X^I_{iu})+(Y_d^2-\frac{g^2}{2})X^C_{jd}(X^R_{id}+\imath X^I_{id})\right]X^U_{kL}\left(X^D_{lL}\right)^*\right.\hspace{1.1cm}\null\\
 &+Y_uY_d\left[X^C_{ju}(X^R_{id}-\imath X^I_{id})+X^C_{jd}(X^R_{iu}+\imath X^I_{iu})\right]\,X^U_{kR}\left(X^D_{lR}\right)^*\\
 &\left.+Y_u\lambda e^{-\imath\varphi_{\lambda}}X^C_{jd}(X^R_{is}-\imath X^I_{is})X^U_{kR}\left(X^D_{lL}\right)^*
+Y_d\lambda  e^{\imath\varphi_{\lambda}}X^C_{ju}(X^R_{is}+\imath X^I_{is})X^U_{kL}\left(X^D_{lR}\right)^*\right\}\\
 &\null\hspace{-0.4cm}=\left(g^{H^-S_i^0D_l^*U_k}\right)^*
\end{align*}
\begin{align*}
g^{S_i^0H^+_jN_L^*E_l}= &-\frac{1}{\sqrt{2}}\left\{\left[(Y_e^2-\frac{g^2}{2})X^C_{jd} (X^R_{id}+\imath X^I_{id})-\frac{g^2}{2}X^C_{ju} (X^R_{iu}-\imath X^I_{iu})\right] \left(X^E_{lL}\right)^*\right.\hspace{3.3cm}\null\\
 &\null\hspace{5cm}\left.+Y_e\lambda e^{\imath\varphi_{\lambda}}X^C_{ju}(X^R_{is}+\imath X^I_{is})\left(X^E_{lR}\right)^*\right\}=\left(g^{H^-S_i^0E_l^*N_L}\right)^*
\end{align*}
\begin{align*}
g^{H^+_iH^-_jU_k^*U_l}= &\left[Y_d^2|V_{CKM}^{ud}|^2X^C_{id}X^C_{jd}+\frac{1}{4}\left(\frac{g'^2}{3}+g^2\right)(X^C_{iu}X^C_{ju}-X^C_{id}X^C_{jd})\right]X^U_{kL}\left(X^U_{lL}\right)^*\hspace{2.8cm}\null\\
 &\hspace{5cm}+\left[Y_u^2X^C_{iu}X^C_{ju}-\frac{g'^2}{3}(X^C_{iu}X^C_{ju}-X^C_{id}X^C_{jd})\right]X^U_{kR}\left(X^U_{lR}\right)^*
\end{align*}
\begin{align*}
g^{H^+_iH^-_jD_k^*D_l}= &\left[Y_u^2|V_{CKM}^{ud}|^2X^C_{iu}X^C_{ju}+\frac{1}{4}\left(\frac{g'^2}{3}-g^2\right)(X^C_{iu}X^C_{ju}-X^C_{id}X^C_{jd})\right]X^D_{kL}\left(X^D_{lL}\right)^*\hspace{2.6cm}\null\\
 &\hspace{5cm}+\left[Y_d^2X^C_{id}X^C_{jd}+\frac{g'^2}{6}(X^C_{iu}X^C_{ju}-X^C_{id}X^C_{jd})\right]X^D_{kR}\left(X^D_{lR}\right)^*
\end{align*}
\begin{align*}
g^{H^+_iH^-_jE_k^*E_l}= &-\frac{g'^2+g^2}{4}\left[X^C_{iu}X^C_{ju}-X^C_{id}X^C_{jd}\right]X^E_{kL}\left(X^E_{lL}\right)^*\hspace{6.9cm}\null\\
 &\hspace{5cm}+\left[Y_e^2X^C_{id}X^C_{jd}+\frac{g'^2}{2}(X^C_{iu}X^C_{ju}-X^C_{id}X^C_{jd})\right]X^E_{kR}\left(X^E_{lR}\right)^*
\end{align*}
\begin{displaymath}
g^{H^+_iH^-_jN_L^*N_L}= Y_e^2X^C_{id}X^C_{jd}+\frac{1}{4}\left(-g'^2+g^2\right)\left[X^C_{iu}X^C_{ju}-X^C_{id}X^C_{jd}\right]\hspace{6.cm}\null
\end{displaymath}

\subsubsection{Higgs-charginos/neutralinos}\label{aphichacoup}
As for the Higgs-fermion couplings:
\begin{align*}
g_L^{S_i^0\overline{\chi^+}_k\chi^+_l}= &\frac{1}{\sqrt{2}}\left[g(X^R_{iu}-\imath X^I_{iu})U^*_{k1}V^*_{l2}+g(X^R_{id}-\imath X^I_{id})U^*_{k2}V^*_{l1}+\lambda e^{\imath\varphi_{\lambda}}(X^R_{is}+\imath X^I_{is})U^*_{k2}V^*_{l2}\right]\\
g_L^{S_i^0\overline{\chi^0}_k\chi^0_l}= &\frac{1}{\sqrt{2}}\left\{\frac{g'}{\sqrt{2}}\left[(X^R_{iu}-\imath X^I_{iu})\left(N^*_{kb}N^*_{lu}+N^*_{lb}N^*_{ku}\right)-(X^R_{id}-\imath X^I_{id})\left(N^*_{kb}N^*_{ld}+N^*_{lb}N^*_{kd}\right)\right]\right.\\
 &\null\hspace{0.8cm}-\frac{g}{\sqrt{2}}\left[(X^R_{iu}-\imath X^I_{iu})\left(N^*_{kw}N^*_{lu}+N^*_{lw}N^*_{ku}\right)-(X^R_{id}-\imath X^I_{id})\left(N^*_{kw}N^*_{ld}+N^*_{lw}N^*_{kd}\right)\right]\\
 &\null\hspace{0.8cm}+\lambda e^{\imath \varphi_{\lambda}}\left[(X^R_{iu}+\imath X^I_{iu})\left(N^*_{ks}N^*_{ld}+N^*_{ls}N^*_{kd}\right)+(X^R_{id}+\imath X^I_{id})\left(N^*_{ks}N^*_{lu}+N^*_{ls}N^*_{ku}\right)\right.\\
 &\null\hspace{8cm}\left.+(X^R_{is}+\imath X^I_{is})\left(N^*_{ku}N^*_{ld}+N^*_{lu}N^*_{kd}\right)\right]\\
 &\null\hspace{0.2cm}\left.\phantom{\frac{g'}{\sqrt{2}}}+\kappa e^{\imath \varphi_{\kappa}}(X^R_{is}+\imath X^I_{is})\left[N^*_{ks}N^*_{ls}+N^*_{ls}N^*_{ks}\right]\right\}\\
g_L^{H^+_i\overline{\chi^0}_k\chi^-_l}= &-\frac{g'}{\sqrt{2}}X^C_{id} N^*_{kb}U^*_{l2}+gX^C_{id}\left(N^*_{kd}U^*_{l1}-\frac{1}{\sqrt{2}}N^*_{kw}U^*_{l2}\right)+\lambda e^{\imath\varphi_{\lambda}}X^C_{iu} N^*_{ks}U^*_{l2}\\
g_L^{H^-_i\overline{\chi^0}_k\chi^+_l}= &\frac{g'}{\sqrt{2}}X^C_{iu} N^*_{kb}V^*_{l2}+gX^C_{iu}\left(N^*_{ku}V^*_{l1}+\frac{1}{\sqrt{2}}N^*_{kw}V^*_{l2}\right)+\lambda e^{\imath\varphi_{\lambda}}X^C_{id} N^*_{ks}V^*_{l2}
\end{align*}

\subsubsection{Higgs-to-Higgs couplings}\label{apHi2Hicoup}
From the tree-level potential of Eq.\ref{tlHiggspot2}, one may derive the trilinear and quartic Higgs couplings:
\begin{align*}
 g^{S_i^0S_j^0S_k^0}= &\frac{1}{\sqrt{2}}\left\{\frac{g'^2+g^2}{4}\left[v_u\left(\Pi^{S\,uuu}_{ijk}+\Pi^{A\,uuu}_{ijk}-\Pi^{S\,udd}_{ijk}-\Pi^{A\,udd}_{ijk}\right)+v_d\left(\Pi^{S\,ddd}_{ijk}+\Pi^{A\,ddd}_{ijk}-\Pi^{S\,duu}_{ijk}-\Pi^{A\,duu}_{ijk}\right)\right]\right.\\
 &\null\hspace{0.5cm}+\lambda^2\left[s\left(\Pi^{S\,suu}_{ijk}+\Pi^{A\,suu}_{ijk}+\Pi^{S\,sdd}_{ijk}+\Pi^{A\,sdd}_{ijk}\right)+v_u\left(\Pi^{S\,udd}_{ijk}+\Pi^{A\,udd}_{ijk}+\Pi^{S\,uss}_{ijk}+\Pi^{A\,uss}_{ijk}\right)\right.\\
 &\null\hspace{1cm}\left.+v_d\left(\Pi^{S\,duu}_{ijk}+\Pi^{A\,duu}_{ijk}+\Pi^{S\,dss}_{ijk}+\Pi^{A\,dss}_{ijk}\right)\right]-\lambda A_{\lambda}\cos\varphi_1\left[\Pi^{S\,sud}_{ijk}-\Pi^{A\,sud}_{ijk}-\Pi^{A\,uds}_{ijk}-\Pi^{A\,dus}_{ijk}\right]\\
 &\null\hspace{0.5cm}-\lambda\kappa\cos(\varphi_{\lambda}-\varphi_{\kappa})\left[2s\left(\Pi^{S\,uds}_{ijk}+\Pi^{A\,uds}_{ijk}+\Pi^{A\,dus}_{ijk}-\Pi^{A\,sud}_{ijk}\right)+v_d\left(\Pi^{S\,uss}_{ijk}-\Pi^{A\,uss}_{ijk}+2\Pi^{A\,sus}_{ijk}\right)\right.\\
 &\null\hspace{1cm}\left.+v_u\left(\Pi^{S\,dss}_{ijk}-\Pi^{A\,dss}_{ijk}+2\Pi^{A\,sds}_{ijk}\right)\right]+\frac{\kappa}{3}A_{\kappa}\cos\varphi_2\left[\Pi^{S\,sss}_{ijk}-3\Pi^{A\,sss}_{ijk}\right]+2\kappa^2s\left[\Pi^{S\,sss}_{ijk}+\Pi^{A\,sss}_{ijk}\right]\\
 &\null\hspace{0.5cm}+\lambda\kappa\sin(\varphi_{\lambda}-\varphi_{\kappa})\left[s\left(\Pi^{P\,uds}_{ijk}+\Pi^{P\,dus}_{ijk}-3\Pi^{P\,sud}_{ijk}+3\Pi^{I\,sud}_{ijk}\right)+v_d\left(\Pi^{P\,uss}_{ijk}-2\Pi^{P\,sus}_{ijk}-\Pi^{I\,uss}_{ijk}\right)\right.\\
 &\null\hspace{4cm}\left.+v_u\left(\Pi^{P\,dss}_{ijk}-2\Pi^{P\,sds}_{ijk}-\Pi^{I\,dss}_{ijk}\right)+\frac{v_uv_d}{s}\left(3\Pi^{P\,sss}_{ijk}-\Pi^{I\,sss}_{ijk}\right)\right]\bigg\}
\end{align*}
where:
\begin{align*}
 \left(\Pi^S\right)^{a,b,c}_{i,j,k} =&X^{R}_{ia}X^{R}_{jb}X^{R}_{kc}+X^{R}_{ib}X^{R}_{jc}X^{R}_{ka}+X^{R}_{ic}X^{R}_{ja}X^{R}_{kb}+X^{R}_{ia}X^{R}_{jc}X^{R}_{kb}+X^{R}_{ic}X^{R}_{jb}X^{R}_{ka}+X^{R}_{ib}X^{R}_{ja}X^{R}_{kc}\\
 \left(\Pi^A\right)^{a,b,c}_{i,j,k} =&X^{R}_{ia}\left(X^{I}_{jb}X^{I}_{kc}+X^{I}_{jc}X^{I}_{kb}\right)+X^{R}_{ja}\left(X^{I}_{ib}X^{I}_{kc}+X^{I}_{ic}X^{I}_{kb}\right)+X^{R}_{ka}\left(X^{I}_{ib}X^{I}_{jc}+X^{I}_{ic}X^{I}_{jb}\right)\\
 \left(\Pi^P\right)^{a,b,c}_{i,j,k} =&X^{I}_{ia}\left(X^{R}_{jb}X^{R}_{kc}+X^{R}_{jc}X^{R}_{kb}\right)+X^{I}_{ja}\left(X^{R}_{ib}X^{R}_{kc}+X^{R}_{ic}X^{R}_{kb}\right)+X^{I}_{ka}\left(X^{R}_{ib}X^{R}_{jc}+X^{R}_{ic}X^{R}_{jb}\right)\\
 \left(\Pi^I\right)^{a,b,c}_{i,j,k} =&X^{I}_{ia}X^{I}_{jb}X^{I}_{kc}+X^{I}_{ib}X^{I}_{jc}X^{I}_{ka}+X^{I}_{ic}X^{I}_{ja}X^{I}_{kb}+X^{I}_{ia}X^{I}_{jc}X^{I}_{kb}+X^{I}_{ic}X^{I}_{jb}X^{I}_{ka}+X^{I}_{ib}X^{I}_{ja}X^{I}_{kc}
\end{align*}
\begin{align*}
 g^{S_i^0H_j^+H_k^-}= &\frac{1}{\sqrt{2}}\left\{\left[2\lambda^2sX^{R}_{is}+\frac{g'^2+g^2}{2}v_uX^{R}_{iu}+\frac{g^2-g'^2}{2}v_dX^{R}_{id}\right]X^{C}_{ju}X^{C}_{ku}\right.\\
 &\null\hspace{0.5cm}+\left[2\lambda^2sX^{R}_{is}+\frac{g^2-g'^2}{2}v_uX^{R}_{iu}+\frac{g'^2+g^2}{2}v_dX^{R}_{id}\right]X^{C}_{jd}X^{C}_{kd}\\
 &\null\hspace{0.5cm}+\left[\lambda\left(A_{\lambda}\cos\varphi_1+2\kappa s\cos(\varphi_{\lambda}-\varphi_{\kappa})\right)X^R_{is}+3\lambda\kappa s \sin(\varphi_{\lambda}-\varphi_{\kappa})X^I_{is}-\left(\lambda^2-\frac{g^2}{2}\right)\left(v_uX^R_{id}+v_dX^R_{iu}\right)\right]\\
 &\null\hspace{11cm}\times\left(X^{C}_{ju}X^{C}_{kd}+X^{C}_{jd}X^{C}_{ku}\right)\\
 &\null\hspace{0.5cm}+\imath\left[\lambda\kappa s \sin(\varphi_{\lambda}-\varphi_{\kappa})X^R_{is}+\lambda\left(A_{\lambda}\cos\varphi_1-2\kappa s\cos(\varphi_{\lambda}-\varphi_{\kappa})\right)X^I_{is}+\left(\lambda^2-\frac{g^2}{2}\right)\left(v_uX^I_{id}+v_dX^I_{iu}\right)\right]\\
 &\null\hspace{11cm}\times\left(X^{C}_{ju}X^{C}_{kd}-X^{C}_{jd}X^{C}_{ku}\right)\bigg\}
\end{align*}
\begin{align*}
 g^{S_i^0S_j^0S_k^0S_l^0}= &\frac{1}{4}\left\{\frac{g'^2+g^2}{8}\left[\Pi^{S\,uuuu}_{ijkl}+\Pi^{S\,dddd}_{ijkl}-2\Pi^{S\,uudd}_{ijkl}+\Pi^{P\,uuuu}_{ijkl}+\Pi^{P\,dddd}_{ijkl}-2\Pi^{P\,uudd}_{ijkl}\right.\right.\\
 &\null\hspace{4cm}\left.+2\Pi^{S\,uu\,P\,uu}_{ijkl}+2\Pi^{S\,dd\,P\,dd}_{ijkl}-2\Pi^{S\,uu\,P\,dd}_{ijkl}-2\Pi^{S\,dd\,P\,uu}_{ijkl}\right]\\
 &\null\hspace{0.5cm}+\lambda^2\left[\Pi^{S\,uudd}_{ijkl}+\Pi^{S\,uuss}_{ijkl}+\Pi^{S\,ddss}_{ijkl}+\Pi^{P\,uudd}_{ijkl}+\Pi^{P\,uuss}_{ijkl}+\Pi^{P\,ddss}_{ijkl}\right.\\
 &\null\hspace{4cm}\left.+\Pi^{S\,uu\,P\,dd}_{ijkl}+\Pi^{S\,dd\,P\,uu}_{ijkl}+\Pi^{S\,uu\,P\,ss}_{ijkl}+\Pi^{S\,dd\,P\,ss}_{ijkl}+\Pi^{S\,ss\,P\,uu}_{ijkl}+\Pi^{S\,ss\,P\,dd}_{ijkl}\right]\\
 &\null\hspace{0.5cm}-2\lambda\kappa\cos(\varphi_{\lambda}-\varphi_{\kappa})\left[\Pi^{S\,ssud}_{ijkl}+\Pi^{P\,ssud}_{ijkl}-\Pi^{S\,ss\,P\,ud}_{ijkl}-\Pi^{S\,ud\,P\,ss}_{ijkl}+2\Pi^{S\,su\,P\,sd}_{ijkl}+2\Pi^{S\,sd\,P\,su}_{ijkl}\right]\\
 &\null\hspace{0.5cm}+2\lambda\kappa\sin(\varphi_{\lambda}-\varphi_{\kappa})\left[\Pi^{S\,ssu\,P\,d}_{ijkl}+\Pi^{S\,ssd\,P\,u}_{ijkl}-\Pi^{S\,u\,P\,ssd}_{ijkl}-\Pi^{S\,d\,P\,ssu}_{ijkl}-2\Pi^{S\,uds\,P\,s}_{ijkl}+2\Pi^{S\,s\,P\,sud}_{ijkl}\right]\\
 &\null\hspace{0.5cm}+\kappa^2\left[\Pi^{S\,ssss}_{ijkl}+\Pi^{P\,ssss}_{ijkl}+2\Pi^{S\,ss\,P\,ss}_{ijkl}\right]\bigg\}
\end{align*}
where:
\begin{align*}
 &\Pi^{S\,abcd}_{ijkl}=\sum_{\sigma\in S_4}X^{R}_{\sigma(i)a}X^{R}_{\sigma(j)b}X^{R}_{\sigma(k)c}X^{R}_{\sigma(l)d}\ \ \ \ \ ;\ \ \ \Pi^{P\,abcd}_{ijkl}=\sum_{\sigma\in S_4}X^{I}_{\sigma(i)a}X^{I}_{\sigma(j)b}X^{I}_{\sigma(k)c}X^{I}_{\sigma(l)d}\\
 &\Pi^{S\,ab\,P\,cd}_{ijkl}=\sum_{\sigma\in S_4}X^{R}_{\sigma(i)a}X^{R}_{\sigma(j)b}X^{I}_{\sigma(k)c}X^{I}_{\sigma(l)d}\ \ \ ;\ \ \ \Pi^{S\,a\,P\,bcd}_{ijkl}=\sum_{\sigma\in S_4}X^{R}_{\sigma(i)a}X^{I}_{\sigma(j)b}X^{I}_{\sigma(k)c}X^{I}_{\sigma(l)d}\\
 &\Pi^{S\,abc\,P\,d}_{ijkl}=\sum_{\sigma\in S_4}X^{R}_{\sigma(i)a}X^{R}_{\sigma(j)b}X^{R}_{\sigma(k)c}X^{I}_{\sigma(l)d}
\end{align*}
\begin{align*}
 g^{S_i^0S_j^0H_k^+H_l^-}= &\frac{g'^2}{4}\left[X^{R}_{iu}X^{R}_{ju}+X^{I}_{iu}X^{I}_{ju}-X^{R}_{id}X^{R}_{jd}-X^{I}_{id}X^{I}_{jd}\right]\left(X^{C}_{ku}X^{C}_{lu}-X^{C}_{kd}X^{C}_{ld}\right)\\
 &+\frac{g^2}{4}\left[X^{R}_{iu}X^{R}_{ju}+X^{I}_{iu}X^{I}_{ju}+X^{R}_{id}X^{R}_{jd}+X^{I}_{id}X^{I}_{jd}\right]\left(X^{C}_{ku}X^{C}_{lu}+X^{C}_{kd}X^{C}_{ld}\right)\\
 &+\lambda^2\left[X^{R}_{is}X^{R}_{js}+X^{I}_{is}X^{I}_{js}\right]\left(X^{C}_{ku}X^{C}_{lu}+X^{C}_{kd}X^{C}_{ld}\right)\\
 &-\frac{1}{2}\left(\lambda^2-\frac{g^2}{2}\right)\left[X^{R}_{iu}X^{R}_{jd}+X^{R}_{id}X^{R}_{ju}-X^{I}_{iu}X^{I}_{jd}-X^{I}_{id}X^{I}_{ju}\right]\left(X^{C}_{ku}X^{C}_{ld}+X^{C}_{kd}X^{C}_{lu}\right)\\
 &+\frac{\imath}{2}\left(\lambda^2-\frac{g^2}{2}\right)\left[X^{R}_{iu}X^{I}_{jd}+X^{R}_{id}X^{I}_{ju}+X^{I}_{iu}X^{R}_{jd}+X^{I}_{id}X^{R}_{ju}\right]\left(X^{C}_{ku}X^{C}_{ld}-X^{C}_{kd}X^{C}_{lu}\right)\\
 &+\lambda\kappa\left[\cos(\varphi_{\lambda}-\varphi_{\kappa})\left(X^{R}_{is}X^{R}_{js}-X^{I}_{is}X^{I}_{js}\right)+\sin(\varphi_{\lambda}-\varphi_{\kappa})\left(X^{R}_{is}X^{I}_{js}+X^{I}_{is}X^{R}_{js}\right)\right]\left(X^{C}_{ku}X^{C}_{ld}+X^{C}_{kd}X^{C}_{lu}\right)\\
 &+\imath\lambda\kappa\left[\sin(\varphi_{\lambda}-\varphi_{\kappa})\left(X^{R}_{is}X^{R}_{js}-X^{I}_{is}X^{I}_{js}\right)-\cos(\varphi_{\lambda}-\varphi_{\kappa})\left(X^{R}_{is}X^{I}_{js}+X^{I}_{is}X^{R}_{js}\right)\right]\left(X^{C}_{ku}X^{C}_{ld}-X^{C}_{kd}X^{C}_{lu}\right)
\end{align*}

\subsection{Other couplings}

\subsubsection{Chargino - Sfermion - SM fermion}\label{apcoupchasff}
For each fermion / sfermion generation (with the convention of Eq.\ref{scalfermccoup}):
\begin{align*}
& g_L^{\tilde{U}^*_j\chi^+_id}=Y_uX^{\tilde{U}}_{jR}V^*_{iu}-gX^{\tilde{U}}_{jL}V^*_{iw} & & g_R^{\tilde{U}^*_j\chi^+_id}=Y_dX^{\tilde{U}}_{jL}U_{id}\\
& g_L^{\tilde{D}^*_j\chi^-_iu}=Y_dX^{\tilde{D}}_{jR}U^*_{id}-gX^{\tilde{D}}_{jL}U^*_{iw} & & g_R^{\tilde{D}^*_j\chi^-_iu}=Y_uX^{\tilde{D}}_{jL}V_{iu}\\
& g_L^{\tilde{N}^*\chi^+_ie}=-gV^*_{iw} & & g_R^{\tilde{N}^*\chi^+_ie}=Y_eU_{id}\\
& g_L^{\tilde{E}^*_j\chi^-_i\nu}=Y_eX^{\tilde{E}}_{jR}U^*_{id}-gX^{\tilde{E}}_{jL}U^*_{iw} & & g_R^{\tilde{E}^*_j\chi^-_i\nu}=0
\end{align*}

\subsubsection{Neutralino - Sfermion - SM fermion}\label{apcoupneusff}
For each fermion / sfermion generation:
\begin{align*}
& g_L^{\tilde{U}^*_j\chi^0_iu}=-Y_uX^{\tilde{U}}_{jR}N^*_{iu}-\frac{1}{\sqrt{2}}\left(\frac{g'}{3}N^*_{ib}+gN^*_{iw}\right)X^{\tilde{U}}_{jL} & & g_R^{\tilde{U}^*_j\chi^0_iu}=-Y_uX^{\tilde{U}}_{jL}N_{iu}+2\sqrt{2}g_1X^{\tilde{U}}_{jR}N_{ib}\\
& g_L^{\tilde{D}^*_j\chi^0_id}=-Y_dX^{\tilde{D}}_{jR}N^*_{id}-\frac{1}{\sqrt{2}}\left(\frac{g'}{3}N^*_{ib}-gN^*_{iw}\right)X^{\tilde{D}}_{jL} & & g_R^{\tilde{D}^*_j\chi^0_id}=-Y_dX^{\tilde{D}}_{jL}N_{id}-\frac{\sqrt{2}}{3}g'X^{\tilde{D}}_{jR}N_{ib}\\
& g_L^{\tilde{N}^*\chi^0_i\nu}=\frac{1}{\sqrt{2}}\left(g'N^*_{ib}-gN^*_{iw}\right) & & g_R^{\tilde{N}^*\chi^0_i\nu}=0\\
& g_L^{\tilde{E}^*_j\chi^0_ie}=-Y_eX^{\tilde{E}}_{jR}N^*_{id}+\frac{1}{\sqrt{2}}\left(g'N^*_{ib}+gN^*_{iw}\right)X^{\tilde{E}}_{jL} & & g_R^{\tilde{E}^*_j\chi^0_ie}=-Y_eX^{\tilde{E}}_{jL}N_{id}-\sqrt{2}g'X^{\tilde{E}}_{jR}N_{ib}
\end{align*}

\subsubsection{Chargino and Neutralino gauge couplings}
Using the notation:
\begin{equation}
  {\cal V}_f\ni V_{\mu}\bar{f}\gamma^{\mu}\left[g_L^{V\bar{f}f'}P_L+g_R^{V\bar{f}f'}P_R\right]f'\ \ \ ;\ \ \ g_L^{V^*\bar{f}'f}=\left(g_L^{V\bar{f}f'}\right)^*\ \ ,\ \ g_R^{V^*\bar{f}'f}=\left(g_R^{V\bar{f}f'}\right)^*
\end{equation}
the chargino and neutralino gauge couplings may be written:
\begin{align*}
 & g_L^{\gamma\bar{\chi}_i^-\chi_j^+}=\frac{gg'}{\sqrt{g^2+g'^2}} & & g_R^{\gamma\bar{\chi}_i^-\chi_j^+}=\frac{gg'}{\sqrt{g^2+g'^2}}\\
 & g_L^{Z\bar{\chi}_i^-\chi_j^+}=\sqrt{g^2+g'^2}\left(V_{iw}V_{jw}^*+\frac{1}{2}V_{iu}V_{ju}^*-s_W^2\delta_{ij}\right) & & g_R^{Z\bar{\chi}_i^-\chi_j^+}=\sqrt{g^2+g'^2}\left(U_{iw}U_{jw}^*+\frac{1}{2}U_{id}U_{jd}^*-s_W^2\delta_{ij}\right)\\
 & g_L^{Z\bar{\chi}_i^0\chi_j^0}=\frac{\sqrt{g^2+g'^2}}{2}\left(N_{id}N_{id}^*-N_{iu}N_{iu}^*\right) & & g_R^{Z\bar{\chi}_i^0\chi_j^0}=\frac{\sqrt{g^2+g'^2}}{2}\left(N_{iu}N_{iu}^*-N_{id}N_{id}^*\right)\\
 & g_L^{W^-\bar{\chi}_i^0\chi_j^+}=g\left(\frac{1}{\sqrt{2}}N_{iu}V_{iu}^*-N_{iw}V_{iw}^*\right) & & g_R^{W^-\bar{\chi}_i^0\chi_j^+}=-g\left(\frac{1}{\sqrt{2}}U_{id}N_{id}^*+U_{iw}N_{iw}^*\right)
\end{align*}


\section{Radiative corrections to the supersymmetric spectrum}

\subsection{Electroweak gauginos and higgsinos}\label{aploopchaneu}

We follow the approach of \cite{Pierce:1996zz} and consider the loops involving sfermions / fermions, higgsinos or gauginos / Higgs and gauge bosons 
in the self energies of the gauginos and higgsinos, under the assumption that the gauge eigenstates are approximately mass states. Taking the complex 
phases $\phi_{M_i}$ and $\varphi_{\lambda,\kappa}$ into account, we find the following corrections to the gaugino and higgsino masses:
\begin{align*}
 \left(\frac{\Delta M_1}{M_1}\right)=& -\frac{g'^2}{16\pi^2}\Big\{11B_1(M_1,0,M_{Q_1})+9B_1(M_1,0,M_{L_1})+B_1(M_1,\mu,M_A)+B_1(M_1,\mu,M_Z)\\
 &\null\hspace{2cm}+\frac{\mu}{M_1}\sin{2\beta}\cos(\phi_{M_1}+\varphi_{\lambda})\left[B_0(M_1,\mu,M_A)-B_0(M_1,\mu,M_Z)\right]\Big\}\\
  \left(\frac{\Delta M_2}{M_2}\right)=&-\frac{g^2}{16\pi^2}\Big\{9B_1(M_2,0,M_{Q_1})+3B_1(M_2,0,M_{L_1})+B_1(M_2,\mu,M_A)+B_1(M_2,\mu,M_Z)\\
 &\null\hspace{2cm}+\frac{\mu}{M_2}\sin{2\beta}\cos(\phi_{M_2}+\varphi_{\lambda})\left[B_0(M_2,\mu,M_A)-B_0(M_2,\mu,M_Z)\right]\\
 &\null\hspace{2cm}+4B_1(M_2,M_2,M_W)-8B_0(M_2,M_2,M_W)\Big\}\\
  \left(\frac{\Delta \mu}{\mu}\right)=&-\frac{3}{32\pi^2}\Big\{(Y_t^2+Y_b^2)B_1(\mu,0,M_{Q_3})+Y_t^2B_1(\mu,0,M_{U_3})+Y_b^2B_1(\mu,\mu,M_{D_3})\Big\}\\
 &-\frac{g'^2}{64\pi^2}\Big\{2B_1(\mu,\mu,M_Z)-4B_0(\mu,\mu,M_Z)+B_1(\mu,M_1,M_A)+B_1(\mu,M_1,M_Z)\\
 &\null\hspace{2cm}+\frac{M_1}{\mu}\sin{2\beta}\cos(\phi_{M_1}+\varphi_{\lambda})\left[B_0(\mu,M_1,M_A)-B_0(\mu,M_1,M_Z)\right]\Big\}\\
 &-\frac{3g^2}{64\pi^2}\Big\{2B_1(\mu,\mu,M_Z)-4B_0(\mu,\mu,M_Z)+B_1(\mu,M_2,M_A)+B_1(\mu,M_2,M_Z)\\
 &\null\hspace{2cm}+\frac{M_2}{\mu}\sin{2\beta}\cos(\phi_{M_2}+\varphi_{\lambda})\left[B_0(\mu,M_2,M_A)-B_0(\mu,M_2,M_Z)\right]\Big\}\\
 &-\frac{\lambda^2}{32\pi^2}\Big\{B_1(\mu,m_{\tilde{s}},M_A)+B_1(\mu,m_{\tilde{s}},M_Z)\\
 &\null\hspace{2cm}-\frac{m_{\tilde{s}}}{\mu}\sin{2\beta}\cos(\varphi_{\lambda}-\varphi_{\kappa})\left[B_0(\mu,m_{\tilde{s}},M_A)-B_0(\mu,m_{\tilde{s}},M_Z)\right]\Big\}\\
 \left(\frac{\Delta m_{\tilde{s}}}{m_{\tilde{s}}}\right)=& -\frac{\lambda^2}{8\pi^2}\Big\{B_1(m_{\tilde{s}},\mu,M_A)+B_1(m_{\tilde{s}},\mu,M_Z)\Big\}\\
 & -\frac{\kappa^2}{8\pi^2}\Big\{B_1(m_{\tilde{s}},m_{\tilde{s}},m_{h_S^0})+B_1(m_{\tilde{s}},m_{\tilde{s}},m_{a_S^0})-B_0(m_{\tilde{s}},m_{\tilde{s}},m_{h_S^0})+B_0(M_1,m_{\tilde{s}},m_{a_S^0})\Big\}
\end{align*}
We took over the 
notations of \cite{Pierce:1996zz} to designate the approximate masses of the particles in the loops; note that $\mu\equiv\lambda s$ stands for the 
doublet higgsino mass, $m_{\tilde{s}}\equiv2\kappa s$ for the singlino one and $m_{h_S^0,a_S^0}$ for the singlet (pseudo)scalar masses.

\subsection{Sfermions}\label{aploopsferm}

$O(\alpha_S)$ corrections to the squark masses are generated by gluon / squark and gluino / quark loops. Another source are squark self-couplings, as 
these receive a contribution from the $SU(3)_c$ D-term. We use the following expressions to correct the squark squared masses $m_{\tilde{Q}}^2$:
\begin{align*}
 \Delta m^2_{\tilde{Q}}= &-\frac{\alpha_S}{3\pi}\Bigg\{2m^2_{\tilde{Q}}\left[B_1(m_{\tilde{Q}},0,m_{\tilde{Q}})-2B_0(m_{\tilde{Q}},0,m_{\tilde{Q}})\right]-\sum_{j=1}^2|X^{\tilde{Q}}_{\tilde{Q}L}X^{\tilde{Q}\,*}_{jL}-X^{\tilde{Q}}_{\tilde{Q}R}X^{\tilde{Q}\,*}_{jR}|^2A_0(m_{\tilde{Q}_j})\\
 &\null\hspace{0.9cm}+4\left[m^2_{\tilde{Q}}B_1(m_{\tilde{Q}},m_{\tilde{g}},m_q)-A_0(m_q)-m_{\tilde{g}}^2\left(1+2\frac{m_q}{m_{\tilde{g}}}\mbox{Re}(e^{-\imath\phi_{M_3}}X^{\tilde{Q}}_{\tilde{Q}L}X^{\tilde{Q}\,*}_{\tilde{Q}R})\right)B_0(m_{\tilde{Q}},m_{\tilde{g}},m_q)\right]\Bigg\}
\end{align*}
We recover the results of \cite{Kraml:1999qd} in the CP-conserving limit.

\subsection{Gluino}\label{aploopgluino}

We follow \cite{Pierce:1996zz} to include the $O(\alpha_S)$ corrections to the gluino mass: these involve the gluon /gluino and the quark / squark 
loops. The latter depend on squark and gluino phases via the quark / squark / gluino couplings. We obtain:
\begin{align*}
\left(\frac{\Delta m_{\tilde{g}}}{m_{\tilde{g}}}\right)=&-\frac{\alpha_S}{4\pi}\bigg\{6\left[B_1(m_{\tilde{g}},m_{\tilde{g}},0)-2B_0(m_{\tilde{g}},m_{\tilde{g}},0)\right]\\
 &\null\hspace{0.9cm}+\sum_{q;i=1,2}\left[B_1(m_{\tilde{g}},m_q,m_{\tilde{Q}_i})+\frac{m_q}{m_{\tilde{g}}}\mbox{Re}(e^{-\imath\phi_{M_3}}X^{\tilde{Q}}_{iL}X^{\tilde{Q}\,*}_{iR})B_0(m_{\tilde{g}},m_q,m_{\tilde{Q}_i})\right]\bigg\}
\end{align*}

Note that the one-loop corrections to the NMSSM spectrum have also been presented in \cite{Staub:2010ty}.

\section{Radiative corrections to the Higgs spectrum}

\subsection{Wave-function renormalization}\label{aploopwave}
We summarize the discussion of section \ref{secwavefunc}. Remember that $\mu_H=125$~GeV replaces the external momentum.
\begin{align}
Z_{H_u} &=1+\frac{1}{16\pi^2}\left\{N_cY_u^2B_0(\mu_H,m_u,m_u)+\frac{g'^2}{2}B_0(\mu_H,M_1,\mu)+\frac{3g^2}{2}B_0(\mu_H,M_2,\mu)+\lambda^2B_0(\mu_H,\mu,m_{\tilde{s}})\right.\nonumber\\
 &\null\hspace{3cm}\left.-\sin^2\beta\left[g^2B_0(\mu_H,M_W,M_W)+\frac{g'^2+g^2}{2}B_0(\mu_H,M_Z,M_Z)\right]\right\}\nonumber\\
Z_{H_d} &=1+\frac{1}{16\pi^2}\Big\{N_cY_d^2B_0(\mu_H,m_d,m_d)+Y_e^2B_0(\mu_H,m_e,m_e)\nonumber\\
 &\null\hspace{3cm}\left.+\frac{g'^2}{2}B_0(\mu_H,M_1,\mu)+\frac{3g^2}{2}B_0(\mu_H,M_2,\mu)+\lambda^2B_0(\mu_H,\mu,m_{\tilde{s}})\right\}\\
Z_{S}\hspace{0.15cm} &=1+\frac{1}{8\pi^2}\left\{\lambda^2B_0(\mu_H,\mu,\mu)+\kappa^2B_0(\mu_H,m_{\tilde{s}},m_{\tilde{s}})\right\}\nonumber
\end{align}

\subsection{One-loop contributions to the effective potential}
\subsubsection{SM-fermions}\label{aploopsmferm}

The squared-bilinear matrices of the fermions of third generation have been provided in Eq.\ref{sqmquarks}, \ref{sqmleptons}. One observes that they  
split into (at-most) $2\times2$ blocks corresponding to the left-handed quark fields, the right-handed ones, the left-handed leptons and the right-handed 
one. The following eigenvalues can be derived:
\begin{align*}
 & m_{t_L}^2(H)=\frac{1}{2}\left[Y_t^2|H_u|^2+Y_b^2|H_d|^2+\sqrt{Y_t^4|H_u|^4+Y_b^4|H_d|^4+2Y_t^2Y_b^2\left(|H_u|^2|H_d|^2-2|H_u\cdot H_d|^2\right)}\right]=m_{t_R}^2(H)\\
 & m_{b_L}^2(H)=\frac{1}{2}\left[Y_t^2|H_u|^2+Y_b^2|H_d|^2-\sqrt{Y_t^4|H_u|^4+Y_b^4|H_d|^4+2Y_t^2Y_b^2\left(|H_u|^2|H_d|^2-2|H_u\cdot H_d|^2\right)}\right]=m_{b_R}^2(H)\\
 & m_{\nu_L}^2(H)=0\\
 & m_{\tau_L}^2(H)=Y_{\tau}^2|H_d|^2=m_{\tau_R}^2(H)
\end{align*}

One can then deduce the following (non-vanishing) contributions to the Higgs mass matrices (with $N_c=3$ the number of colors):
\begin{align*}
& \delta\left<{\cal M}^2_{H^{\pm}}\right>_{11}=\Omega_f\frac{v_d}{v_u} & &\delta\left<{\cal M}^2_{H^0}\right>_{11}=-\frac{N_c}{4\pi^2}Y_t^4v_u^2\ln\frac{m_t^2}{Q^2}\\
& \delta\left<{\cal M}^2_{H^{\pm}}\right>_{22}=\Omega_f\frac{v_u}{v_d} & &\delta\left<{\cal M}^2_{H^0}\right>_{22}=-\frac{N_c}{4\pi^2}Y_b^4v_d^2\ln\frac{m_b^2}{Q^2}-\frac{1}{4\pi^2}Y_{\tau}^4v_d^2\ln\frac{m_{\tau}^2}{Q^2}\\
& \delta\left<{\cal M}^2_{H^{\pm}}\right>_{12}=\Omega_f=\delta\left<{\cal M}^2_{H^{\pm}}\right>_{21} & &\Omega_f\equiv-\frac{N_c}{16\pi^2}Y_t^2Y_b^2v_uv_d{\cal F}_1(m_t^2,m_b^2)
\end{align*}

We also note the following contributions to the trilinear Higgs couplings:
\begin{align*}
 \delta g^{S_i^0S^0_jS^0_k}=&-\frac{N_c}{8\sqrt{2}\pi^2}\left\{Y_t^4v_u\left[\ln\frac{Y_t^2v_u^2}{Q^2}+\frac{2}{3}\right]\left(\Pi^S\right)^{u,u,u}_{i,j,k}
+Y_b^4v_d\left[\ln\frac{Y_b^2v_d^2}{Q^2}+\frac{2}{3}\right]\left(\Pi^S\right)^{d,d,d}_{i,j,k}\right.\\
 &\null\hspace{3cm}\left.+Y_t^4v_u\ln\frac{Y_t^2v_u^2}{Q^2}\left(\Pi^A\right)^{u,u,u}_{i,j,k}+Y_b^4v_d\ln\frac{Y_b^2v_d^2}{Q^2}\left(\Pi^A\right)^{d,d,d}_{i,j,k}\right\}\\
 &-\frac{Y_{\tau}^4v_d}{8\sqrt{2}\pi^2}\left\{\left[\ln\frac{Y_{\tau}^2v_d^2}{Q^2}+\frac{2}{3}\right]\left(\Pi^S\right)^{d,d,d}_{i,j,k}+\ln\frac{Y_{\tau}^2v_d^2}{Q^2}\left(\Pi^A\right)^{d,d,d}_{i,j,k}\right\}\\
 \delta g^{S_i^0H^+H^-}=&\frac{-N_c}{8\sqrt{2}\pi^2\left(Y_t^2v_u^2-Y_b^2v_d^2\right)^2}\bigg\{Y_b^2X^R_{id}\bigg[Y_b^4v_d^4\left((Y_b^2v_d\sin^2\beta+Y_t^2v_d)\ln\frac{Y_b^2v_d^2}{Q^2}+Y_t^2v_u\sin\beta\cos\beta\right)\\
 &\null\hspace{3cm}+Y_t^4v_u^4\left(Y_b^2v_d\sin^2\beta+Y_t^2v_d\right)\left(\ln\frac{Y_t^2v_u^2}{Q^2}-1\right)\\
 &\null\hspace{-1cm}-2Y_t^2Y_b^2v_u^2v_d^2\left(Y_t^2v_u\sin\beta\cos\beta\left(3\ln\frac{Y_b^2v_d^2}{Q^2}-\ln\frac{Y_t^2v_u^2}{Q^2}\right)+(Y_b^2v_d\sin^2\beta+Y_t^2v_d\cos^2\beta)\left(\ln\frac{Y_b^2v_d^2}{Q^2}-\frac{1}{2}\right)\right)\bigg]\\
 &\null\hspace{1.5cm}+Y_t^2X^R_{iu}\bigg[Y_b^4v_d^4\left(Y_b^2v_u+Y_t^2v_u\cos^2\beta\right)\left(\ln\frac{Y_b^2v_d^2}{Q^2}-1\right)\\
 &\null\hspace{3cm}+Y_t^4v_u^4\left((Y_t^2v_u\cos^2\beta+Y_b^2v_u)\ln\frac{Y_t^2v_u^2}{Q^2}+Y_b^2v_d\sin\beta\cos\beta\right)\\
 &\null\hspace{-1cm}-2Y_t^2Y_b^2v_u^2v_d^2\left(Y_b^2v_d\sin\beta\cos\beta\left(3\ln\frac{Y_t^2v_u^2}{Q^2}-\ln\frac{Y_b^2v_d^2}{Q^2}\right)+(Y_t^2v_u\cos^2\beta+Y_b^2v_u\sin^2\beta)\left(\ln\frac{Y_t^2v_u^2}{Q^2}-\frac{1}{2}\right)\right)\bigg]\bigg\}\\
 &\null\hspace{-1cm}-\frac{Y_{\tau}^4v_dX^R_{id}\sin^2\beta}{8\sqrt{2}\pi^2}\ln\frac{Y_{\tau}^2v_d^2}{Q^2}
\end{align*}

\subsubsection{Electroweak gauge bosons}\label{aploopsmgauge}

Using Eq.\ref{massgauge}, we obtain the following contribution to the effective potential (where we have performed an expansion in terms of the 
charged-Higgs fields):
\begin{align*}
 \delta{\cal V}^G_H=&\frac{3}{64\pi^2}\left\{\frac{(g'^2+g^2)^2}{4}\left(|H_u^0|^2+|H_d^0|^2\right)^2\left[\ln\frac{(g'^2+g^2)(|H_u^0|^2+|H_d^0|^2)}{2Q^2}-\frac{3}{2}\right]\right.\\
 &\null\hspace{1cm}+\frac{g^4}{2}\left(|H_u^0|^2+|H_d^0|^2\right)^2\left[\ln\frac{g^2(|H_u^0|^2+|H_d^0|^2)}{2Q^2}-\frac{3}{2}\right]\\
 &\null\hspace{1cm}+\left[\frac{(g'^2+g^2)^2}{2}\left(|H_u^0|^2+|H_d^0|^2\right)\left(H_u^+H_u^-+H_d^+H_d^-\right)\right.\\
 &\null\hspace{1.5cm}-2g^2g'^2\left(|H_d^0|^2H_u^+H_u^-+|H_u^0|^2H_d^+H_d^-+2\mbox{Re}(H_u^{0\,*}H_d^{0\,*}H_u^+H_d^-)\right)\Big]\\
 &\null\hspace{5cm}\times\left[\ln\frac{(g'^2+g^2)(|H_u^0|^2+|H_d^0|^2)}{2Q^2}-1\right]\\
 &\left.\null\hspace{1cm}+g^4\left(|H_u^0|^2+|H_d^0|^2\right)\left(H_u^+H_u^-+H_d^+H_d^-\right)\left[\ln\frac{g^2(|H_u^0|^2+|H_d^0|^2)}{2Q^2}-1\right]+O\left((H^+H^-)^2\right)\right\}
\end{align*}
One derives the following (non-vanishing) contributions to the mass-matrices:
\begin{align*}
 & \delta\left<{\cal M}^2_{H^{\pm}}\right>_{11}=\Omega_G\frac{v_d}{v_u} & &\delta\left<{\cal M}^2_{H^0}\right>_{11}=\tilde{\Omega}_Gv_u^2\\
 & \delta\left<{\cal M}^2_{H^{\pm}}\right>_{22}=\Omega_G\frac{v_u}{v_d} & &\delta\left<{\cal M}^2_{H^0}\right>_{22}=\tilde{\Omega}_Gv_d^2\\
 & \delta\left<{\cal M}^2_{H^{\pm}}\right>_{12}=\Omega_G=\delta\left<{\cal M}^2_{H^{\pm}}\right>_{21} & &\delta\left<{\cal M}^2_{H^0}\right>_{12}=\tilde{\Omega}_Gv_uv_d=\delta\left<{\cal M}^2_{H^0}\right>_{21}\\
 & \Omega_G\equiv-\frac{3g^2g'^2}{32\pi^2}v_uv_d\left[\ln\frac{M_Z^2}{Q^2}-1\right] & &\tilde{\Omega}_G\equiv\frac{3}{64\pi^2}\left[2g^4\ln\frac{M_W^2}{Q^2}+(g'^2+g^2)^2\ln\frac{M_Z^2}{Q^2}\right]
\end{align*}
Similarly, one can derive the corrections to the trilinear Higgs couplings:
\begin{align*}
 \delta g^{S_i^0S^0_jS^0_k}=&\frac{3}{64\sqrt{2}\pi^2}\left\{\left[\frac{(g'^2+g^2)^2}{2}\ln\frac{M_Z^2}{Q^2}+g^4\ln\frac{M_W^2}{Q^2}\right]\times\left(v_u\left[\left(\Pi^S\right)^{u,u,u}_{i,j,k}+\left(\Pi^S\right)^{u,d,d}_{i,j,k}+\left(\Pi^A\right)^{u,u,u}_{i,j,k}+\left(\Pi^A\right)^{u,d,d}_{i,j,k}\right]\right.\right.\\
 &\null\hspace{7cm}\left.+v_d\left[\left(\Pi^S\right)^{d,d,d}_{i,j,k}+\left(\Pi^S\right)^{d,u,u}_{i,j,k}+\left(\Pi^A\right)^{d,d,d}_{i,j,k}+\left(\Pi^A\right)^{d,u,u}_{i,j,k}\right]\right)\\
 &\null\hspace{1.5cm}\left.+\frac{(g'^2+g^2)^2+2g^4}{3(v_u^2+v_d^2)}\left[v_u^3\left(\Pi^S\right)^{u,u,u}_{i,j,k}+3v_u^2v_d\left(\Pi^S\right)^{d,u,u}_{i,j,k}+3v_uv_d^2\left(\Pi^S\right)^{u,d,d}_{i,j,k}+v_d^3\left(\Pi^S\right)^{d,d,d}_{i,j,k}\right]\right\}\\
 \delta g^{S_i^0H^+H^-}=&\frac{3}{64\sqrt{2}\pi^2}\left[\frac{(g'^2-g^2)^2}{2}\ln\frac{M_Z^2}{Q^2}+g^4\ln\frac{M_W^2}{Q^2}\right]\left[v_uX^R_{iu}+v_dX^R_{id}\right]
\end{align*}

\subsubsection{Sfermions}\label{aploophisferm}

Considering the sfermion mass-matrices of Eq.\ref{massup}, \ref{massdown}, \ref{massqmix}, \ref{massneut}, \ref{massel}, \ref{massemix} and focussing 
on the neutral-Higgs dependence first, one obtains decoupling $2\times2$ blocks -- $1\times1$ in the case of the sneutrinos -- so that eigenvalues may be expressed
as $m^2_{\tilde{F}_{m}}=\frac{1}{2}\left[T_{\tilde{F}}(H^0)+(-1)^m\sqrt{R^2_{\tilde{F}}(H^0)}\right]$, $m=1,2$. Corresponding contributions to the
neutral Higgs mass-matrix thus read (we denote as $E_{ijk}$ the coefficients coming from the tadpole equations -- see Eq.\ref{masstadpole}):
\begin{align*}
\delta\left<{\cal M}^2_{H^0}\right>_{ij}=  & \frac{N_c}{64\pi^2}\left\{\left<\frac{\partial^2T_{\tilde{F}}}{\partial S_i^0/\sqrt{2}\partial S_j^0/\sqrt{2}}-E_{ijk}\frac{\partial T_{\tilde{F}}}{\partial S^0_k/\sqrt{2}}\right>\left[m_{\tilde{F}_1}^2\left(\ln\frac{m_{\tilde{F}_1}^2}{Q_2}-1\right)+m_{\tilde{F}_2}^2\left(\ln\frac{m_{\tilde{F}_2}^2}{Q_2}-1\right)\right]\right.\\
 &\null\hspace{1cm}+\frac{1}{2}\left<\frac{\partial T_{\tilde{F}}}{\partial S^0_i/\sqrt{2}}\frac{\partial T_{\tilde{F}}}{\partial S^0_j/\sqrt{2}}\right>\ln\frac{m_{\tilde{F}_1}^2m_{\tilde{F}_2}^2}{Q^4}\\
 &\null\hspace{1cm}+\frac{1}{4}\left<\frac{\partial T_{\tilde{F}}}{\partial S^0_i/\sqrt{2}}\frac{\partial R^2_{\tilde{F}}}{\partial S^0_j/\sqrt{2}}+\frac{\partial R^2_{\tilde{F}}}{\partial S^0_i/\sqrt{2}}\frac{\partial T_{\tilde{F}}}{\partial S^0_j/\sqrt{2}}\right>\frac{1}{m_{\tilde{F}_2}^2-m_{\tilde{F}_1}^2}\ln\frac{m_{\tilde{F}_2}^2}{m_{\tilde{F}_1}^2}\\
 &\null\hspace{1cm}+\frac{1}{8}\left<\frac{\partial R^2_{\tilde{F}}}{\partial S^0_i/\sqrt{2}}\frac{\partial R^2_{\tilde{F}}}{\partial S^0_j/\sqrt{2}}\right>\frac{1}{(m_{\tilde{F}_2}^2-m_{\tilde{F}_1}^2)^2}\left[\frac{m_{\tilde{F}_2}^2+m_{\tilde{F}_1}^2}{m_{\tilde{F}_2}^2-m_{\tilde{F}_1}^2}\ln\frac{m_{\tilde{F}_1}^2}{m_{\tilde{F}_2}^2}+2\right]\\
 &\null\hspace{1cm}\left.+\frac{1}{2}\left<\frac{\partial^2R^2_{\tilde{F}}}{\partial S_i^0/\sqrt{2}\partial S_j^0/\sqrt{2}}-E_{ijk}\frac{\partial R^2_{\tilde{F}}}{\partial S^0_k/\sqrt{2}}\right>{\cal F}_1(m_{\tilde{F}_1}^2,m_{\tilde{F}_2}^2)\right\}
\end{align*}
with:
\begin{align*}
\null\hspace{-0.35cm} T_{\tilde{U}}=&m_{Q}^2+m_U^2+2Y_u^2|H_u^0|^2-\frac{g'^2+g^2}{4}(|H_u^0|^2-|H_d^0|^2)\\
\null\hspace{-0.35cm} R^2_{\tilde{U}}=&\left[m_{Q}^2-m_U^2+\frac{1}{4}\left(\frac{5}{3}g'^2-g^2\right)(|H_u^0|^2-|H_d^0|^2)\right]^2+4Y_u^2\left[A_u^2|H_u^0|^2+\lambda^2|S|^2|H_d^0|^2-2\lambda A_u\mbox{Re}(e^{\imath(\varphi_{A_u}+\varphi_{\lambda})}SH_u^0H_d^0)\right]\\
\null\hspace{-0.35cm} T_{\tilde{D}}=&m_{Q}^2+m_D^2+2Y_d^2|H_d^0|^2+\frac{g'^2+g^2}{4}(|H_u^0|^2-|H_d^0|^2)\\
\null\hspace{-0.35cm} R^2_{\tilde{D}}=&\left[m_{Q}^2-m_D^2+\frac{1}{4}\left(-\frac{g'}{3}-g^2\right)(|H_u^0|^2-|H_d^0|^2)\right]^2+4Y_d^2\left[A_d^2|H_d^0|^2+\lambda^2|S|^2|H_u^0|^2-2\lambda A_d\mbox{Re}(e^{\imath(\varphi_{A_d}+\varphi_{\lambda})}SH_u^0H_d^0)\right]\\
\null\hspace{-0.35cm} T_{\tilde{E}}=&m_{L}^2+m_E^2+2Y_e^2|H_d^0|^2+\frac{g'^2+g^2}{4}(|H_u^0|^2-|H_d^0|^2)\\
\null\hspace{-0.35cm} R^2_{\tilde{E}}=&\left[m_{L}^2-m_E^2+\frac{1}{4}\left(-2g'^2+g^2\right)(|H_u^0|^2-|H_d^0|^2)\right]^2+4Y_e^2\left[A_e^2|H_d^0|^2+\lambda^2|S|^2|H_u^0|^2-2\lambda A_e\mbox{Re}(e^{\imath(\varphi_{A_e}+\varphi_{\lambda})}SH_u^0H_d^0)\right]
\end{align*}
Additionally, the sneutrino mass reads: $m^2_{\tilde{N}}=m_{L}^2-\frac{g'^2+g^2}{4}(|H_u^0|^2-|H_d^0|^2)$.

To derive the corrections to the charged-Higgs masses and Higgs-to-Higgs couplings, one is confronted to the task of diagonalizing the matrix-system 
of Eq.\ref{massup}, \ref{massdown}, \ref{massqmix}, \ref{massneut}, \ref{massel}, \ref{massemix}. This can be performed perturbatively, as an expansion
in the Higgs-doublet fields, which amounts to a series in $\frac{v}{M_{\mbox{\tiny SUSY}}}$. We confine to a precision of order 
$O\left(v^2\right)$ at the level of the masses, which means that we compute the potential up to terms of $H^4$-order
($H$ standing for any Higgs-doublet field) and freeze singlet fields to their v.e.v.\ $s$ for terms of $H^4$-order (they are kept explicitly for
terms of lower order in the expansion). The ensuing corrections to the Higgs potential can be matched onto Eq.\ref{potential} and we may then use the 
results of section \ref{asimppotmass}, e.g.\ for the charged-Higgs mass (Eq.\ref{cHiggsmass}) or the Higgs couplings. For squarks of each generation
(note that we neglect the Yukawa couplings of the two first families):
{\small\begin{align*}
  &\delta {\cal V}_0=\frac{N_c}{32\pi^2}\left\{2m_Q^4\left[\ln\frac{m_Q^2}{Q^2}-\frac{3}{2}\right]+m_U^4\left[\ln\frac{m_U^2}{Q^2}-\frac{3}{2}\right]+m_D^4\left[\ln\frac{m_D^2}{Q^2}-\frac{3}{2}\right]\right\}\\
 &\delta M_u^2=\frac{2N_c}{32\pi^2}\left\{\left(Y_u^2+\frac{g'^2}{6}\right){\cal F}_0(m_Q^2)+\left(Y_u^2-\frac{g'^2}{3}\right){\cal F}_0(m_U^2)+\frac{g'^2}{6}{\cal F}_0(m_D^2)+Y_u^2A_u^2{\cal F}_1(m_Q^2,m_U^2)+Y_d^2\lambda^2|S|^2{\cal F}_1(m_Q^2,m_D^2)\right\}\\
 &\delta M_d^2=\frac{2N_c}{32\pi^2}\left\{\left(Y_d^2-\frac{g'^2}{6}\right){\cal F}_0(m_Q^2)+\frac{g'^2}{3}{\cal F}_0(m_U^2)+\left(Y_d^2-\frac{g'^2}{6}\right){\cal F}_0(m_D^2)+Y_u^2\lambda^2|S|^2{\cal F}_1(m_Q^2,m_U^2)+Y_d^2A_d^2{\cal F}_1(m_Q^2,m_D^2)\right\}\\
 &\delta A_{ud}e^{\imath\varphi_{A_{ud}}}=\frac{2N_c}{32\pi^2}\lambda\left\{Y_u^2A_ue^{\imath(\varphi_{A_u}+\varphi_{\lambda})}{\cal F}_1(m_Q^2,m_U^2)+Y_d^2A_de^{\imath(\varphi_{A_d}+\varphi_{\lambda})}{\cal F}_1(m_Q^2,m_D^2)\right\}
\end{align*}
\begin{align*}
 &\delta\lambda_u=\frac{2N_c}{32\pi^2}\left\{\left[Y_u^4+2Y_u^2\left(\frac{g'^2}{12}-\frac{g^2}{4}\right)+2\left(\frac{g'^2}{12}\right)^2+2\left(\frac{g^2}{4}\right)^2\right]\ln\frac{m_Q^2}{Q^2}+\left(Y_u^2-\frac{g'^2}{3}\right)^2\ln\frac{m_U^2}{Q^2}+\left(\frac{g'^2}{6}\right)^2\ln\frac{m_D^2}{Q^2}\right.\\
 &\left.\null\hspace{2cm}+2Y_u^2A_u^2\left[Y_u^2+\frac{1}{4}\left(\frac{g'^2}{3}-g^2\right)\right]{\cal F}_3(m_U^2,m_Q^2)+2Y_u^2A_u^2\left(Y_u^2-\frac{g'^2}{3}\right){\cal F}_3(m_Q^2,m_U^2)+Y_u^4A_u^4{\cal F}_7(m_Q^2,m_U^2)\right.\\
 &\left.\null\hspace{2cm}+2Y_d^2\lambda^2|S|^2\frac{1}{4}\left(\frac{g'^2}{3}+g^2\right){\cal F}_3(m_D^2,m_Q^2)+2Y_d^2\lambda^2|S|^2\frac{g'^2}{6}{\cal F}_3(m_Q^2,m_D^2)+Y_d^4\lambda^4|S|^4{\cal F}_7(m_Q^2,m_D^2)\right\}\\
 &\delta\lambda_d=\frac{2N_c}{32\pi^2}\left\{\left[Y_d^4-2Y_d^2\left(\frac{g'^2}{12}+\frac{g^2}{4}\right)+2\left(\frac{g'^2}{12}\right)^2+2\left(\frac{g^2}{4}\right)^2\right]\ln\frac{m_Q^2}{Q^2}+\left(\frac{g'^2}{3}\right)^2\ln\frac{m_U^2}{Q^2}+\left(Y_d^2-\frac{g'^2}{6}\right)^2\ln\frac{m_D^2}{Q^2}\right.\\
 &\left.\null\hspace{2cm}+2Y_u^2\lambda^2|S|^2\frac{1}{4}\left(-\frac{g'^2}{3}+g^2\right){\cal F}_3(m_U^2,m_Q^2)+2Y_u^2\lambda^2|S|^2\frac{g'^2}{3}{\cal F}_3(m_Q^2,m_U^2)+Y_u^4\lambda^4|S|^4{\cal F}_7(m_Q^2,m_U^2)\right.\\
 &\left.\null\hspace{2cm}+2Y_d^2A_d^2\left[Y_d^2-\frac{1}{4}\left(\frac{g'^2}{3}+g^2\right)\right]{\cal F}_3(m_D^2,m_Q^2)+2Y_d^2A_d^2\left(Y_d^2-\frac{g'^2}{6}\right){\cal F}_3(m_Q^2,m_D^2)+Y_d^4A_d^4{\cal F}_7(m_Q^2,m_D^2)\right\}
\end{align*}
\begin{align*}
 &\delta\lambda_3=\frac{2N_c}{32\pi^2}\left\{\left[Y_u^2Y_d^2-Y_u^2\left(\frac{g'^2}{12}+\frac{g^2}{4}\right)-Y_d^2\left(-\frac{g'^2}{12}+\frac{g^2}{4}\right)-2\left(\frac{g'^2}{12}\right)^2+2\left(\frac{g^2}{4}\right)^2\right]\ln\frac{m_Q^2}{Q^2}+\frac{g'^2}{3}\left(Y_u^2-\frac{g'^2}{3}\right)\ln\frac{m_U^2}{Q^2}\right.\\
 &\left.\null\hspace{1.5cm}+\frac{g'^2}{6}\left(Y_d^2-\frac{g'^2}{6}\right)\ln\frac{m_D^2}{Q^2}+Y_u^2Y_d^2{\cal F}_1(m_U^2,m_D^2)+Y_u^4A_u^2\lambda^2|S|^2{\cal F}_7(m_Q^2,m_U^2)+Y_d^4A_d^2\lambda^2|S|^2{\cal F}_7(m_Q^2,m_D^2)\right.\\
 &\left.\null\hspace{1.5cm}+Y_u^2\left[Y_d^2A_u^2+\frac{1}{4}\left(\frac{g'^2}{3}+g^2\right)\left(\lambda^2|S|^2-A_u^2\right)\right]{\cal F}_3(m_U^2,m_Q^2)+Y_u^2\left[Y_u^2\lambda^2|S|^2+\frac{g'^2}{3}\left(A_u^2-\lambda^2|S|^2\right)\right]{\cal F}_3(m_Q^2,m_U^2)\right.\\
 &\left.\null\hspace{1.5cm}+Y_d^2\left[Y_u^2A_d^2+\frac{1}{4}\left(\frac{g'^2}{3}-g^2\right)\left(A_d^2-\lambda^2|S|^2\right)\right]{\cal F}_3(m_D^2,m_Q^2)+Y_d^2\left[Y_d^2\lambda^2|S|^2+\frac{g'^2}{6}\left(A_d^2-\lambda^2|S|^2\right)\right]{\cal F}_3(m_Q^2,m_D^2)\right.\\
 &\left.\null\hspace{1.5cm}+2Y_u^2Y_d^2\left[A_uA_d\cos{(\varphi_{A_u}-\varphi_{A_d})}-\lambda^2|S|^2\right]{\cal F}_5(m_Q^2,m_U^2,m_D^2)\right.\\
 &\left.\null\hspace{5cm}\phantom{\frac{1^1}{1^1}}+Y_u^2Y_d^2\left[A_u^2A_d^2+\lambda^4|S|^4-2A_uA_d\lambda^2|S|^2\cos{(\varphi_{A_u}-\varphi_{A_d})}\right]{\cal F}_6(m_Q^2,m_U^2,m_D^2)\right\}\\
 &\delta\lambda_4=\frac{2N_c}{32\pi^2}\left\{-\left(Y_u^2-\frac{g^2}{2}\right)\left(Y_d^2-\frac{g^2}{2}\right)\ln\frac{m_Q^2}{Q^2}-Y_u^2Y_d^2{\cal F}_1(m_U^2,m_D^2)\right.\\
 &\left.\null\hspace{0.7cm}+Y_u^2\left[\left(Y_u^2-\frac{g^2}{2}\right)\lambda^2|S|^2-\left(Y_d^2-\frac{g^2}{2}\right)A_u^2\right]{\cal F}_3(m_U^2,m_Q^2)+Y_d^2\left[\left(Y_d^2-\frac{g^2}{2}\right)\lambda^2|S|^2-\left(Y_u^2-\frac{g^2}{2}\right)A_d^2\right]{\cal F}_3(m_D^2,m_Q^2)\right.\\
 &\left.\null\hspace{0.7cm}-2Y_u^2Y_d^2\left[A_uA_d\cos(\varphi_{A_u}-\varphi_{A_d})-\lambda^2|S|^2\right]{\cal F}_5(m_Q^2,m_U^2,m_D^2)+Y_u^4A_u^2\lambda^2|S|^2{\cal F}_7(m_Q^2,m_U^2)+Y_d^4A_d^2\lambda^2|S|^2{\cal F}_7(m_Q^2,m_D^2)\right.\\
 &\left.\null\hspace{5cm}\phantom{\frac{1^1}{1^1}}-Y_u^2Y_d^2\left[A_u^2A_d^2+\lambda^4|S|^4-2A_uA_d\lambda^2|S|^2\cos(\varphi_{A_u}-\varphi_{A_d})\right]{\cal F}_6(m_Q^2,m_U^2,m_D^2)\right\}
\end{align*}
\begin{align*}
 &\delta\lambda_5=\frac{2N_c}{32\pi^2}\lambda^2S^2\left\{Y_u^4A_u^2e^{2\imath(\varphi_{A_u}+\varphi_{\lambda})}{\cal F}_7(m_Q^2,m_U^2)+Y_d^4A_d^2e^{2\imath(\varphi_{A_d}+\varphi_{\lambda})}{\cal F}_7(m_Q^2,m_D^2)\right\}\\
 &\delta\lambda_6=\frac{2N_c}{32\pi^2}\lambda S\left\{Y_u^2A_ue^{\imath(\varphi_{A_u}+\varphi_{\lambda})}\left[\left(Y_u^2+\frac{g'^2}{12}-\frac{g^2}{4}\right){\cal F}_3(m_U^2,m_Q^2)+\left(Y_u^2-\frac{g'^2}{3}\right){\cal F}_3(m_Q^2,m_U^2)+Y_u^2A_u^2{\cal F}_7(m_Q^2,m_U^2)\right]\right.\\
 &\left.\null\hspace{2cm}+Y_d^2A_de^{\imath(\varphi_{A_d}+\varphi_{\lambda})}\left[\left(\frac{g'^2}{12}+\frac{g^2}{4}\right){\cal F}_3(m_D^2,m_Q^2)+\frac{g'^2}{6}{\cal F}_3(m_Q^2,m_D^2)+Y_d^2\lambda^2|S|^2{\cal F}_7(m_Q^2,m_D^2)\right]\right\}\\
 &\delta\lambda_7=\frac{2N_c}{32\pi^2}\lambda S\left\{Y_u^2A_ue^{\imath(\varphi_{A_u}+\varphi_{\lambda})}\left[\left(-\frac{g'^2}{12}+\frac{g^2}{4}\right){\cal F}_3(m_U^2,m_Q^2)+\frac{g'^2}{3}{\cal F}_3(m_Q^2,m_U^2)+Y_u^2\lambda^2|S|^2{\cal F}_7(m_Q^2,m_U^2)\right]\right.\\
 &\left.\null\hspace{2cm}+Y_d^2A_de^{\imath(\varphi_{A_d}+\varphi_{\lambda})}\left[\left(Y_d^2-\frac{g'^2}{12}-\frac{g^2}{4}\right){\cal F}_3(m_D^2,m_Q^2)+\left(Y_d^2-\frac{g'^2}{6}\right){\cal F}_3(m_Q^2,m_D^2)+Y_d^2A_d^2{\cal F}_7(m_Q^2,m_D^2)\right]\right\}
\end{align*}}
For the sleptons:
{\small\begin{align*}
  &\delta {\cal V}_0=\frac{1}{32\pi^2}\left\{2m_L^4\left[\ln\frac{m_L^2}{Q^2}-\frac{3}{2}\right]+m_E^4\left[\ln\frac{m_E^2}{Q^2}-\frac{3}{2}\right]\right\}\\
 &\delta M_u^2=\frac{2}{32\pi^2}\left\{\frac{g'^2}{2}\left[{\cal F}_0(m_E^2)-{\cal F}_0(m_Q^2)\right]+Y_e^2\lambda^2|S|^2{\cal F}_1(m_L^2,m_E^2)\right\}\\
 &\delta M_d^2=\frac{2}{32\pi^2}\left\{\left(Y_e^2+\frac{g'^2}{2}\right){\cal F}_0(m_L^2)+\left(Y_e^2-\frac{g'^2}{2}\right){\cal F}_0(m_E^2)+Y_e^2A_e^2{\cal F}_1(m_L^2,m_E^2)\right\}\\
 &\delta A_{ud}e^{\imath\varphi_{A_{ud}}}=\frac{2}{32\pi^2}\lambda Y_e^2A_ee^{\imath(\varphi_{A_e}+\varphi_{\lambda})}{\cal F}_1(m_L^2,m_E^2)\\
 &\delta\lambda_u=\frac{2}{32\pi^2}\left\{2\left[\left(\frac{g'^2}{2}\right)^2+\left(\frac{g^2}{2}\right)^2\right]\ln\frac{m_L^2}{Q^2}+\left(\frac{g'^2}{2}\right)^2\ln\frac{m_E^2}{Q^2}\right.\\
 &\left.\null\hspace{2cm}+2Y_e^2\lambda^2|S|^2\frac{-g'^2+g^2}{4}{\cal F}_3(m_E^2,m_L^2)+2Y_e^2\lambda^2|S|^2\frac{g'^2}{2}{\cal F}_3(m_L^2,m_E^2)+Y_e^4\lambda^4|S|^4{\cal F}_7(m_L^2,m_E^2)\right\}\\
 &\delta\lambda_d=\frac{2}{32\pi^2}\left\{\left[Y_e^4+2Y_e^2\frac{g'^2-g^2}{4}+2\left(\frac{g'^2}{4}\right)^2+2\left(\frac{g^2}{4}\right)^2\right]\ln\frac{m_L^2}{Q^2}+\left(Y_e^2-\frac{g'^2}{2}\right)^2\ln\frac{m_E^2}{Q^2}\right.\\
 &\left.\null\hspace{2cm}+2Y_e^2A_e^2\left(Y_e^2+\frac{g'^2-g^2}{4}\right){\cal F}_3(m_E^2,m_L^2)+2Y_e^2A_e^2\left(Y_e^2-\frac{g'^2}{2}\right){\cal F}_3(m_L^2,m_E^2)+Y_e^4A_e^4{\cal F}_7(m_L^2,m_E^2)\right\}
\end{align*}
\begin{align*}
 &\delta\lambda_3=\frac{2}{32\pi^2}\left\{-\frac{g'^2+g^2}{4}\left(Y_e^2+\frac{g'^2-g^2}{2}\right)\ln\frac{m_L^2}{Q^2}+\frac{g'^2}{2}\left(Y_e^2-\frac{g'^2}{2}\right)\ln\frac{m_E^2}{Q^2}+Y_e^4A_e^2\lambda^2|S|^2{\cal F}_7(m_L^2,m_E^2)\right.\\
 &\left.\null\hspace{2cm}+Y_e^2\frac{g'^2+g^2}{4}\left(\lambda^2|S|^2-A_e^2\right){\cal F}_3(m_E^2,m_L^2)+Y_e^2\left[Y_e^2\lambda^2|S|^2+\frac{g'^2}{2}\left(A_e^2-\lambda^2|S|^2\right)\right]{\cal F}_3(m_L^2,m_E^2)\right\}\\
 &\delta\lambda_4=\frac{2}{32\pi^2}\left\{\frac{g^2}{2}\left(Y_e^2-\frac{g^2}{2}\right)\ln\frac{m_L^2}{Q^2}+Y_e^2\left[Y_e^2\lambda^2|S|^2+\frac{g^2}{2}\left(A_e^2-\lambda^2|S|^2\right)\right]{\cal F}_3(m_E^2,m_L^2)+Y_e^4A_e^2\lambda^2|S|^2{\cal F}_7(m_L^2,m_E^2)\right\}\\
 &\delta\lambda_5=\frac{2}{32\pi^2}\lambda^2S^2Y_e^4A_e^2e^{2\imath(\varphi_{A_e}+\varphi_{\lambda})}{\cal F}_7(m_L^2,m_E^2)\\
 &\delta\lambda_6=\frac{2}{32\pi^2}\lambda SY_e^2A_ee^{\imath(\varphi_{A_e}+\varphi_{\lambda})}\left\{\frac{-g'^2+g^2}{4}{\cal F}_3(m_E^2,m_L^2)+\frac{g'^2}{2}{\cal F}_3(m_L^2,m_E^2)+Y_e^2\lambda^2|S|^2{\cal F}_7(m_L^2,m_E^2)\right\}\\
 &\delta\lambda_7=\frac{2}{32\pi^2}\lambda SY_e^2A_ee^{\imath(\varphi_{A_e}+\varphi_{\lambda})}\left\{\left(Y_e^2+\frac{g'^2-g^2}{4}\right){\cal F}_3(m_E^2,m_L^2)+\left(Y_e^2-\frac{g'^2}{2}\right){\cal F}_3(m_L^2,m_E^2)+Y_e^2A_e^2{\cal F}_7(m_L^2,m_E^2)\right\}
\end{align*}}

\subsubsection{Charginos and neutralinos}\label{aploophiinos}

To include the chargino and neutralino contributions to the effective Higgs potential, we turn exclusively to the method that we have just presented 
in the case of the sfermions. In other words, we diagonalize the matrix system of Eq.\ref{masschaneu} perturbatively, in an expansion of doublet Higgs-fields
and we match the ensuing potential to the form of Eq.\ref{potential}. One can then work out the contributions to the Higgs mass-matrix and couplings.
Note, however, that instead of diagonalizing directly the $9\times9$ (squared) bilinear matrix of Eq.\ref{masschaneu}, it is easier to consider the dependence
on neutral Higgs fields only (that is replacing charged fields by $0$), as the corresponding matrix then splits into various blocks. All the 
couplings of Eq.\ref{potential} can be identified from the neutral potential, with the exception of $\lambda_{3,4}$, which only appear in terms of the 
sum $\lambda_3+\lambda_4$. It is a straightforward task, however, to compute $\lambda_4$ in a second step, from the charged couplings of 
the full potential. Another remark accompanies the observation that, as higgsino and singlino masses depend on the singlet Higgs field, the coefficients
in our matching procedure depend on the singlet fields as well (those would correspond to operators of dimension $>4$), which leads to 
additional (but straightforward) terms, with respect to the results of section \ref{asimppotmass}. This $S$-dependence can be neglected for terms of
order $H^4$, as keeping it would produce terms of higher order in $\frac{v^2}{M_{\mbox{\tiny SUSY}}^2}$. For $M_u^2$ and $M_d^2$, the $S$-dependence
is largely absorbed (at least at leading logarithmic order) by the coefficients $\lambda_P^u$ and $\lambda_P^d$. We obtain:
{\small\begin{align*}
 &\delta {\cal V}_0=\frac{-1}{32\pi^2}\left\{M_1^4\left[\ln\frac{M_1^2}{Q^2}-\frac{3}{2}\right]+3M_2^4\left[\ln\frac{M_2^2}{Q^2}-\frac{3}{2}\right]+4(\lambda^2|S|^2)^2\left[\ln\frac{\lambda^2|S|^2}{Q^2}-\frac{3}{2}\right]+(4\kappa^2|S|^2)^2\left[\ln\frac{4\kappa^2|S|^2}{Q^2}-\frac{3}{2}\right]\right\}\\
 &\delta M_{u,d}^2=\frac{-1}{32\pi^2}\left\{g'^2{\cal F}_0(M_1^2)+3g^2{\cal F}_0(M_2^2)+(2\lambda^2+g'^2+3g^2){\cal F}_0(\lambda^2|S|^2)+2\lambda^2{\cal F}_0(4\kappa^2|S|^2)\right.\\
 &\left.\null\hspace{1.cm}+g'^2\left(M_1^2+\lambda^2|S|^2\right){\cal F}_1(M_1^2,\lambda^2|S|^2)+3g^2\left(M_2^2+\lambda^2|S|^2\right){\cal F}_1(M_2^2,\lambda^2|S|^2)+2\lambda^2\left(4\kappa^2+\lambda^2\right)|S|^2{\cal F}_1(4\kappa^2|S|^2,\lambda^2|S|^2)\right\}\\
 &\delta \lambda_P^{u,d}=\frac{-\lambda^2}{32\pi^2}\left\{g'^2\left[\ln\frac{\lambda^2|S|^2}{Q^2}+{\cal F}_1(M_1^2,\lambda^2|S|^2)+(M_1^2+\lambda^2|S|^2){\cal F}_3(M_1^2,\lambda^2|S|^2)\right]\right.\\
 &\null\hspace{2.cm}+3g^2\left[\ln\frac{\lambda^2|S|^2}{Q^2}+{\cal F}_1(M_2^2,\lambda^2|S|^2)+(M_2^2+\lambda^2|S|^2){\cal F}_3(M_2^2,\lambda^2|S|^2)\right]\\
 &\null\hspace{2.cm}+2\lambda^2\left[\ln\frac{\lambda^2|S|^2}{Q^2}+{\cal F}_1(4\kappa^2|S|^2,\lambda^2|S|^2)+(4\kappa^2+\lambda^2)|S|^2{\cal F}_3(4\kappa^2|S|^2,\lambda^2|S|^2)\right]\\
 &\null\hspace{2.cm}+8\kappa^2\left[\ln\frac{4\kappa^2|S|^2}{Q^2}+{\cal F}_1(4\kappa^2|S|^2,\lambda^2|S|^2)+(4\kappa^2+\lambda^2)|S|^2{\cal F}_3(\lambda^2|S|^2,4\kappa^2|S|^2)\right]\bigg\}\\
 &\delta A_{ud}e^{\imath\varphi_{A_{ud}}}=\frac{-1}{32\pi^2}(-2\lambda)\left\{g'^2M_1e^{\imath(\varphi_{M_1}+\varphi_{\lambda})}{\cal F}_1(M_1^2,\lambda^2|S|^2)+3g^2M_2e^{\imath(\varphi_{M_2}+\varphi_{\lambda})}{\cal F}_1(M_2^2,\lambda^2|S|^2)\right\}\\
 &\delta \lambda_P^Me^{\imath\varphi_M}=\frac{-1}{32\pi^2}8\kappa\lambda^3e^{\imath(\varphi_{\lambda}-\varphi_{\kappa})}{\cal F}_1(4\kappa^2|S|^2,\lambda^2|S|^2)
\end{align*}
\begin{align*}
 &\delta\lambda_{u,d}=\frac{-1}{32\pi^2}\left\{\frac{g'^4}{2}\ln\frac{M_1^2}{Q^2}+\frac{5g^4}{2}\ln\frac{M_2^2}{Q^2}+\left(2\lambda^4+\frac{g'^4+5g^4+2g'^2g^2}{2}\right)\ln\frac{\lambda^2|S|^2}{Q^2}+2\lambda^4\ln\frac{4\kappa^2|S|^2}{Q^2}+g'^2g^2{\cal F}_1(M_1^2,M_2^2)\right.\\
 &\left.\null\hspace{2cm}+g'^4\left(M_1^2+\lambda^2|S|^2\right){\cal F}_3(\lambda^2|S|^2,M_1^2)+g'^2\left[(g'^2+g^2)M_1^2+2\lambda^4|S|^2\right]{\cal F}_3(M_1^2,\lambda^2|S|^2)\right.\\
 &\left.\null\hspace{2cm}+g^4\left(M_2^2+5\lambda^2|S|^2\right){\cal F}_3(\lambda^2|S|^2,M_2^2)+g^2\left[(g'^2+5g^2)M_2^2+2\lambda^4|S|^2\right]{\cal F}_3(M_2^2,\lambda^2|S|^2)\right.\\
 &\left.\null\hspace{2cm}+4\lambda^4\left(\lambda^2+4\kappa^2\right)|S|^2{\cal F}_3(\lambda^2|S|^2,4\kappa^2|S|^2)+2\lambda^4|S|^2\left(g'^2+g^2+8\kappa^2\right){\cal F}_3(4\kappa^2|S|^2,\lambda^2|S|^2)\right.\\
 &\left.\null\hspace{2cm}+2g'^2g^2\left[M_1M_2\cos{(\varphi_{M_1}-\varphi_{M_2})}+\lambda^2|S|^2\right]{\cal F}_5(\lambda^2|S|^2,M_1^2,M_2^2)+\frac{g'^4}{2}\left(M_1^2+\lambda^2|S|^2\right)^2{\cal F}_7(M_1^2,\lambda^2|S|^2)\right.\\
 &\left.\null\hspace{2cm}+\frac{g^4}{2}\left(5M_2^4+5\lambda^4|S|^4+2M_2^2\lambda^2|S|^2\right){\cal F}_7(M_2^2,\lambda^2|S|^2)+2\lambda^4\left(\lambda^2+4\kappa^2\right)^2|S|^4{\cal F}_7(4\kappa^2|S|^2,\lambda^2|S|^2)\right.\\
 &\left.\null\hspace{2cm}+g'^2g^2\left[M_1^2M_2^2+\lambda^4|S|^4+2M_1M_2\lambda^2|S|^2\cos{(\varphi_{M_1}-\varphi_{M_2})}\right]{\cal F}_6(\lambda^2|S|^2,M_1^2,M_2^2)\right.\\
 &\left.\null\hspace{2cm}+2g'^2\lambda^2\left[\left(M_1^2+4\kappa^2|S|^2\right)\lambda^2|S|^2+4\kappa\lambda^2M_1\mbox{Re}\left(e^{\imath(\varphi_{M_1}+\varphi_{\kappa})}S^3\right)\right]{\cal F}_6(\lambda^2|S|^2,M_1^2,4\kappa^2|S|^2)\right.\\
 &\left.\null\hspace{2cm}+2g^2\lambda^2\left[\left(M_2^2+4\kappa^2|S|^2\right)\lambda^2|S|^2+4\kappa\lambda^2M_2\mbox{Re}\left(e^{\imath(\varphi_{M_2}+\varphi_{\kappa})}S^3\right)\right]{\cal F}_6(\lambda^2|S|^2,M_2^2,4\kappa^2|S|^2)\right\}
\end{align*}
\begin{align*}
 &\delta\lambda_3=\frac{-1}{32\pi^2}\left\{\frac{g'^4}{2}\ln\frac{M_1^2}{Q^2}+\frac{5g^4}{2}\ln\frac{M_2^2}{Q^2}+\left(2\lambda^4+\frac{g'^4+5g^4-2g'^2g^2}{2}\right)\ln\frac{\lambda^2|S|^2}{Q^2}+2\lambda^4\ln\frac{4\kappa^2|S|^2}{Q^2}-g'^2g^2{\cal F}_1(M_1^2,M_2^2)\right.\\
 &\left.\null\hspace{2cm}+g'^4\left(M_1^2+\lambda^2|S|^2\right){\cal F}_3(\lambda^2|S|^2,M_1^2)+g'^2\left[(g'^2-g^2)M_1^2-2\left(\lambda^2-2g^2\right)\lambda^2|S|^2\right]{\cal F}_3(M_1^2,\lambda^2|S|^2)\right.\\
 &\left.\null\hspace{2cm}+g^4\left(M_2^2+5\lambda^2|S|^2\right){\cal F}_3(\lambda^2|S|^2,M_2^2)+g^2\left[(-g'^2+5g^2)M_2^2+2\left(\lambda^2+2g'^2\right)\lambda^2|S|^2\right]{\cal F}_3(M_2^2,\lambda^2|S|^2)\right.\\
 &\left.\null\hspace{2cm}+4\lambda^4\left(\lambda^2+4\kappa^2\right)|S|^2{\cal F}_3(\lambda^2|S|^2,4\kappa^2|S|^2)+2\lambda^4|S|^2\left(8\kappa^2-g'^2+g^2\right){\cal F}_3(4\kappa^2|S|^2,\lambda^2|S|^2)\right.\\
 &\left.\null\hspace{2cm}-2g'^2g^2\left[M_1M_2\cos{(\varphi_{M_1}-\varphi_{M_2})}+\lambda^2|S|^2\right]{\cal F}_5(\lambda^2|S|^2,M_1^2,M_2^2)+\frac{g'^4}{2}\left(M_1^2+\lambda^2|S|^2\right)^2{\cal F}_7(M_1^2,\lambda^2|S|^2)\right.\\
 &\left.\null\hspace{2cm}+\frac{g^4}{2}\left(5M_2^4+5\lambda^4|S|^4+2M_2^2\lambda^2|S|^2\right){\cal F}_7(M_2^2,\lambda^2|S|^2)+2\lambda^4\left(\lambda^2+4\kappa^2\right)^2|S|^4{\cal F}_7(4\kappa^2|S|^2,\lambda^2|S|^2)\right.\\
 &\left.\null\hspace{2cm}-g'^2g^2\left[M_1^2M_2^2+\lambda^4|S|^4-2\lambda^2|S|^2(2M_1^2+2M_2^2-M_1M_2\cos(\varphi_{M_1}-\varphi_{M_2}))\right]{\cal F}_6(\lambda^2|S|^2,M_1^2,M_2^2)\right.\\
 &\left.\null\hspace{2cm}-2g'^2\lambda^2\left[\left(M_1^2+4\kappa^2|S|^2\right)\lambda^2|S|^2+4\kappa\lambda^2M_1\mbox{Re}\left(e^{\imath(\varphi_{M_1}+\varphi_{\kappa})}S^3\right)\right]{\cal F}_6(\lambda^2|S|^2,M_1^2,4\kappa^2|S|^2)\right.\\
 &\left.\null\hspace{2cm}+2g^2\lambda^2\left[\left(M_2^2+4\kappa^2|S|^2\right)\lambda^2|S|^2+4\kappa\lambda^2M_2\mbox{Re}\left(e^{\imath(\varphi_{M_2}+\varphi_{\kappa})}S^3\right)\right]{\cal F}_6(\lambda^2|S|^2,M_2^2,4\kappa^2|S|^2)\right\}
\end{align*}
\begin{align*}
 &\delta\lambda_4=\frac{-1}{32\pi^2}\left\{-2g^4\left[\ln\frac{M_2^2}{Q^2}+\ln\frac{\lambda^2|S|^2}{Q^2}\right]+2g'^2g^2\left[\ln\frac{\lambda^2|S|^2}{Q^2}+{\cal F}_1(M_1^2,M_2^2)\right]\right.\\
 &\left.\null\hspace{2cm}+g'^2\left[g^2M_1^2+\left(g'^2-g^2\right)\lambda^2|S|^2\right]{\cal F}_3(M_1^2,\lambda^2|S|^2)+4g^4\left(M_2^2-\lambda^2|S|^2\right){\cal F}_3(\lambda^2|S|^2,M_2^2)\right.\\
 &\left.\null\hspace{2cm}+2g^2\left[(-g'^2+2g^2)M_2^2-\left(2\lambda^2+g'^2-3g^2\right)\lambda^2|S|^2\right]{\cal F}_3(M_2^2,\lambda^2|S|^2)\right.\\
 &\left.\null\hspace{2cm}+4\lambda^4\left(2\lambda^2-g^2\right)|S|^2{\cal F}_3(4\kappa^2|S|^2,\lambda^2|S|^2)+4g'^2g^2\left[M_1M_2\cos{(\varphi_{M_1}-\varphi_{M_2})}+\lambda^2|S|^2\right]{\cal F}_5(\lambda^2|S|^2,M_1^2,M_2^2)\right.\\
 &\left.\null\hspace{2cm}+2g'^4M_1^2\lambda^2|S|^2{\cal F}_7(M_1^2,\lambda^2|S|^2)+2g^4\left(-M_2^4-\lambda^4|S|^4+5M_2^2\lambda^2|S|^2\right){\cal F}_7(M_2^2,\lambda^2|S|^2)\right.\\
 &\left.\null\hspace{2cm}+32\kappa^2\lambda^6|S|^4{\cal F}_7(4\kappa^2|S|^2,\lambda^2|S|^2)\right.\\
 &\left.\null\hspace{2cm}+2g'^2g^2\left[M_1^2M_2^2+\lambda^4|S|^4-\lambda^2|S|^2(M_1^2+M_2^2-2M_1M_2\cos(\varphi_{M_1}-\varphi_{M_2}))\right]{\cal F}_6(\lambda^2|S|^2,M_1^2,M_2^2)\right.\\
 &\left.\null\hspace{2cm}-4g^2\lambda^2\left[\left(M_2^2+4\kappa^2|S|^2\right)\lambda^2|S|^2+4\kappa\lambda^2M_2\mbox{Re}\left(e^{\imath(\varphi_{M_2}+\varphi_{\kappa})}S^3\right)\right]{\cal F}_6(\lambda^2|S|^2,M_2^2,4\kappa^2|S|^2)\right\}
\end{align*}
\begin{align*}
 &\delta\lambda_5=\frac{-2}{32\pi^2}\left\{g'^4M_1^2\lambda^2S^2e^{2\imath(\varphi_{M_1}+\varphi_{\lambda})}{\cal F}_7(M_1^2,\lambda^2|S|^2)+3g^4M_2^2\lambda^2S^2e^{2\imath(\varphi_{M_2}+\varphi_{\lambda})}{\cal F}_7(M_2^2,\lambda^2|S|^2)\right.\\
 &\left.\null\hspace{2cm}+16\kappa^2\lambda^6S^{*\,4}e^{2\imath(\varphi_{\lambda}-\varphi_{\kappa})}{\cal F}_7(4\kappa^2|S|^2,\lambda^2|S|^2)+2g'^2g^2M_1M_2\lambda^2S^2e^{\imath(\varphi_{M_1}+\varphi_{M_2}+2\varphi_{\lambda})}{\cal F}_6(\lambda^2|S|^2,M_1^2,M_2^2)\right\}\\
 &\delta\lambda_{6,7}=\frac{-1}{32\pi^2}\left\{-g'^4M_1\lambda Se^{\imath(\varphi_{M_1}+\varphi_{\lambda})}{\cal F}_3(\lambda^2|S|^2,M_1^2)-g'^2(g'^2+g^2)M_1\lambda Se^{\imath(\varphi_{M_1}+\varphi_{\lambda})}{\cal F}_3(M_1^2,\lambda^2|S|^2)\right.\\
 &\left.\null\hspace{2cm}-3g^4M_2\lambda Se^{\imath(\varphi_{M_2}+\varphi_{\lambda})}{\cal F}_3(\lambda^2|S|^2,M_2^2)-g^2(g'^2+3g^2)M_2\lambda Se^{\imath(\varphi_{M_2}+\varphi_{\lambda})}{\cal F}_3(M_2^2,\lambda^2|S|^2)\right.\\
 &\left.\null\hspace{2cm}+8\kappa\lambda^5S^{*\,2}e^{\imath(\varphi_{\lambda}-\varphi_{\kappa})}\left[{\cal F}_3(\lambda^2|S|^2,4\kappa^2|S|^2)+{\cal F}_3(4\kappa^2|S|^2,\lambda^2|S|^2)\right]\right.\\
 &\left.\null\hspace{2cm}-g'^2g^2\lambda S\left[M_1e^{\imath(\varphi_{M_1}+\varphi_{\lambda})}+M_2e^{\imath(\varphi_{M_2}+\varphi_{\lambda})}\right]{\cal F}_5(\lambda^2|S|^2,M_1^2,M_2^2)\right.\\
 &\left.\null\hspace{2cm}-g'^4\left(M_1^2+\lambda^2|S|^2\right)M_1\lambda Se^{\imath(\varphi_{M_1}+\varphi_{\lambda})}{\cal F}_7(M_1^2,\lambda^2|S|^2)-3g^4\left(M_2^2+\lambda^2|S|^2\right)M_2\lambda Se^{\imath(\varphi_{M_2}+\varphi_{\lambda})}{\cal F}_7(M_2^2,\lambda^2|S|^2)\right.\\
 &\left.\null\hspace{2cm}+8\kappa\lambda^5\left(\lambda^2+4\kappa^2\right)|S|^2S^{*\,2}e^{\imath(\varphi_{\lambda}-\varphi_{\kappa})}{\cal F}_7(4\kappa^2|S|^2,\lambda^2|S|^2)\right.\\
 &\left.\null\hspace{2cm}-g'^2g^2\lambda S\left[\left(M_1^2+\lambda^2|S|^2\right)M_2e^{\imath(\varphi_{M_2}+\varphi_{\lambda})}+\left(M_2^2+\lambda^2|S|^2\right)M_1e^{\imath(\varphi_{M_1}+\varphi_{\lambda})}\right]{\cal F}_6(\lambda^2|S|^2,M_1^2,M_2^2)\right\}
\end{align*}}

\subsubsection{Higgs-to-Higgs contributions}\label{aploophihiggs}
Instead of diagonalizing the Higgs bilinear terms, we compute the Higgs self-energy ($\Pi$) and tadpole ($T$) diagrams mediated by gauge and Higgs particles in the 
Feynmann gauge. We then set the external momentum to zero to determine the potential contributions to the Higgs mass matrices and subtract the pure
gauge effects in the Landau gauge -- from the results in appendix \ref{aploopsmgauge}. Finally we identify the corrections to the $\mathbb{Z}_3$-conserving parameters of 
Eq.\ref{potential}.

\vspace{0.2cm}
{\em i) Pure-gauge contributions to the Higgs self-energies and tadpoles}\newline
For external neutral Higgs:
\begin{displaymath}
 \begin{cases}
  \Pi^{V}_{S^0_iS^0_j}(p^2)=\frac{1}{16\pi^2}\left\{\frac{7}{2}\left[g^2M_W^2B_0(p,M_W,M_W)+\frac{g'^2+g^2}{2}M_Z^2B_0(p,M_Z,M_Z)\right]\right.\\
\null\hspace{4cm}\times\left[\sin^2\beta X_{iu}^RX_{ju}^R+\cos^2\beta X_{id}^RX_{jd}^R+\sin\beta\cos\beta (X_{iu}^RX_{jd}^R+X_{id}^RX_{ju}^R)\right]\\
\null\hspace{2.8cm}\left.+2\left[g^2A_0(M_W)+\frac{g'^2+g^2}{2}A_0(M_Z)\right]\left[X_{iu}^RX_{ju}^R+X_{id}^RX_{jd}^R+X_{iu}^IX_{ju}^I+X_{id}^IX_{jd}^I\right]\right\}\\
T^V_{S^0_i}=\frac{1}{16\pi^2}\frac{3}{2}\left[g^2A_0(M_W)+\frac{g'^2+g^2}{2}A_0(M_Z)\right]\left[v_uX_{iu}^R+v_dX_{id}^R\right]
 \end{cases}
\end{displaymath}
For an external charged Higgs (we consider only the physical state):
\begin{displaymath}
  \Pi^{V}_{H^+H^-}(p^2)=\frac{2}{16\pi^2}\left[g^2A_0(M_W)+\frac{(g'^2-g^2)^2}{2(g'^2+g^2)}A_0(M_Z)\right]
\end{displaymath}

\vspace{0.2cm}
{\em ii) Higgs / gauge diagrams}\newline
The neutral self energy receives contributions from hybrid Higgs (Goldstone) / vector diagrams:
{\small\begin{align*}\textstyle
  \Pi^{SV}_{S^0_iS^0_j}(p^2)=&\frac{1}{32\pi^2}\left\{\left[g^2B_{SV}(p,M_W,M_W)+\frac{g'^2+g^2}{2}B_{SV}(p,M_Z,M_Z)\right]\right.\\
&\null\hspace{3cm}\times\left[\sin^2\beta X_{iu}^RX_{ju}^R+\cos^2\beta X_{id}^RX_{jd}^R+\sin\beta\cos\beta (X_{iu}^RX_{jd}^R+X_{id}^RX_{ju}^R)\right]\\
&\null\hspace{1cm}+g^2B_{SV}(p,M_W,M_W)\left[\sin^2\beta X_{iu}^IX_{ju}^I+\cos^2\beta X_{id}^IX_{jd}^I-\sin\beta\cos\beta (X_{iu}^IX_{jd}^I+X_{id}^IX_{ju}^I)\right]\\
&\null\hspace{1cm}+g^2B_{SV}(p,m_{H^{\pm}},M_W)\left[\cos^2\beta (X_{iu}^RX_{ju}^R+X_{iu}^IX_{ju}^I)+\sin^2\beta (X_{id}^RX_{jd}^R+X_{id}^IX_{jd}^I)\right.\\
&\null\hspace{5cm}\left.+\sin\beta\cos\beta (X_{iu}^RX_{jd}^R+X_{id}^RX_{ju}^R-X_{iu}^IX_{jd}^I-X_{id}^IX_{ju}^I)\right]\\
&\null\hspace{1cm}+\frac{g'^2+g^2}{2}\sum_{k=1}^5B_{SV}(p,m_{S^0_k},M_Z)\times\\
&\null\hspace{2cm}\left(X_{iu}^IX_{ku}^R-X_{iu}^RX_{ku}^I+X_{id}^RX_{kd}^I-X_{id}^IX_{kd}^R\right)\left(X_{ju}^IX_{ku}^R-X_{ju}^RX_{ku}^I+X_{jd}^RX_{kd}^I-X_{jd}^IX_{kd}^R\right)\bigg\}
\end{align*}}
For the charged Higgs self-energy:
{\small\begin{align*}\textstyle
  \Pi^{SV}_{H^+H^-}(p^2)=&\frac{1}{32\pi^2}\left\{\frac{2g'^2g^2}{g'^2+g^2}B_{SV}(p,m_{H^{\pm}},0)+\frac{(g'^2-g^2)^2}{2(g'^2+g^2)}B_{SV}(p,m_{H^{\pm}},M_Z)+\frac{g^2}{2}\sum_{k=1}^5B_{SV}(p,m_{S^0_k},M_W)\right\}\\
&\null\hspace{2cm}\times\left[\cos^2\beta(X_{ku}^{R\,2}+ X_{ku}^{I\,2})+\sin^2\beta (X_{kd}^{R\,2}+ X_{kd}^{I\,2})-\sin\beta\cos\beta(X_{ku}^RX_{kd}^R-X_{ku}^IX_{kd}^I)\right]
\end{align*}}

\vspace{0.2cm}
{\em iii) Pure Higgs loops}\newline
The loops including only Higgs bosons (including the Goldstone bosons, with mass $M_W$ and $M_Z$) read:
{\begin{align*}\textstyle
\Pi^{S}_{S^0_iS^0_j}(p^2)=&\frac{1}{16\pi^2}\left\{\sum_{m,n=1}^2g^{S^0_iH_m^+H_n^-}g^{S^0_jH_n^+H_m^-}B_0(p,m_{H_m^{\pm}},m_{H_n^{\pm}})+\sum_{m=1}^2g^{S^0_iS^0_jH^+_mH^-_m}A_0(m_{H_m^{\pm}})\right.\\
&\null\hspace{1cm}\left.+\frac{1}{2}\sum_{m,n=1}^6g^{S^0_iS^0_mS^0_n}g^{S^0_jS^0_nS^0_m}B_0(p,m_{S^0_m},m_{S^0_n})+\frac{1}{2}\sum_{m=1}^6g^{S^0_iS^0_jS^0_mS^0_m}A_0(m_{S^0_m})\right\}\\
\Pi^{S}_{H^+H^-}(p^2)=&\frac{1}{16\pi^2}\left\{\sum_{m=1}^2g^{H^+H^-H^+_mH^-_m}A_0(m_{H_m^{\pm}})+\frac{1}{2}\sum_{m=1}^6g^{S^0_mS^0_mH^+H^-}A_0(m_{S^0_m})\right.\\
&\null\hspace{4cm}\left.+\sum_{m=1,6}^{n=1,2}g^{S^0_mH^+H_n^-}g^{S^0_mH^+_nH^-}B_0(p,m_{S^0_m},m_{H^{\pm}_n})\right\}\\
T^S_{S^0_i}=&\frac{1}{16\sqrt{2}\pi^2}\left\{\sum_{m=1}^2g^{S^0_iH^+_mH^-_m}A_0(m_{H_m^{\pm}})+\frac{1}{2}\sum_{m=1}^6g^{S^0_iS^0_mS^0_m}A_0(m_{S^0_m})\right\}
\end{align*}}
where the Higgs-to-Higgs couplings can be found in appendix \ref{apHi2Hicoup}.

\vspace{0.2cm}
{\em iv) Contributions to the Higgs mass-matrices}\newline
Since we are interested in the contributions from the effective potential, we take the limit $p^2=0$, which simply induces the replacement 
$B_0(p,m,M)\to-{\cal F}_1(m^2,M^2)$. The contribution to the mass matrices of the Higgs states reads:
\begin{displaymath}
 \delta\left<{\cal M}^2_{H^0}\right>_{ij}=-\left[\Pi^{V+SV+S}_{S_i^0S_j^0}(0)-E_{ijk}\,T^{V+S}_{S_k^0}\right]\hspace{0.7cm};\hspace{0.7cm}\delta m^2_{H^{\pm}}=-\left[\Pi^{V+SV+S}_{H^+H^-}(0)-\cos^2\beta\, T_{h_u^0}^{V+S}-\sin^2\beta\, T_{h_d^0}^{V+S}\right]
\end{displaymath}
where the coefficients $E_{ijk}$ are the same as in appendix \ref{aploophisferm}, that is, they correspond to the tadpole coefficients of Eq.\ref{masstadpole}.

\vspace{0.2cm}
{\em v) Reconstruction of the Higgs contributions to the potential}\newline
After subtracting the pure gauge contributions from appendix \ref{aploopsmgauge}, one can reconstruct the $\mathbb{Z}_3$-conserving parameters of the 
potential of Eq.\ref{potential} induced by Higgs corrections. We start by rotating away the neutral Goldstone boson, obtaining thus a $5\times5$ matrix
for the Higgs corrections to the neutral Higgs mass-matrix: $\delta\left<\tilde{\cal M}^2_{H^0}\right>_{ij}$. We then employ the method that was 
outlined in \cite{Chalons:2012qe}:
\begin{align*}
&\delta A_S\cos\varphi_2=-\frac{1}{3s}\left\{\delta\left<\tilde{\cal M}^2_{H^0}\right>_{55}+\frac{v}{2s}\sin{2\beta}\left[\delta\left<\tilde{\cal M}^2_{H^0}\right>_{45}-\frac{v}{s}\sin{2\beta}\,\delta\left<\tilde{\cal M}^2_{H^0}\right>_{44}\right]\right\}\\
&\delta {\cal V}_0(|S|^2)=\frac{1}{4s^2}\left\{\delta\left<\tilde{\cal M}^2_{H^0}\right>_{33}+\frac{1}{3}\delta\left<\tilde{\cal M}^2_{H^0}\right>_{55}-\frac{v^2}{3s^2}\sin^2{2\beta}\,\delta\left<\tilde{\cal M}^2_{H^0}\right>_{44}\right\}|S|^4\\
&\delta A_{ud}\cos\varphi_{A_{ud}}=\frac{1}{3v}\left\{\delta\left<\tilde{\cal M}^2_{H^0}\right>_{45}+\frac{v}{s}\sin{2\beta}\,\delta\left<\tilde{\cal M}^2_{H^0}\right>_{44}\right\}\\
&\delta \lambda_P^M\cos\varphi_M=-\frac{1}{3sv}\left\{\delta\left<\tilde{\cal M}^2_{H^0}\right>_{45}-\frac{v}{2s}\sin{2\beta}\,\delta\left<\tilde{\cal M}^2_{H^0}\right>_{44}\right\}\\
&\delta \lambda_P^M\sin\varphi_M=\frac{2}{3v^2\sin{2\beta}}\left\{\delta\left<\tilde{\cal M}^2_{H^0}\right>_{35}-\frac{v}{s}\sin{2\beta}\,\delta\left<\tilde{\cal M}^2_{H^0}\right>_{34}\right\}\\
&\delta \lambda_P^u=\frac{1}{2sv_u}\left\{\delta\left<\tilde{\cal M}^2_{H^0}\right>_{13}-\frac{\cos\beta}{3}\left[\delta\left<\tilde{\cal M}^2_{H^0}\right>_{45}-\frac{2v}{s}\sin{2\beta}\,\delta\left<\tilde{\cal M}^2_{H^0}\right>_{44}\right]\right\}\\
&\delta \lambda_P^d=\frac{1}{2sv_d}\left\{\delta\left<\tilde{\cal M}^2_{H^0}\right>_{23}-\frac{\sin\beta}{3}\left[\delta\left<\tilde{\cal M}^2_{H^0}\right>_{45}-\frac{2v}{s}\sin{2\beta}\,\delta\left<\tilde{\cal M}^2_{H^0}\right>_{44}\right]\right\}\\
&\delta \lambda_u=\frac{1}{2v_u^2}\left\{\delta\left<\tilde{\cal M}^2_{H^0}\right>_{11}-\cos^2\beta\,\delta\left<\tilde{\cal M}^2_{H^0}\right>_{44}\right\}\\
&\delta \lambda_d=\frac{1}{2v_d^2}\left\{\delta\left<\tilde{\cal M}^2_{H^0}\right>_{22}-\sin^2\beta\,\delta\left<\tilde{\cal M}^2_{H^0}\right>_{44}\right\}\\
&\delta \lambda_3=\frac{1}{2v_uv_d}\left\{\delta\left<\tilde{\cal M}^2_{H^0}\right>_{12}+\sin2\beta\left[\delta m^2_{H^{\pm}}-\frac{1}{2}\delta\left<\tilde{\cal M}^2_{H^0}\right>_{44}\right]\right\}\\
&\delta \lambda_4=\frac{1}{v^2}\left\{\delta\left<\tilde{\cal M}^2_{H^0}\right>_{44}-\delta m^2_{H^{\pm}}\right\}
\end{align*}

\subsection{\boldmath Leading two-loop effects $O(Y_{t,b}^4\alpha_{(S)},Y_{t,b}^6)$}\label{aptwoloop}
We follow \cite{NMSSM,Ellwanger:1999ji}:
\begin{align*}
 \delta\lambda_u=&\frac{3Y_t^4}{256\pi^4}\left\{\left[16g_3^2+\frac{4}{3}g'^2-3\sin^2\beta Y_t^2+3\cos^2\beta Y_b^2\right]\ln^2\frac{Q^2}{m_t^2}\right.\\
 &\null\hspace{5cm}\left.+\left[3\cos^2\beta Y_t^2+(3\cos^2\beta+1)Y_b^2\right]\left(\ln^2\frac{M_A^2}{m_t^2}-\ln^2\frac{Q^2}{m_t^2}\right)\right\}\\
 \delta\lambda_d=&\frac{3Y_b^4}{256\pi^4}\left\{\left[16g_3^2-\frac{2}{3}g'^2+3\sin^2\beta Y_t^2-3\cos^2\beta Y_b^2\right]\ln^2\frac{Q^2}{m_t^2}\right.\\
 &\left.\null\hspace{5cm}+\left[3\sin^2\beta Y_b^2+(3\sin^2\beta+1)Y_t^2\right]\left(\ln^2\frac{M_A^2}{m_t^2}-\ln^2\frac{Q^2}{m_t^2}\right)\right\}
\end{align*}
Note that these leading effects are conveyed by the SM-fermion and gauge sector, as well as the doublet Higgs sector, so that the new-physics phases do
not intervene. The contribution of sfermions or gauginos is merely reduced to the cutoff $Q$ in the logarithms.

\subsection{Pole corrections}\label{appolecorr}
Here we compute the shifts in the Higgs self-energies, which then allow to evaluate the pole corrections to the $\overline{DR}$ Higgs masses. We use the 
notation $\Delta f (p^2)\equiv f(p^2)-f(0)$ for any function $f$ of the external momentum. We are still working in the Feynmann gauge.

\vspace{0.2cm}
{\em i) Contributions from SM fermions}
\begin{align*}
 \Delta\Pi_{S_i^0S_j^0}(p^2)=&\frac{1}{16\pi^2}\left\{N_cY_u^2\left[(X_{iu}^RX_{ju}^R+X_{iu}^IX_{ju}^I)\Delta B_{FF}(p,m_u,m_u)-2m_u^2(X_{iu}^RX_{ju}^R-X_{iu}^IX_{ju}^I)\Delta B_0(p,m_u,m_u)\right]\right.\\
 &\null\hspace{0.7cm}+N_cY_d^2\left[(X_{id}^RX_{jd}^R+X_{id}^IX_{jd}^I)\Delta B_{FF}(p,m_d,m_d)-2m_d^2(X_{id}^RX_{jd}^R-X_{id}^IX_{jd}^I)\Delta B_0(p,m_d,m_d)\right]\\
 &\null\hspace{1cm}\left.+Y_e^2\left[(X_{id}^RX_{jd}^R+X_{id}^IX_{jd}^I)\Delta B_{FF}(p,m_e,m_e)-2m_e^2(X_{id}^RX_{jd}^R-X_{id}^IX_{jd}^I)\Delta B_0(p,m_e,m_e)\right]\right\}\\
 \Delta\Pi_{H^+H^-}(p^2)=&\frac{1}{16\pi^2}\left\{N_C\left[(Y_u^2\cos^2\beta+Y_d^2\sin^2\beta)\Delta B_{FF}(p,m_u,m_d)-2Y_uY_dm_um_d\sin{2\beta}\,\Delta B_0(p,m_u,m_d)\right]\right.\\
 &\null\hspace{1cm}\left.+Y_e^2\sin^2\beta\,\Delta B_{FF}(p,0,m_e)\right\}
\end{align*}

\vspace{0.2cm}
{\em ii) Contributions from gauginos and higgsinos}
\begin{align*}
 \Delta\Pi_{S_i^0S_j^0}(p^2)=&\frac{1}{16\pi^2}\left\{\left[\frac{g'^2}{2}\Delta B_{FF}(p,M_1,\mu)+\frac{3g^2}{2}\Delta B_{FF}(p,M_2,\mu)+\lambda^2\Delta B_{FF}(p,m_{\tilde{s}},\mu)\right]\right.\\
&\null\hspace{6cm}\times(X_{iu}^RX_{ju}^R+X_{id}^RX_{jd}^R+X_{iu}^IX_{ju}^I+X_{id}^IX_{jd}^I)\\
&\null\hspace{1cm}+2\left[\lambda^2\Delta B_{FF}(p,\mu,\mu)+\kappa^2\Delta B_{FF}(p,m_{\tilde{s}},m_{\tilde{s}})\right](X_{is}^RX_{js}^R+X_{is}^IX_{js}^I)\bigg\}\\
 \Delta\Pi_{H^+H^-}(p^2)=&\frac{1}{16\pi^2}\left\{\frac{g'^2}{2}\Delta B_{FF}(p,M_1,\mu)+\frac{3g^2}{2}\Delta B_{FF}(p,M_2,\mu)+\lambda^2\Delta B_{FF}(p,m_{\tilde{s}},\mu)\right\}
\end{align*}

\vspace{0.2cm}
{\em iii) Contributions exclusively from the electroweak gauge sector}
\begin{align*}
 \Delta\Pi_{S_i^0S_j^0}(p^2)=&\frac{1}{16\pi^2}\frac{7}{2}\left\{g^2M_W^2\Delta B_0(p,M_W,M_W)+\frac{g'^2+g^2}{2}M_Z^2\Delta B_0(p,M_Z,M_Z)\right\}\\
&\null\hspace{3cm}\times\left[\sin^2\beta X_{iu}^RX_{ju}^R+\cos^2\beta X_{id}^RX_{jd}^R+\sin\beta\cos\beta(X_{iu}^RX_{jd}^R+X_{id}^RX_{ju}^R)\right]\\
 \Delta\Pi_{H^+H^-}(p^2)=&0
\end{align*}

\vspace{0.2cm}
{\em iv) Contributions from the gauge / Higgs diagrams}
{\small\begin{align*}
 \Delta\Pi_{S_i^0S_j^0}(p^2)=&\frac{1}{32\pi^2}\left\{\left[g^2\Delta B_{SV}(p,M_W,M_W)+\frac{g'^2+g^2}{2}\Delta B_{SV}(p,M_Z,M_Z)\right]\right.\\
&\null\hspace{3cm}\times\left[\sin^2\beta X_{iu}^RX_{ju}^R+\cos^2\beta X_{id}^RX_{jd}^R+\sin\beta\cos\beta (X_{iu}^RX_{jd}^R+X_{id}^RX_{ju}^R)\right]\\
&\null\hspace{0.8cm}+g^2\Delta B_{SV}(p,M_W,M_W)\left[\sin^2\beta X_{iu}^IX_{ju}^I+\cos^2\beta X_{id}^IX_{jd}^I-\sin\beta\cos\beta (X_{iu}^IX_{jd}^I+X_{id}^IX_{ju}^I)\right]\\
&\null\hspace{0.8cm}+g^2\Delta B_{SV}(p,m_{H^{\pm}},M_W)\left[\cos^2\beta (X_{iu}^RX_{ju}^R+X_{iu}^IX_{ju}^I)+\sin^2\beta (X_{id}^RX_{jd}^R+X_{id}^IX_{jd}^I)\right.\\
&\null\hspace{5cm}\left.+\sin\beta\cos\beta (X_{iu}^RX_{jd}^R+X_{id}^RX_{ju}^R-X_{iu}^IX_{jd}^I-X_{id}^IX_{ju}^I)\right]\\
&\null\hspace{0.8cm}+\frac{g'^2+g^2}{2}\sum_{k=1}^5\Delta B_{SV}(p,m_{S^0_k},M_Z)\times\\
&\null\hspace{1.5cm}\left(X_{iu}^IX_{ku}^R-X_{iu}^RX_{ku}^I+X_{id}^RX_{kd}^I-X_{id}^IX_{kd}^R\right)\left(X_{ju}^IX_{ku}^R-X_{ju}^RX_{ku}^I+X_{jd}^RX_{kd}^I-X_{jd}^IX_{kd}^R\right)\bigg\}\\
 \Delta\Pi_{H^+H^-}(p^2)=&\frac{1}{32\pi^2}\left\{\frac{2g'^2g^2}{g'^2+g^2}\Delta B_{SV}(p,m_{H^{\pm}},0)+\frac{(g'^2-g^2)^2}{2(g'^2+g^2)}\Delta B_{SV}(p,m_{H^{\pm}},M_Z)+\frac{g^2}{2}\sum_{k=1}^5\Delta B_{SV}(p,m_{S^0_k},M_W)\right\}\\
&\null\hspace{2cm}\times\left[\cos^2\beta(X_{ku}^{R\,2}+ X_{ku}^{I\,2})+\sin^2\beta (X_{kd}^{R\,2}+ X_{kd}^{I\,2})-\sin\beta\cos\beta(X_{ku}^RX_{kd}^R-X_{ku}^IX_{kd}^I)\right]
\end{align*}}

\vspace{0.2cm}
{\em v) Contributions exclusively from the Higgs / Goldstone sector}
{\small\begin{align*}
\Delta\Pi_{S^0_iS^0_j}(p^2)=&\frac{1}{16\pi^2}\left\{\sum_{m,n=1}^2g^{S^0_iH_m^+H_n^-}g^{S^0_jH_n^+H_m^-}\Delta B_0(p,m_{H_m^{\pm}},m_{H_n^{\pm}})+\frac{1}{2}\sum_{m,n=1}^6g^{S^0_iS^0_mS^0_n}g^{S^0_jS^0_nS^0_m}\Delta B_0(p,m_{S^0_m},m_{S^0_n})\right\}\\
\Delta\Pi_{H^+H^-}(p^2)=&\frac{1}{16\pi^2}\sum_{m=1,6}^{n=1,2}g^{S^0_mH^+H_n^-}g^{S^0_mH^+_nH^-}\Delta B_0(p,m_{S^0_m},m_{H^{\pm}_n})
\end{align*}}
The tree-level Higgs-to-Higgs couplings are given in appendix \ref{apHi2Hicoup}.

\vspace{0.2cm}
{\em vi) Contributions from the sfermions}
{\small\begin{align*}
\Delta\Pi_{S^0_iS^0_j}(p^2)=&\frac{1}{16\pi^2}\left\{\sum_{m,n=1}^2\left[N_cg^{S^0_i\tilde{U}_m^*\tilde{U}_n}g^{S^0_j\tilde{U}_n^*\tilde{U}_m}\Delta B_0(p,m_{\tilde{U}_m},m_{\tilde{U}_n})+N_cg^{S^0_i\tilde{D}_m^*\tilde{D}_n}g^{S^0_j\tilde{D}_n^*\tilde{D}_m}\Delta B_0(p,m_{\tilde{D}_m},m_{\tilde{D}_n})\right.\right.\\
&\null\hspace{2.5cm}\left.+g^{S^0_i\tilde{E}_m^*\tilde{E}_n}g^{S^0_j\tilde{E}_n^*\tilde{E}_m}\Delta B_0(p,m_{\tilde{E}_m},m_{\tilde{E}_n})\right]+g^{S^0_i\tilde{N}^*\tilde{N}}g^{S^0_j\tilde{N}^*\tilde{N}}\Delta B_0(p,m_{\tilde{N}},m_{\tilde{N}})\bigg\}\\
\Delta\Pi_{H^+H^-}(p^2)=&\frac{1}{16\pi^2}\left\{\sum_{m,n=1}^2N_cg^{H^+\tilde{U}_m^*\tilde{D}_n}g^{H^-\tilde{D}_n^*\tilde{U}_m}\Delta B_0(p,m_{\tilde{U}_m},m_{\tilde{D}_n})+\sum_{m=1}^2g^{H^+\tilde{N}^*\tilde{E}_m}g^{H^-\tilde{E}^*_m\tilde{N}}\Delta B_0(p,m_{\tilde{N}},m_{\tilde{E}_m})\right\}
\end{align*}}
The Higgs sfermion couplings are given in appendix \ref{apHiSFcoup}.

\section{Simplified effective potential}\label{apsimppot}
In this appendix, we study the simplified effective Higgs potential of Eq.\ref{potential}, or more precisely the following and slightly modified
version:
\begin{align}
 \tilde{\cal V}_{\mbox{\tiny eff}}= &M_S^2|S|^2+\frac{A_S}{3}\left[e^{\imath\varphi_{A_S}}S^3+h.c.\right]+{\cal V}_0(|S|^2)\label{simppot}\\
\null & +(M_u^2+\lambda_P^u|S|^2)|H_u|^2+(M^2_d+\lambda_P^d|S|^2)|H_d|^2+\left[\left(A_{ud} e^{\imath\varphi_{A_{ud}}}S+\lambda_P^Me^{\imath\varphi_M}S^{*2}\right)H_u\cdot H_d+h.c.\right]\nonumber\\
\null & +\frac{\lambda_u}{2}|H_u|^4+\frac{\lambda_d}{2}|H_d|^4+ \lambda_3|H_u|^2|H_d|^2+\lambda_4|H_u\cdot H_d|^2\nonumber\\
\null & +\left[\frac{\lambda_5}{2}e^{\imath\varphi_5}\frac{S^2}{s^2}(H_u\cdot H_d)^2+(\lambda_6e^{\imath \varphi_6}|H_u|^2+\lambda_7e^{\imath \varphi_7}|H_d|^2)\frac{S}{s}H_u\cdot H_d+h.c.\right]\nonumber
\end{align}
This simplified potential is meant as an expansion of the effective potential -- see Eq.\ref{effpot} -- up to quartic order in the doublet fields. It
slightly differs from Eq.\ref{potential} in that the $\mathbb{Z}_3$-symmetry has been explicitly restored in the terms of the last line. Note that this way
of restoring the $\mathbb{Z}_3$-symmetry is just the simplest educated guess, while any additional factor $f(|S|^2,S^3,S^{*3})$ could intervene. Therefore,
the factors of $S/s$ appearing in the last line are just chosen as such because they will provide improved results numerically. Formally
however, the associated corrections will remain of subleading order in the expansion in the doublet v.e.v.'s.

\subsection{Matching the tree-level Higgs potential}\label{apsimppotmatchtree}
The tree-level Higgs potential (Eq.\ref{tlHiggspot}) matches straightforwardly on Eq.\ref{simppot}:
\begin{align*}
 & M_S^2=m_{S}^2 &  & M_u^2=m_{H_u}^2 &  & \lambda_u=\frac{g'^2+g^2}{4}=\lambda_d\\
 & A_{S} e^{\imath\varphi_{A_{S}}}=\kappa A_{\kappa}e^{\imath\varphi_{2}}  &  & M_d^2=m_{H_d}^2 &  & \lambda_3=\frac{-g'^2+g^2}{4}\\
 & {\cal V}_0(|S|^2)=\kappa^2|S|^4 &  & \lambda_P^u=\lambda^2=\lambda_P^d &  & \lambda_4=\lambda^2-\frac{g^2}{2}\\
 & &  & A_{ud} e^{\imath\varphi_{A_{ud}}}=\lambda A_{\lambda}e^{\imath\varphi_{1}} &  & \lambda_5e^{\imath\varphi_5}=0=\lambda_6e^{\imath\varphi_6}=\lambda_7e^{\imath\varphi_7}\\
 &   & & \lambda_P^M=\kappa\lambda e^{\imath(\varphi_{\lambda}-\varphi_{\kappa})}&  &  
\end{align*}

\subsection{Minimization conditions}
\begin{align*}
 &M_u^2=\left[\left(A_{ud}\cos\varphi_{A_{ud}}+\lambda_P^Ms\cos\varphi_{M}\right)s+3\lambda_6\cos\varphi_6v_u^2+\lambda_7\cos\varphi_7v_d^2\right]\frac{v_d}{v_u}-\left[\lambda_P^us^2+\lambda_uv_u^2+(\lambda_3+\lambda_4+\lambda_5\cos\varphi_5)v_d^2\right]\\
 &M_d^2=\left[\left(A_{ud}\cos\varphi_{A_{ud}}+\lambda_P^Ms\cos\varphi_{M}\right)s+\lambda_6\cos\varphi_6v_u^2+3\lambda_7\cos\varphi_7v_d^2\right]\frac{v_u}{v_d}-\left[\lambda_P^ds^2+\lambda_dv_d^2+(\lambda_3+\lambda_4+\lambda_5\cos\varphi_5)v_u^2\right]\\
 &M_s^2=-\left<{\cal V}_0'\right>-A_Ss\cos\varphi_{A_S}-\lambda_P^uv_u^2-\lambda_P^dv_d^2+\left(A_{ud}\cos\varphi_{A_{ud}}+2\lambda_P^Ms\cos\varphi_M\right)\frac{v_uv_d}{s}\\
 &A_{ud}\sin\varphi_{A_{ud}}=-\lambda_P^Ms\sin\varphi_M+\frac{1}{s}\left[\lambda_5\sin\varphi_5v_uv_d-\lambda_6\sin\varphi_6v_u^2-\lambda_7\sin\varphi_7v_d^2\right]\\
 &A_S\sin\varphi_{A_S}=\left(A_{ud}\sin\varphi_{A_{ud}}-2\lambda_P^Ms\sin\varphi_M\right)\frac{v_uv_d}{s^2}\simeq-3\lambda_P^M\sin\varphi_M\frac{v_uv_d}{s}
\end{align*}
$O(v^4)$ terms have been neglected. $\left<{\cal V}_0'\right>\equiv\left.\frac{d{\cal V}_0}{d|S|^2}\right|_{|S|^2=s^2}$.

\subsection{Higgs mass matrices}\label{asimppotmass}
Charged Higgs in the base $(H^{\pm}_u,H_d^{\pm})$:
\begin{multline}
 \left<{\cal M}^2_{H^{\pm}}\right>_{11}=\Omega_{\pm}\frac{v_d}{v_u}\ \ \ \ \ \ \ \ \ \ \ \ \left<{\cal M}^2_{H^{\pm}}\right>_{22}=\Omega_{\pm}\frac{v_u}{v_d}\ \ \ \ \ \ \ \ \ \ \ \ \left<{\cal M}^2_{H^{\pm}}\right>_{12}=\Omega_{\pm}=\left<{\cal M}^2_{H^{\pm}}\right>_{21}\\
\Omega_{\pm}\equiv\mbox{Re}\left(A_{ud}\cos\varphi_{A_{ud}}+\lambda_P^Ms\cos\varphi_{M}\right)s-\left(\lambda_4+\lambda_5\cos\varphi_5\right)v_uv_d+\lambda_6\cos\varphi_6v_u^2+\lambda_7\cos\varphi_7v_d^2\label{cHiggsmass}
\end{multline}
Neutral Higgs in the base $(h_u^0,h_d^0,h_s^0,a_u^0,a_d^0,a_s^0)$; $O(v^4)$ terms are neglected.:
\begin{align*}
  \left<{\cal M}^2_{H^0}\right>_{11}=&\left[\left(A_{ud}\cos\varphi_{A_{ud}}+\lambda_P^Ms\cos\varphi_{M}\right)s-3\lambda_6\cos\varphi_6v_u^2+\lambda_7\cos\varphi_7v_d^2\right]\frac{v_d}{v_u}+2\lambda_uv_u^2\nonumber\\
  \left<{\cal M}^2_{H^0}\right>_{12}=&-\left[\left(A_{ud}\cos\varphi_{A_{ud}}+\lambda_P^Ms\cos\varphi_{M}\right)s+3\lambda_6\cos\varphi_6v_u^2+3\lambda_7\cos\varphi_7v_d^2\right]+2\left[\lambda_3+\lambda_4+\lambda_5\cos\varphi_5\right]v_uv_d\\
  \left<{\cal M}^2_{H^0}\right>_{22}=&\left[\left(A_{ud}\cos\varphi_{A_{ud}}+\lambda_P^Ms\cos\varphi_{M}\right)s+\lambda_6\cos\varphi_6v_u^2-3\lambda_7\cos\varphi_7v_d^2\right]\frac{v_u}{v_d}+2\lambda_dv_d^2\nonumber\\
  \left<{\cal M}^2_{H^0}\right>_{13}=& -\left[A_{ud}\cos\varphi_{A_{ud}}+2\lambda_P^Ms\cos\varphi_{M}+\frac{1}{s}\left(3\lambda_6\cos\varphi_6v_u^2+\lambda_7\cos\varphi_7v_d^2-2\lambda_5\cos\varphi_5v_uv_d\right)\right]v_d+2\lambda_P^usv_u\nonumber\\
  \left<{\cal M}^2_{H^0}\right>_{23}=& -\left[A_{ud}\cos\varphi_{A_{ud}}+2\lambda_P^Ms\cos\varphi_{M}+\frac{1}{s}\left(\lambda_6\cos\varphi_6v_u^2+3\lambda_7\cos\varphi_7v_d^2-2\lambda_5\cos\varphi_5v_uv_d\right)\right]v_u+2\lambda_P^dsv_d\nonumber\\
  \left<{\cal M}^2_{H^0}\right>_{33}=& sA_S\cos\varphi_{A_S}+2s^2\left<{\cal V}_0''\right>+A_{ud}\cos\varphi_{A_{ud}}\frac{v_uv_d}{s}\nonumber
\end{align*}
\begin{align*}
  \left<{\cal M}^2_{H^0}\right>_{44}=&\Omega_0\frac{v_d}{v_u}\ \ \ \ \ \ ;\ \ \ \ \ \ \left<{\cal M}^2_{H^0}\right>_{45}=\Omega_0\ \ \ \ \ \ ;\ \ \ \ \ \ \left<{\cal M}^2_{H^0}\right>_{55}=\Omega_0\frac{v_u}{v_d}\\
&\null\hspace{2.8cm}\Omega_0\equiv\left(A_{ud}\cos\varphi_{A_{ud}}+\lambda_P^Ms\cos\varphi_{M}\right)s-2\lambda_5\cos\varphi_5v_uv_d+\lambda_6\cos\varphi_6v_u^2+\lambda_7\cos\varphi_7v_d^2\nonumber\\
  \left<{\cal M}^2_{H^0}\right>_{46}=& \left[A_{ud}\cos\varphi_{A_{ud}}-2\lambda_P^Ms\cos\varphi_{M}+\frac{1}{s}\left(\lambda_6\cos\varphi_6v_u^2+\lambda_7\cos\varphi_7v_d^2-2\lambda_5\cos\varphi_5v_uv_d\right)\right]v_d\nonumber\\
  \left<{\cal M}^2_{H^0}\right>_{56}=& \left[A_{ud}\cos\varphi_{A_{ud}}-2\lambda_P^Ms\cos\varphi_{M}+\frac{1}{s}\left(\lambda_6\cos\varphi_6v_u^2+\lambda_7\cos\varphi_7v_d^2-2\lambda_5\cos\varphi_5v_uv_d\right)\right]v_u\nonumber\\
  \left<{\cal M}^2_{H^0}\right>_{66}=& -3sA_S\cos\varphi_{A_S}+\left(A_{ud}\cos\varphi_{A_{ud}}+4\lambda_P^Ms\cos\varphi_{M}\right)\frac{v_uv_d}{s}\nonumber
\end{align*}
\begin{align*}
  \left<{\cal M}^2_{H^0}\right>_{14}=&\left(2\lambda_6\sin\varphi_6v_u-\lambda_5\sin\varphi_5v_d\right)v_d\nonumber\\
  \left<{\cal M}^2_{H^0}\right>_{15}=&\left(2\lambda_6\sin\varphi_6v_u-\lambda_5\sin\varphi_5v_d\right)v_u\\
  \left<{\cal M}^2_{H^0}\right>_{16}=&\left[-3\lambda_P^Ms\sin\varphi_M+\frac{1}{s}\left(2\lambda_6\sin\varphi_6v_u^2-\lambda_5\sin\varphi_5v_uv_d\right)\right]v_d\nonumber\\
  \left<{\cal M}^2_{H^0}\right>_{24}=&\left(2\lambda_7\sin\varphi_7v_d-\lambda_5\sin\varphi_5v_u\right)v_d\nonumber\\
  \left<{\cal M}^2_{H^0}\right>_{25}=&\left(2\lambda_7\sin\varphi_7v_d-\lambda_5\sin\varphi_5v_u\right)v_u\nonumber\\
  \left<{\cal M}^2_{H^0}\right>_{26}=&\left[-3\lambda_P^Ms\sin\varphi_M+\frac{1}{s}\left(2\lambda_7\sin\varphi_7v_d^2-\lambda_5\sin\varphi_5v_uv_d\right)\right]v_u\nonumber\\
  \left<{\cal M}^2_{H^0}\right>_{34}=&v_d\left[\lambda_P^Ms\sin\varphi_M-\frac{v_uv_d}{s}\lambda_5\sin\varphi_5\right]\nonumber\\
  \left<{\cal M}^2_{H^0}\right>_{35}=&v_u\left[\lambda_P^Ms\sin\varphi_M-\frac{v_uv_d}{s}\lambda_5\sin\varphi_5\right]\nonumber\\
  \left<{\cal M}^2_{H^0}\right>_{36}=&4\lambda_P^M\sin\varphi_Mv_uv_d\nonumber
\end{align*}

\subsection{Trilinear Higgs couplings}\label{apsimpeffpotcoup}
\begin{align*}
 g^{S_i^0S_j^0S_k^0}= &\frac{1}{\sqrt{2}}\bigg\{\lambda_uv_u\left[\left(\Pi^S\right)^{u,u,u}_{i,j,k}+\left(\Pi^A\right)^{u,u,u}_{i,j,k}\right]
+\lambda_dv_d\left[\left(\Pi^S\right)^{d,d,d}_{i,j,k}+\left(\Pi^A\right)^{d,d,d}_{i,j,k}\right]\\
 &\null\hspace{0.3cm}+\left(\lambda_3+\lambda_4\right)\left[v_u\left(\left(\Pi^S\right)^{u,d,d}_{i,j,k}+\left(\Pi^A\right)^{u,d,d}_{i,j,k}\right)+v_d\left(\left(\Pi^S\right)^{d,u,u}_{i,j,k}+\left(\Pi^A\right)^{d,u,u}_{i,j,k}\right)\right]\\
 &\null\hspace{0.3cm}+\lambda_5\cos\varphi_5\left[v_u\left(\left(\Pi^S\right)^{u,d,d}_{i,j,k}-\left(\Pi^A\right)^{u,d,d}_{i,j,k}-2\left(\Pi^A\right)^{d,u,d}_{i,j,k}\right)+v_d\left(\left(\Pi^S\right)^{d,u,u}_{i,j,k}-\left(\Pi^A\right)^{d,u,u}_{i,j,k}-2\left(\Pi^A\right)^{u,d,u}_{i,j,k}\right)\right]\\
 &\null\hspace{0.3cm}+\lambda_5\sin\varphi_5\left[v_u\left(\left(\Pi^I\right)^{u,d,d}_{i,j,k}-\left(\Pi^P\right)^{u,d,d}_{i,j,k}-2\left(\Pi^P\right)^{d,u,d}_{i,j,k}\right)+v_d\left(\left(\Pi^I\right)^{d,u,u}_{i,j,k}-\left(\Pi^P\right)^{d,u,u}_{i,j,k}-2\left(\Pi^P\right)^{u,d,u}_{i,j,k}\right)\right]\\
 &\null\hspace{0.3cm}-\lambda_6\cos\varphi_6\left[v_u\left(3\left(\Pi^S\right)^{d,u,u}_{i,j,k}+\left(\Pi^A\right)^{d,u,u}_{i,j,k}-2\left(\Pi^A\right)^{u,u,d}_{i,j,k}\right)+v_d\left(\left(\Pi^S\right)^{u,u,u}_{i,j,k}+\left(\Pi^A\right)^{u,u,u}_{i,j,k}\right)\right]\\
 &\null\hspace{0.3cm}+\lambda_6\sin\varphi_6\left[v_u\left(3\left(\Pi^P\right)^{d,u,u}_{i,j,k}+2\left(\Pi^P\right)^{u,u,d}_{i,j,k}+\left(\Pi^I\right)^{u,u,d}_{i,j,k}\right)+v_d\left(\left(\Pi^P\right)^{u,u,u}_{i,j,k}+\left(\Pi^I\right)^{u,u,u}_{i,j,k}\right)\right]\\
 &\null\hspace{0.3cm}-\lambda_7\cos\varphi_7\left[v_u\left(\left(\Pi^S\right)^{d,d,d}_{i,j,k}+\left(\Pi^A\right)^{d,d,d}_{i,j,k}\right)+v_d\left(3\left(\Pi^S\right)^{u,d,d}_{i,j,k}+\left(\Pi^A\right)^{u,d,d}_{i,j,k}-2\left(\Pi^A\right)^{d,d,u}_{i,j,k}\right)\right]\\
 &\null\hspace{0.3cm}+\lambda_7\sin\varphi_7\left[v_u\left(\left(\Pi^P\right)^{d,d,d}_{i,j,k}+\left(\Pi^I\right)^{d,d,d}_{i,j,k}\right)+v_d\left(3\left(\Pi^P\right)^{u,d,d}_{i,j,k}+2\left(\Pi^P\right)^{d,d,u}_{i,j,k}+\left(\Pi^I\right)^{u,d,d}_{i,j,k}\right)\right]\\
 &\null\hspace{0.3cm}-A_{ud}\cos\varphi_1\left[\left(\Pi^S\right)^{s,u,d}_{i,j,k}-\left(\Pi^A\right)^{s,u,d}_{i,j,k}-\left(\Pi^A\right)^{u,s,d}_{i,j,k}-\left(\Pi^A\right)^{d,u,s}_{i,j,k}\right]\\
 &\null\hspace{0.3cm}-\lambda_P^M\cos\varphi_M\left[2s\left(\left(\Pi^S\right)^{s,u,d}_{i,j,k}-\left(\Pi^A\right)^{s,u,d}_{i,j,k}+\left(\Pi^A\right)^{u,s,d}_{i,j,k}+\left(\Pi^A\right)^{d,u,s}_{i,j,k}\right)\right.\\
 & \left.\null\hspace{2cm}+v_u\left(\left(\Pi^S\right)^{s,s,d}_{i,j,k}+2\left(\Pi^A\right)^{s,s,d}_{i,j,k}-\left(\Pi^A\right)^{d,s,s}_{i,j,k}\right)+v_d\left(\left(\Pi^S\right)^{s,s,u}_{i,j,k}+2\left(\Pi^A\right)^{s,s,u}_{i,j,k}-\left(\Pi^A\right)^{u,s,s}_{i,j,k}\right)\right]\\
 &\null\hspace{0.3cm}+\lambda_P^M\sin\varphi_M\left[s\left(3\left(\Pi^I\right)^{s,u,d}_{i,j,k}-3\left(\Pi^P\right)^{s,u,d}_{i,j,k}+\left(\Pi^P\right)^{u,s,d}_{i,j,k}+\left(\Pi^P\right)^{d,u,s}_{i,j,k}\right)-\frac{v_uv_d}{s}\left(\left(\Pi^I\right)^{s,s,s}_{i,j,k}-3\left(\Pi^P\right)^{s,s,s}_{i,j,k}\right)\right.\\
 & \left.\null\hspace{2cm}-v_u\left(\left(\Pi^I\right)^{s,s,d}_{i,j,k}+2\left(\Pi^P\right)^{s,s,d}_{i,j,k}-\left(\Pi^P\right)^{d,s,s}_{i,j,k}\right)-v_d\left(\left(\Pi^I\right)^{s,s,u}_{i,j,k}+2\left(\Pi^P\right)^{s,s,u}_{i,j,k}-\left(\Pi^P\right)^{u,s,s}_{i,j,k}\right)\right]\\
 &\null\hspace{0.3cm}+\lambda_P^u\left[s\left(\left(\Pi^S\right)^{s,u,u}_{i,j,k}+\left(\Pi^A\right)^{s,u,u}_{i,j,k}\right)+v_u\left(\left(\Pi^S\right)^{u,s,s}_{i,j,k}+\left(\Pi^A\right)^{u,s,s}_{i,j,k}\right)\right]\\
 &\null\hspace{0.3cm}+\lambda_P^d\left[s\left(\left(\Pi^S\right)^{s,d,d}_{i,j,k}+\left(\Pi^A\right)^{s,d,d}_{i,j,k}\right)+v_d\left(\left(\Pi^S\right)^{d,s,s}_{i,j,k}+\left(\Pi^A\right)^{d,s,s}_{i,j,k}\right)\right]\\
 &\null\hspace{0.3cm}+\frac{A_{S}\cos\varphi_2}{3\sqrt{2}}\left[\left(\Pi^S\right)^{s,s,s}_{i,j,k}-3\left(\Pi^A\right)^{s,s,s}_{i,j,k}\right] 
+\frac{2}{3}s^3\left<{{\cal V}_0}'''\right>\left(\Pi^S\right)^{s,s,s}_{i,j,k}+s\left<{{\cal V}_0}''\right>\left[\left(\Pi^S\right)^{s,s,s}_{i,j,k}+\left(\Pi^A\right)^{s,s,s}_{i,j,k}\right]\bigg\}
\end{align*}
\begin{align*}
 g^{S_i^0H^+_jH^-_k}= &\frac{1}{\sqrt{2}}\Big\{\lambda_uv_u X_{iu}^RX_{ju}^CX_{ku}^C+\lambda_dv_d X_{id}^RX_{jd}^CX_{kd}^C+\lambda_3\left[v_u X_{iu}^RX_{jd}^CX_{kd}^C+v_d\cos^2\beta X_{id}^RX_{ju}^CX_{ku}^C\right]\\
 &\null\hspace{0.5cm}+\frac{1}{2}\left[-(\lambda_4+\lambda_5\cos\varphi_5)\left(v_u X_{id}^R+v_d X_{iu}^R\right)+\lambda_5\sin\varphi_5\left(v_u X_{id}^I+v_d X_{iu}^I\right)\right]\left(X_{ju}^CX_{kd}^C+X_{jd}^CX_{ku}^C\right)\\
 &\null\hspace{0.5cm}+\left[\lambda_6\cos\varphi_6v_u X^R_{iu}+\lambda_7\cos\varphi_7v_dX^R_{id}\right]\left(X_{ju}^CX_{kd}^C+X_{jd}^CX_{ku}^C\right)\\
 &\null\hspace{0.5cm}+\left[-\lambda_6\cos\varphi_6\left(v_u X_{id}^R+v_d X_{iu}^R\right)+\lambda_6\sin\varphi_6\left(v_u X_{id}^I+v_d X_{iu}^I\right)\right]X_{ju}^CX_{ku}^C\\
 &\null\hspace{0.5cm}+\left[-\lambda_7\cos\varphi_7\left(v_u X_{id}^R+v_d X_{iu}^R\right)+\lambda_7\sin\varphi_7\left(v_u X_{id}^I+v_d X_{iu}^I\right)\right]X_{jd}^CX_{kd}^C\\
 &\null\hspace{0.5cm}+\left[\lambda_P^uX_{ju}^CX_{ku}^C+\lambda_P^dX_{jd}^CX_{kd}^C\right]sX_{is}^R \\
 &\null\hspace{0.5cm}+\frac{1}{2}\left[\left(A_{ud}\cos\varphi_{A_{ud}}+2\lambda_P^Ms\cos\varphi_M\right)X^R_{is}+3\lambda_P^Ms\sin\varphi_MX^I_{is}\right]\left(X_{ju}^CX_{kd}^C+X_{jd}^CX_{ku}^C\right)\\
 &\null\hspace{0.5cm}+\frac{\imath}{2}\left[(\lambda_4-\lambda_5\cos\varphi_5)\left(v_u X_{id}^I+v_d X_{iu}^I\right)-\lambda_5\sin\varphi_5\left(v_u X_{id}^R+v_d X_{iu}^R\right)\right.\\
 &\null\hspace{2cm}+2\left(\lambda_6\sin\varphi_6v_uX_{iu}^R+\lambda_7\sin\varphi_7v_dX_{id}^R\right)\\
 &\null\hspace{2cm}\left.+\left(A_{ud}\cos\varphi_{A_{ud}}-2\lambda_P^Ms\cos\varphi_M\right)X^I_{is}+\lambda_P^Ms\sin\varphi_MX^R_{is}\right]\left(X_{ju}^CX_{kd}^C-X_{jd}^CX_{ku}^C\right)\Big\}
\end{align*}
We omit the quartic couplings.

\section{\boldmath Extension to $\mathbb{Z}_3$-violating terms}\label{Z3violation}
A departure from the $\mathbb{Z}_3$-conserving NMSSM is motivated by cosmological considerations (Domain-Wall problem). Turning to the most general
singlet extension of the MSSM, the solution to the $\mu$-problem becomes less obvious, as several dimensionful parameters now enter the superpotential.
However, this difficulty is lifted when these supersymmetric masses appear as a by-product of the supersymmetry-breaking mechanism -- see e.g.\ 
\cite{Ellwanger:2008py,Lee:2011dya}. This motivates the inclusion of $\mathbb{Z}_3$-violating terms. Here we discuss how our results can be extended
to this more general version of NMSSM.

Eqs.\ref{supo} and \ref{soft} are supplemented with the following operators:
\begin{align}
 & \Delta W=\mu e^{\imath\phi_{\mu}}\hat{H}_u\cdot \hat{H}_d+\xi_F e^{\imath\varphi_{F}}\hat{S}+\frac{\mu'}{2} e^{\imath\varphi_{\mu'}}\hat{S}^2\label{supoZ3}\\
 & -\Delta{\cal L}_{\mbox{\tiny soft}}=m_3^2e^{\imath\varphi_3}H_u\cdot H_d+\xi_S  e^{\imath\varphi_{S}}S+\frac{{m_S'}^2}{2} e^{\imath\varphi_{m'}}S^2+h.c.\label{softZ3}
\end{align}
However, without loss of generality, the $\mu$-term (for instance) can be set to $0$ by means of a shift of the superfield $\hat{S}$ and a re-definition 
of the other dimensionful parameters:
\begin{align}
 &\hat{S}\leftarrow\hat{S}-\frac{\mu}{\lambda} e^{\imath(\phi_{\mu}-\varphi_{\lambda})}\hspace{4.55cm}m_3^2e^{\imath\varphi_3}\leftarrow m_3^2e^{\imath\varphi_3}-\mu A_{\lambda}e^{\imath(\varphi_1+\phi_{\mu}-\varphi_{\lambda})}\\
 &\mu' e^{\imath\varphi_{\mu'}}\leftarrow\mu' e^{\imath\varphi_{\mu'}}-2\frac{\kappa\mu}{\lambda} e^{\imath(\phi_{\mu}+\varphi_{\kappa}-\varphi_{\lambda})}\hspace{2cm}{m_S'}^2 e^{\imath\varphi_{m'}}\leftarrow\frac{{m_S'}^2}{2} e^{\imath\varphi_{m'}}-2\frac{\kappa\mu}{\lambda}A_{\kappa} e^{\imath(\varphi_{2}+\phi_{\mu}-\varphi_{\lambda})}\nonumber\\
 &\xi_F e^{\imath\varphi_{F}}\leftarrow\xi_F e^{\imath\varphi_{F}}+\frac{\kappa\mu^2}{\lambda^2} e^{\imath(2\phi_{\mu}+\varphi_{\kappa}-2\varphi_{\lambda})}-\frac{\mu'\mu}{\lambda} e^{\imath(\phi_{\mu}+\varphi_{\mu'}-\varphi_{\lambda})}\nonumber\\
 &\xi_S  e^{\imath\varphi_{S}}\leftarrow\xi_S  e^{\imath\varphi_{S}}+\frac{\kappa\mu^2}{\lambda^2}A_{\kappa} e^{\imath(2\phi_{\mu}+\varphi_{2}-2\varphi_{\lambda})}-\frac{\mu}{\lambda} {m_S'}^2e^{\imath(\phi_{\mu}+\varphi_{m'}-\varphi_{\lambda})}-\frac{\mu}{\lambda} m_S^2\,e^{\imath(\varphi_{\lambda}-\phi_{\mu})}\nonumber
\end{align}
This choice simplifies significantly the corrections to the $\mathbb{Z}_3$-conserving case and we thus use this
freedom in the rest of this appendix: $\mu\equiv0$. We then have five new complex parameters with respect to the $\mathbb{Z}_3$-conserving NMSSM. Note
that these only affect the singlino and the Higgs sector so that most of what we derived in the context of the $\mathbb{Z}_3$-conserving NMSSM remains
valid.

The singlino mass-entry in the neutralino mass-matrix -- Eq.\ref{massneu} -- is changed to $\mu'e^{\imath\varphi_{\mu'}}+2\kappa e^{\imath\varphi_{\kappa}}s$
and this is the only modification in the gaugino / higgsino sector at the order of our calculation. Concerning the neutralino loop corrections to the 
Higgs sector, our results of appendix \ref{aploophiinos} can be extended to the $\mathbb{Z}_3$-violating case by the simple substitutions:
$2\kappa e^{\imath\varphi_{\kappa}}S\to\mu'e^{\imath\varphi_{\mu'}}+2\kappa e^{\imath\varphi_{\kappa}}S$ and $4\kappa^2|S|^2\to\mu'^2+
4\mu'\kappa\mbox{Re}\left[e^{\imath(\varphi_{\kappa}-\varphi_{\mu'})}S\right]+4\kappa^2|S|^2$.

The modifications in the Higgs sector are more substantial. The tree-level Higgs potential of Eq.\ref{tlHiggspot} receives the following additions:
\begin{multline}
 \Delta{\cal V}_{H}=\left[m_3^2e^{\imath\varphi_3}+\lambda\left(\xi_Fe^{\imath(\varphi_{\lambda}-\varphi_{F})}+\mu'e^{\imath(\varphi_{\lambda}-\varphi_{\mu'})}S^*\right)\right]\left(H_u^+H_d^--H_u^0H_d^0\right)+h.c.\\
+\left[\xi_Se^{\imath\varphi_{S}}+\xi_F\mu'e^{\imath(\varphi_{\mu'}-\varphi_F)}\right]S+\left[\frac{{m_S'}^2}{2} e^{\imath\varphi_{m'}}+\kappa\xi_F e^{\imath(\varphi_{\kappa}-\varphi_{F})}\right]S^2+\kappa\mu' e^{\imath(\varphi_{\kappa}-\varphi_{\mu'})}S|S|^2+h.c.\\
+\xi_F^2+{\mu'}^2|S|^2
\end{multline}
Correspondingly, the Higgs mass-matrix elements of Eq.\ref{charhiggstreemass} and \ref{neuthiggstreemass} are modified as:
\begin{align}
 & \Delta m_{H^{\pm}}^2=\Delta\Omega\frac{v_u^2+v_d^2}{v_uv_d} \hspace{3.05cm} \Delta\Omega\equiv m_3^2\cos\varphi_3+\lambda\xi_F\cos(\varphi_{\lambda}-\varphi_F)+\lambda s\mu'\cos(\varphi_{\lambda}-\varphi_{\mu'}) \\
 & \Delta \left<{\cal M}^2_{H^{0}}\right>_{11}=\Delta\Omega\frac{v_d}{v_u} \hspace{3.1cm} \Delta \left<{\cal M}^2_{H^{0}}\right>_{12}=-\Delta\Omega \hspace{2.7cm} \Delta \left<{\cal M}^2_{H^{0}}\right>_{22}=\Delta\Omega\frac{v_u}{v_d} \nonumber\\
 & \Delta \left<{\cal M}^2_{H^{0}}\right>_{44}=\Delta\Omega\frac{v_d}{v_u} \hspace{3.1cm} \Delta \left<{\cal M}^2_{H^{0}}\right>_{45}=\Delta\Omega \hspace{2.95cm} \Delta \left<{\cal M}^2_{H^{0}}\right>_{55}=\Delta\Omega\frac{v_u}{v_d} \nonumber\\
 & \Delta \left<{\cal M}^2_{H^{0}}\right>_{13}=-\lambda v_d\mu'\cos(\varphi_{\lambda}-\varphi_{\mu'})\hspace{2.5cm} \Delta \left<{\cal M}^2_{H^{0}}\right>_{23}=-\lambda v_u\mu'\cos(\varphi_{\lambda}-\varphi_{\mu'}) \nonumber\\
 & \Delta \left<{\cal M}^2_{H^{0}}\right>_{33}=-\frac{1}{s}\left[\xi_S\cos\varphi_{S}+\xi_F\mu'\cos(\varphi_{\mu'}-\varphi_F)\right]+\mu'\left[3\kappa s\cos(\varphi_{\kappa}-\varphi_{\mu'})+\lambda\frac{v_uv_d}{s}\cos(\varphi_{\lambda}-\varphi_{\mu'})\right]\nonumber\\
 & \Delta \left<{\cal M}^2_{H^{0}}\right>_{46}=-\lambda v_d\mu'\cos(\varphi_{\lambda}-\varphi_{\mu'})\hspace{2.5cm} \Delta \left<{\cal M}^2_{H^{0}}\right>_{56}=-\lambda v_u\mu'\cos(\varphi_{\lambda}-\varphi_{\mu'}) \nonumber\\
 & \Delta \left<{\cal M}^2_{H^{0}}\right>_{66}=-\frac{1}{s}\left[\xi_S\cos\varphi_{S}+\xi_F\mu'\cos(\varphi_{\mu'}-\varphi_F)\right]+\mu'\left[-\kappa s\cos(\varphi_{\kappa}-\varphi_{\mu'})+\lambda\frac{v_uv_d}{s}\cos(\varphi_{\lambda}-\varphi_{\mu'})\right]\nonumber\\
 &\null\hspace{9cm}-2\left[{m_S'}^2\cos\varphi_{m'}+2\kappa\xi_F \cos(\varphi_{\kappa}-\varphi_{F})\right]\nonumber
\end{align}
\begin{align}
 & \Delta \left<{\cal M}^2_{H^{0}}\right>_{14\ ,\ 15\ ,\ 24\ ,\ 25}=0\nonumber\\
 & \Delta \left<{\cal M}^2_{H^{0}}\right>_{34}=-\left[m_3^2\sin\varphi_3+\lambda\xi_F\sin(\varphi_{\lambda}-\varphi_{F})\right]\frac{v_d}{s}\hspace{1cm}\Delta \left<{\cal M}^2_{H^{0}}\right>_{35}=-\left[m_3^2\sin\varphi_3+\lambda\xi_F\sin(\varphi_{\lambda}-\varphi_{F})\right]\frac{v_u}{s}\nonumber\\
 & \Delta \left<{\cal M}^2_{H^{0}}\right>_{16}=-\left[m_3^2\sin\varphi_3+2\lambda s\mu'\sin(\varphi_{\lambda}-\varphi_{\mu'})+\lambda\xi_F\sin(\varphi_{\lambda}-\varphi_{F})\right]\frac{v_d}{s}\nonumber\\
 & \Delta \left<{\cal M}^2_{H^{0}}\right>_{26}=-\left[m_3^2\sin\varphi_3+2\lambda s\mu'\sin(\varphi_{\lambda}-\varphi_{\mu'})+\lambda\xi_F\sin(\varphi_{\lambda}-\varphi_{F})\right]\frac{v_u}{s}\nonumber\\
 & \Delta \left<{\cal M}^2_{H^{0}}\right>_{36}=\frac{2}{s}\left[\xi_S\sin\varphi_{S}+\xi_F\mu'\sin(\varphi_{\mu'}-\varphi_F)\right]+{m_S'}^2\sin\varphi_{m'}+2\kappa\xi_F\sin(\varphi_{\kappa}-\varphi_{F})\nonumber\\
 &\null\hspace{7cm}+2\frac{v_uv_d}{s^2}\left[m_3^2\sin\varphi_3+\lambda \xi_F\sin(\varphi_{\lambda}-\varphi_{F})+2\lambda s\, \mu'\sin(\varphi_{\lambda}-\varphi_{\mu'})\right]\nonumber
\end{align}
Finally, the trilinear Higgs couplings of appendix \ref{apHi2Hicoup} receive the following changes:
\begin{align}
 g^{S_i^0H_j^+H_k^-}= &\frac{1}{\sqrt{2}}\left\{\left[\lambda\mu'\cos(\varphi_{\lambda}-\varphi_{\mu'})X_{is}^R+\left(\frac{1}{s}[m_3^2\sin\varphi_3+\lambda\xi_F\sin(\varphi_{\lambda}-\varphi_{F})]+2\lambda\mu'\sin(\varphi_{\lambda}-\varphi_{\mu'})\right)X_{is}^I\right]\right.\nonumber\\
 &\null\hspace{9cm}\times\left(X^{C}_{ju}X^{C}_{kd}+X^{C}_{jd}X^{C}_{ku}\right)\\
 &\null\hspace{0.5cm}\left.-\imath\left[\lambda\mu'\cos(\varphi_{\lambda}-\varphi_{\mu'})X_{is}^I+\frac{1}{s}[m_3^2\sin\varphi_3+\lambda\xi_F\sin(\varphi_{\lambda}-\varphi_{F})]\,X_{is}^R\right]\left(X^{C}_{ju}X^{C}_{kd}-X^{C}_{jd}X^{C}_{ku}\right)\right\}\nonumber\\
 g^{S_i^0S_j^0S_k^0}= &\frac{1}{\sqrt{2}}\bigg\{-\lambda\mu'\cos(\varphi_{\lambda}-\varphi_{\mu'})\left[\Pi^{S\,sud}_{ijk}-\Pi^{A\,sud}_{ijk}+\Pi^{A\,uds}_{ijk}+\Pi^{A\,dus}_{ijk}\right]+\kappa\mu'\cos(\varphi_{\kappa}-\varphi_{\mu'})\left[\Pi^{S\,sss}_{ijk}+\Pi^{A\,sss}_{ijk}\right]\nonumber\\
 &\null\hspace{0.3cm}-\frac{1}{s}\left(m_3^2\sin\varphi_3+\lambda\xi_F\sin(\varphi_{\lambda}-\varphi_{F})\right)\left[\Pi^{P\,sud}_{ijk}+\Pi^{P\,uds}_{ijk}+\Pi^{P\,dus}_{ijk}-\Pi^{I\,sud}_{ijk}+\frac{v_uv_d}{3s^2}\left[3\Pi^{P\,sss}_{ijk}-\Pi^{I\,sss}_{ijk}\right]\right]\nonumber\\
 &\null\hspace{0.3cm}-2\lambda\mu'\sin(\varphi_{\lambda}-\varphi_{\mu'})\left[\Pi^{P\,sud}_{ijk}+\frac{v_uv_d}{3s^2}\left[\Pi^{I\,sss}_{ijk}-3\Pi^{P\,sss}_{ijk}\right]\right]-\frac{4}{3}\kappa\mu'\sin(\varphi_{\kappa}-\varphi_{\mu'})\Pi^{I\,sss}_{ijk}\nonumber\\
 &\null\hspace{0.3cm}-\frac{1}{3s}\left(\frac{1}{s}[\xi_S\sin\varphi_S+\xi_F\mu'\sin(\varphi_{\mu'}-\varphi_{F})]+{m_S'}^2\sin\varphi_{m'}\right)\left[\Pi^{I\,sss}_{ijk}-3\Pi^{P\,sss}_{ijk}\right]\bigg\}\nonumber
\end{align}
Up to these modifications, the calculation of Higgs loop corrections to the Higgs masses -- see Appendix \ref{aploophihiggs} ({\em iii, iv\,}) -- remains 
valid. On the other hand, reconstructing the contributions to the effective potential -- i.e.\ the effective couplings -- in the same fashion as in
the $\mathbb{Z}_3$-conserving case -- see Appendix \ref{aploophihiggs} ({\em v\,}) -- becomes problematic: indeed, no argument of symmetry allows to
reduce the number of terms in the generic potential for two Higgs doublets and a singlet and the latter contains too large a number of parameters
in view of the $16$ independent mass entries. Refer to \cite{Chalons:2012qe} for a discussion concerning the generic Higgs potential. It is possible
to use an expansion of the Higgs mass matrices in terms of the doublet fields, however.

\end{document}